\documentclass{article}
\pdfoutput=1
\usepackage[top=2cm,bottom=3cm,left=2cm,right=2cm]{geometry}
\usepackage{amsmath}
\usepackage{amsfonts}
\usepackage{stmaryrd}
\usepackage{bm}
\usepackage{hyperref}
\usepackage{upgreek}
\usepackage[mathscr]{euscript}
\usepackage{graphicx}
\usepackage{mathtools}
\usepackage{soul}

\newcommand{\inverttriangle}{%
               \mathrel{\raisebox{.1em}{%
               \reflectbox{\rotatebox[origin=c]{180}{$\triangle$}}}}\!}
               
\numberwithin{equation}{section}
\numberwithin{figure}{section}

\def\eq#1{(\ref{eq:#1})}
\def\lineup{\!\!\!\!\!\!\!\!\!\!&&}

\def\d{\partial}
\def\eps{\epsilon}
\def\Vspace{\phantom{\bigg(}}

\def\deg{\mathrm{deg}}
\def\rank{\mathrm{rank}}

\def\Q{{\bf Q}}
\def\n{{\bm \upeta}}
\def\m{{\bf m}}

\def\D{{\bf D}}
\def\C{{\bf C}}
\def\U{{\bf \hat{U}}}
\def\del{{\bm \partial}}
\def\llambda{{\bm \uplambda}}

\def\M{{\bf M}}
\def\G{{\bf \hat{G}}}

\def\N{{\bf N}}
\def\F{{\bf \hat{F}}}
\def\g{{\bf \hat{g}}}
\def\mmu{\bm{\upmu}}
\def\b{{\bf b}}
\def\c{{\bf c}}
\def\J{{\bf \hat{J}}}

\def\s{{\bf s}}
\def\Ssigma{\bm{\Sigma}}

\def\PsiR{{\Psi_\mathrm{R}}}
\def\PsiN{{\Psi_\mathrm{NS}}}

\def\PhiR{{\Phi_\mathrm{R}}}
\def\PhiN{{\Phi_\mathrm{NS}}}

\def\H{\mathcal{H}}

\def\[{\big[}
\def\]{\big]}

\begin{document}

\begin{titlepage}
\rightline\today

\begin{center}
\vskip 3.5cm

{\large \bf{Superstring Field Theory and \\ the Wess-Zumino-Witten Action}}

\vskip 1.0cm

{\large {Theodore Erler}}

\vskip 1.0cm

{\it Institute of Physics of the AS{\v C}R, v.v.i.}\\
{\it Na Slovance 2, 182 21 Prague 8, Czech Republic}\\

tchovi@gmail.com\\
\vspace{.5cm}

\vskip 2.0cm

{\bf Abstract}

\end{center}

We describe a notion of ``higher" Wess-Zumino-Witten-like action which is natural in the context of superstring field theories formulated in the large Hilbert space. For the open string, the action is characterized by a pair of commuting cyclic $A_\infty$ structures together with a hierarchy of higher-form potentials analogous to the Maurer-Cartan elements which appear in the conventional Wess-Zumino-Witten action. We apply this formalism to get a better understanding of symmetries of open superstring field theory and the structure of interactions in the Ramond sector, describing an interesting connection between Ramond vertices and Feynman diagrams.

\end{titlepage}

\tableofcontents

\section{Introduction}

One of the most successful formulations of open superstring field theory is due to Berkovits \cite{Berkovits}, and is characterized by a string field $\PhiN$ in the {\it large} Hilbert space whose dynamics is governed by a Wess-Zumino-Witten-like (WZW-like) action. In the string field theory literature, the action is often written in a nonstandard but condensed form~\cite{heterotic}:
\begin{equation}
S = -\int_0^1 dt\, \big\langle A_t,Q A_\eta\big\rangle_L,\label{eq:BerkAct}
\end{equation}
where $\eta$ is the zero mode of the eta ghost, $Q$ is the BRST operator, $A_t$ and $A_\eta$ are Maurer-Cartan elements
\begin{eqnarray}
A_t \lineup \equiv\left(\frac{d}{dt}e^{\PhiN(t)}\right)e^{-\PhiN(t)},\\
A_\eta \lineup \equiv \Big(\eta e^{\PhiN(t)}\Big)e^{-\PhiN(t)},
\end{eqnarray}
and $\PhiN=\PhiN(t)|_{t=1}$ is the NS dynamical string field in the large Hilbert space at ghost and picture number~0. All string fields in the action are multiplied with Witten's associative star product, and $\langle,\rangle_L$ is the BPZ inner product in the large Hilbert space. Recently there has been substantial interest in a different kind of string field theory based on a dynamical field in the {\it small} Hilbert space. Such theories are easier to quantize, and provide a useful framework for taming infrared divergences in superstring perturbation theory \cite{SenRev}. Nevertheless, the Berkovits formulation continues to be intriguing. In many ways, the theory has a more nonperturbative flavor: The action can be expressed in closed form, and the classical equations of motion are amenable to nonperturbative analytic solution \cite{supervac}. The formulation is also more general, in the sense that small Hilbert space string field theories can be derived from partial gauge fixing \cite{INOT,OkWB,WB}. 

However, the Wess-Zumino-Witten action does not appear to have a sufficiently flexible algebraic structure to describe more general superstring field theories in the large Hilbert space. This was already apparent in the formulation of heterotic string field theory \cite{heterotic}, where the action must be generalized in a form where $Q$ and $\eta$ play assymmetrical roles, unlike the holomorphic and antiholomorphic exterior derivatives of the conventional WZW action. An even more puzzling example is the Ramond sector of open string field theory \cite{complete}, where a plausible WZW-like formulation has not been found. With this problem in mind, in this paper we develop a generalized WZW-like action which is sufficient to formulate all known superstring field theories in the large Hilbert space.\footnote{In Type II superstring field theories, it is natural to consider a ``doubly large" Hilbert space which includes $\xi$ zero modes in both the left- and right-moving sectors. This would require further generalization of the action we describe here. Some steps towards such a formulation can be found in \cite{MatsunagaIimori,MatsunagaII}.} The action is characterized by two mutually commuting cyclic $A_\infty$ structures, which we may represent by coderivations $\D$ and~$\C$:
\begin{equation}
\D^2 = \C^2 = [\D,\C]=0.
\end{equation}
The coderivation $\D$ may be interpreted as a nonlinear extension of the BRST operator $Q$, and is responsible for the dynamics of the string field. Therefore $\D$ will be called the {\it dynamical $A_\infty$ structure}. The coderivation $\C$ may be interpreted as a nonlinear extension of the eta zero mode $\eta$. This generalizes the small Hilbert space constraint on the string field $\eta\Psi = 0$. Therefore $\C$ will be called the {\it constraint $A_\infty$ structure}. The action is expressed
\begin{equation}S = \int_0^1 dt\,\omega\left(A_t,\pi_1\D\frac{1}{1- A_C}\right),\label{eq:genWZW}\end{equation}
where $A_t$ and $ A_C$ are functionals of the dynamical field generalizing the Maurer-Cartan elements that appear in the conventional WZW action. Following the notation of \cite{WB}, $\omega$ is a symplectic form on the state space, $\pi_n$ is the projection onto the $n$-string component of the tensor algebra, and $1/(1-A_C)$ is the group-like element generated by taking sums of tensor products of $A_C$. Particularly important are an infinite hierarchy of identities satisfied by $A_t,A_C$ and analogous string fields to be described later. For example,  $A_C$ satisfies 
\begin{equation}
\C\frac{1}{1-A_C} = 0.
\end{equation}
This generalizes an identity satisfied by the Maurer-Cartan element $A_\eta$ in the Berkovits theory:
\begin{equation}
\eta A_\eta - A_\eta^2 = 0.
\end{equation} 
The primary application we have in mind is to get a clearer understanding of the structure of Ramond sector vertices in 
open superstring field theory. The formulation of the WZW-like action in the Ramond sector leads to an interesting connection to the decomposition model of an $A_\infty$ algebra \cite{Kontsevich1,Kajiura}, which gives a potentially useful Feynman-diagram interpretation of the Ramond sector vertices. 

To give the cleanest presentation of the algebraic structure of the theory, we will make extensive use of the coalgebra representation of $A_\infty$ algebras. We will not review the formalism here, which has played an important role in much recent work. See especially \cite{WB}, whose notation and conventions we generally follow. 

This paper is organized as follows. In section \ref{sec:WZW} we motivate the form of the action by considering two examples: the WZW-like action of Berkovits NS open superstring field theory, and a more nonstandard WZW-like action which turns out to be equivalent to NS open superstring field theory formulated in the small Hilbert space. Having gone through these examples we consider the general form of the action. It is characterized by a pair of commuting cyclic $A_\infty$ structures---the {\it dynamical} and {\it constraint} $A_\infty$ structures---together with a hierarchy of higher-form potentials satisfying so-called {\it flatness conditions}. We then discuss symmetries of the theory, which come in three kinds: Symmetry under cyclic $A_\infty$ isomorphism, symmetries that transform the hierarchy of potentials, and ``true" symmetries which transform the dynamical field. The true symmetries include in particular gauge symmetry, and we motivate its general form. In section \ref{sec:KO} we apply the general formalism to formulate a WZW-like action for open superstring field theory including the Ramond sector. Though the construction is in principle independent, it will quickly become clear that the action is in fact equivalent to the one proposed by Kunitomo and Okawa. We start with the naive $A_\infty$ structures defined by the simplest form of the Neveu-Schwarz and Ramond field equations. Noting that these $A_\infty$ structures do not define cyclic Ramond vertices, we ``correct" them by constructing an $A_\infty$ isomorphism $\F$ which maps to new dynamical and constraint $A_\infty$ structures which are cyclic. The most natural choice of this map has the effect of transforming the Ramond sector into the small Hilbert space. We note that the construction of $\F$ has a structural similarity with the decomposition model of an $A_\infty$ algebra. This leads to a remarkably simple characterization of the dynamical $A_\infty$ structure in terms of a Feynman graph expansion, given by the Witten vertices connected by ``propagators" in the form of $\xi$-ghost insertions. Next we describe a systematic procedure for deriving potentials for the WZW-like action consistent with the flatness conditions. We apply this procedure to find explicit expressions for the potentials which are most relevant for open superstring field theory. We describe the form of the gauge transformation, and show that it is equivalent to the gauge transformation derived by Kunitomo and Okawa. Next we show by explicit calculation that the WZW-like action is equivalent to  the action as written by Kunitomo and Okawa. We also consider the relation to open superstring field theory formulated in the small Hilbert space. The $A_\infty$ structure of the small Hilbert space was also derived by an $A_\infty$ isomorphism mapping to the simplest form of the field equations, though the details of that derivation were rather different. Our results imply that this $A_\infty$ isomorphism can be factorized: the first factor consists of $\F$, and defines Ramond vertices consistent with cyclicity; the second factor, denoted $\g$, has the effect of transforming the NS sector into the small Hilbert space. In section \ref{sec:susy} we describe the realization of supersymmetry. The supersymmetry transformation is characterized by a collection of string products which, like the dynamical $A_\infty$ structure, have a simple interpretation in terms of a Feynman graph expansion. We derive an expression for the potential representing supersymmetry, and relate this description to the supersymmetry transformation found by Kunitomo. As a physical application we show that transverse displacement of the reference D-brane, represented as an analytic solution of the field equations, preserves all supersymmetries of the reference D-brane. We end with some concluding remarks, followed by appendices containing detailed derivations of results referred to in the text.

\section{WZW-like Action}
\label{sec:WZW}

Formulating an action principle for open superstring field theory can be complicated. However, the field equations are simple: 
\begin{equation}(Q-\eta)\varphi +\varphi^2 = 0.\label{eq:field_equation}\end{equation}
Different formulations of open superstring field theory can be viewed as alternative approaches to deriving these equations from an action. The string field $\varphi$ is Grassmann odd and carries ghost number 1, but the picture of $\varphi$ may vary depending on the approach.\footnote{In the democratic superstring field theory \cite{democratic}, $\varphi$ carries all pictures.} For present purposes we consider NS open superstring field theory with $\varphi$ at picture~$-1$. The field equation then breaks up into a piece at picture $-1$ and a piece at picture $-2$:
\begin{eqnarray}
Q\varphi \lineup = 0,\label{eq:BerkEOM}\\
\eta\varphi- \varphi^2 \lineup = 0.\label{eq:Berkconst}
\end{eqnarray}
Since $Q$ contains the d'Alembertian operator, the first equation can be viewed as an equation of motion. The second equation can be interpreted as a nonlinear constraint. For example, at the linearized level the second equation is precisely the constraint that $\varphi$ should be in the small Hilbert space. In Berkovits' NS open superstring field theory, the constraint is solved by postulating that $\varphi$ takes the ``pure gauge" form
\begin{equation}
\varphi=  \Big(\eta e^{\PhiN}\Big)e^{-\PhiN},
\end{equation}
where $\PhiN$ is the NS dynamical string field. Substituting into the dynamical field equation gives 
\begin{equation}
Q \left(\Big(\eta e^{\PhiN}\Big)e^{-\PhiN}\right)=0.
\end{equation}
These are the equations of motion of Berkovits' NS open superstring field theory. 

To motivate the generalization we will consider, it is helpful to express these equations in the coalgebra formalism. The dynamical field equation and constraint take the form:
\begin{eqnarray}
\Q\frac{1}{1-\varphi} \lineup = 0,\label{eq:Bdyn}\\
(\n-\m_2)\frac{1}{1-\varphi}\lineup =0.\label{eq:Bconst}
\end{eqnarray}
Recall that the coalgebra formalism uses a shifted grading on the open string state space called {\it degree}. The degree of a string field $A$, denoted $\deg(A)$ is defined to be its Grassmann parity plus one (mod $\mathbb{Z}_2$). Above $\Q,\n$ are the coderivations corresponding to the BRST operator and eta zero mode, $\m_2$ is the coderivation corresponding to the open string star product $m_2(A,B) \equiv (-1)^{\deg(A)}A*B$,\footnote{We often drop the star when writing the open string star product: $AB=A*B$. The product $m_2(A,B)$ differs from $A*B$ by a sign from the shift in grading, but we nevertheless refer to it as the star product as confusion should not arise. All commutators in this paper are graded with respect to degree, except for commutators of string fields computed with the open string star product, which are graded with respect to Grassmann parity.} and $1/(1-\varphi)$ is the group-like element of the tensor algebra generated by the degree even string field $\varphi$. The important point is that the equations of motion and constraint may be characterized by two mutually commuting cyclic $A_\infty$ structures, $\Q$ and $\n-\m_2$:
\begin{equation}
\Q^2 = 0,\ \ \ \ (\n-\m_2)^2= 0,\ \ \ \ \[\Q,\n-\m_2]= 0.
\end{equation}
These equations follow from the fact that $Q$ and $\eta$ are nilpotent and mutually commute, are derivations of the star product, and that the star product is associative. 

From here it seems natural to consider a generalization where $\Q$ and $\n-\m_2$ are replaced by more general cyclic $A_\infty$ structures $\D$ and $\C$: 
\begin{eqnarray}
\D\frac{1}{1-\varphi} \lineup = 0,\label{eq:genEOM}\\
\C\frac{1}{1-\varphi}\lineup =0,\label{eq:genconst}
\end{eqnarray}
where $\D$ and $\C$ satisfy 
\begin{equation}\D^2=0,\ \ \ \ \C^2 = 0,\ \ \ \ [\C,\D] = 0.\end{equation}
We will interpret $\D$ as defining the dynamical field equation, while $\C$ defines the constraint.  We refer to $\D$ as the {\it dynamical} $A_\infty$ structure and $\C$ at the {\it constraint} $A_\infty$ structure. We can equivalently write \eq{genEOM} and \eq{genconst} as
\begin{eqnarray}
D_1\varphi + D_2(\varphi,\varphi) + D_3(\varphi,\varphi,\varphi) +...\lineup = 0,\\
C_1\varphi + C_2(\varphi,\varphi) + C_3(\varphi,\varphi,\varphi) +... \lineup = 0,
\end{eqnarray}
where $D_n$ and $C_n$ are the multi-string products contained in $\D$ and $\C$, respectively. The main goal in what follows is to derive these equations from a generalized form of the Wess-Zumino-Witten action. As before, $\varphi$ will be expressed as a function of the dynamical field in such a way that the constraint \eq{genconst} is satisfied automatically. Then the equation of motion \eq{genEOM} will appear as the Euler-Lagrange equation derived from the action. To formulate the WZW-like action it is necessary to assume that $\D$ and $\C$ are cyclic with respect to the appropriate symplectic form. This condition is responsible for most complications in generalizing the action to the Ramond sector. 

\subsection{Berkovits Theory}
\label{subsec:BerkWZW}

Here we review Berkovits' NS open superstring field theory with an eye towards generalization of the WZW-like action. An important ingredient in the theory is the definition of certain functionals of the dynamical field called {\it potentials}. In Berkovits' NS open superstring field theory, the potentials are Maurer-Cartan elements
\begin{eqnarray}
A_\eta\lineup \equiv\Big(\eta e^{\PhiN(t)}\Big)e^{-\PhiN(t)},\label{eq:Aeta}\\
A_Q\lineup \equiv\Big(Q e^{\PhiN(t)}\Big)e^{-\PhiN(t)},\\
A_t\lineup \equiv\left(\frac{d}{dt} e^{\PhiN(t)}\right)e^{-\PhiN(t)},\label{eq:At}\\
A_\delta\lineup \equiv\Big(\delta e^{\PhiN(t)}\Big)e^{-\PhiN(t)},\label{eq:Adelta}
\end{eqnarray}
where $t\in[0,1]$ is the auxiliary ``third dimension" which appears in the WZW action and $\delta$ denotes a generic variation of the dynamical string field $\PhiN$. The string field $\PhiN(t)$ is a function of $t$ and $\PhiN$,
\begin{equation}\PhiN(t) = f(t,\PhiN),\label{eq:Berkint}\end{equation}
subject to the boundary conditions
\begin{equation}\PhiN(0) = 0,\ \ \ \PhiN(1)=\PhiN.\label{eq:intbc}\end{equation}
We call $\PhiN(t)$ an {\it interpolation} of the dynamical string field $\PhiN$. With these ingredients, the WZW-like action is written:
\begin{equation}
S = -\frac{1}{2}\int_0^1 \left(\frac{d}{dt}\langle A_\eta,A_Q\rangle_L + \langle A_t,[A_Q,A_\eta]\rangle_L\right).
\end{equation}
We would like to compute the variation of this action. A pivotal part of the following development is understanding the ``correct" way to do this: It is best to compute the variation by understanding the algebraic properties of the potentials, rather than expanding the action explicitly in terms of $\PhiN$ \cite{heterotic}. Note that by virtue of the ``pure gauge" form of the potentials, the components of the field strength vanish identically: 
\begin{eqnarray}
F_{\eta\eta}\lineup =2(\eta A_\eta - A_\eta^2) = 0,  \ \ \ \ \ \ \ \ \ \ \ \ \ \ \ \ \ \ \ \ \ \ \ \ \ \ \ \ F_{QQ} =2(QA_Q -A_Q^2) = 0,\nonumber\\
F_{\eta Q}\lineup = \eta A_Q + Q A_\eta - [A_\eta,A_Q] = 0,\ \ \ \ \ \ \ \ \ \ \ \ \ \ \ F_{Qt}= QA_t - \frac{d}{dt}A_Q -[A_Q,A_t] = 0,\nonumber\\
F_{\eta t}\lineup = \eta A_t -\frac{d}{dt}A_\eta -[A_\eta,A_t] = 0, \ \ \ \ \ \ \ \ \ \ \ \ \ \ \ \  F_{Q\delta}=QA_\delta - \delta A_Q - [A_Q,A_\delta] = 0,\nonumber\\
F_{\eta\delta}\lineup = \eta A_\delta - \delta A_\eta -[A_\eta,A_\delta] = 0,\ \ \ \ \ \ \ \ \ \ \ \ \ \ \ \ \ \ F_{t\delta} = \frac{d}{dt}A_\delta - \delta A_t - [A_t,A_\delta] = 0.\label{eq:flat0}
\end{eqnarray}
We call these {\it flatness conditions}. The flatness conditions can be described efficiently by introducing basis 1-forms
\begin{equation}
dx^\eta,\ \ dx^Q,\ \ dt,\ \ dx^\delta
\end{equation}
whose Grassmann parity is opposite that of $\eta,Q,d/dt$ and $\delta$, respectively. The basis 1-forms commute through operators and string fields with a sign given by Grassmann parity. Defining an exterior derivative $d$ and a 1-form potential $A$,
\begin{eqnarray}
d \lineup = dx^\eta \eta+ dx^Q Q + dt\frac{d}{dt}+dx^\delta \delta,\\
A \lineup = dx^\eta A_\eta+ dx^Q A_Q + dt A_t +dx^\delta A_\delta,
\end{eqnarray}
we may write 
\begin{equation}dA - A^2 = 0.\label{eq:flat1}\end{equation}
Using the flatness conditions it is possible to express the variation of the action in terms of $A_\delta$, without requiring specific knowledge of how the potentials depend on $\PhiN$. With some algebra \cite{heterotic}, the result is 
\begin{equation}\delta S = \langle A_\delta,Q A_\eta\rangle_L|_{t=1}.\end{equation}
Note that $A_\delta|_{t=1} = \delta\PhiN$ at leading order in $\PhiN$. Therefore, for sufficiently small $\PhiN$, an arbitrary variation $\delta\PhiN$ can produce an arbitrary value of $A_\delta|_{t=1}$. Since the BPZ inner product is nondegenerate, setting $\delta S=0$ therefore implies
\begin{equation}QA_\eta|_{t=1} = 0.\label{eq:BerkEOM2}\end{equation}
These are the equations of motion of Berkovits' NS open superstring field theory. Returning to the beginning, recall the equation of motion and constraint:
\begin{equation}Q\varphi = 0,\ \ \ \eta \varphi - \varphi^2=0.\end{equation}
If we identify
\begin{equation}\varphi = A_\eta|_{t=1}\end{equation}
then $Q\varphi=0$ is the Berkovits equations of motion, and $\eta\varphi-\varphi^2=0$ is solved identically by virtue of the fact that $A_\eta$ satisfies flatness conditions.

This story works well for the NS open superstring, but does not directly generalize to other forms of superstring field theory. An important insight following the construction of heterotic string field theory \cite{heterotic} is that the WZW-like action can be written in an alternative form:
\begin{equation}
S = -\int_0^1 dt\, \langle A_t,QA_\eta\rangle_L.\label{eq:Bact2}
\end{equation}
This expression for the action has an interesting property: The equations of motion \eq{BerkEOM2} follow even if some field strengths do not vanish. Namely, the components without an index $\eta$,
\begin{equation}F_{QQ},\ \ F_{Qt},\ \ F_{Q\delta},\ \ F_{t\delta},\label{eq:Fnon0}\end{equation}
can be nonzero. Following \cite{WBlarge}, this suggests a rather dramatic reformulation of the flatness conditions. The key idea is to regard $\eta$ and $A_\eta$ as scalar quantities, so that the basis 1-form $dx^\eta$ is absent. The exterior derivative now only contains $dx^Q,dt$ and $dx^\delta$: 
\begin{equation}
d = dx^Q Q+dt\frac{d}{dt}+dx^\delta \delta.
\end{equation}
The scalar $A_\eta$ is called the {\it 0-potential}, while $A_Q,A_t,A_\delta$ are components of a 1-form called the {\it 1-potential}:
\begin{eqnarray}
A^{(0)}\lineup \equiv A_\eta,\\
A^{(1)}\lineup \equiv dx^Q A_Q+dt A_t+dx^\delta A_\delta.
\end{eqnarray}
The formal sum 
\begin{equation}A = A^{(0)}+A^{(1)}\end{equation}
will be called the {\it multi-form potential}. The flatness conditions can be reexpressed
\begin{equation}
(\eta+d)A - A^2 = 0.\label{eq:flat2}
\end{equation}
It is easy to confirm that this is equivalent to the earlier formulation of the flatness conditions \eq{flat1}. The linear independence of $dx^\eta$ has simply been traded for the linear independence of 0-forms and 1-forms.  However, now it is easy see how to generalize when the field strengths \eq{Fnon0} are nonzero: We simply add a 2-form component $A^{(2)}$, called the {\it  2-potential}, to the multiform potential $A$. The flatness conditions \eq{flat2} then imply
\begin{equation} dA^{(1)} -(A^{(1)})^2=-\eta A^{(2)} + [A^{(0)},A^{(2)}].
\end{equation}
The left hand side defines the nonvanishing field strengths \eq{Fnon0}. The 2-potential can be chosen so that the flatness conditions are still obeyed. Continuing, it is clear that we will in general need to introduce a 3-potential $A^{(3)}$, a 4-potential $A^{(4)}$, and so on, so that the multi-form potential $A$ contains components of all form degrees. Since $dx^Q$ is Grassmann even, there is no top degree form. The higher potentials can be set to zero if $A_\eta,A_Q,A_t$ and $A_\delta$ are chosen to be Maurer-Cartan elements. However, the WZW-like action as expressed in \eq{Bact2} in principle allows for a more general choice of potentials.

For later development, it is helpful to rephrase these ideas in the coalgebra formalism. We consider the tensor algebra of string-field-valued differential forms. The {\it degree} of a string-field-valued differential form is given by its Grassmann parity plus one (mod $\mathbb{Z}_2$). We can rewrite the flatness conditions \eq{flat2} in the form
\begin{equation}\del \frac{1}{1-A} = 0,\label{eq:coflat}\end{equation}
where
\begin{equation}\frac{1}{1-A} = 1\, +\,  A\, +\, A\otimes A + A\otimes A\otimes A+ ...\end{equation}
is the group-like element derived from the multiform potential $A$, and $\del$ is a coderivation
\begin{equation}
\del \equiv \n - \m_2 +dx^Q \Q +dt \bm{ \frac{d}{dt}} + dx^\delta {\bm \delta}.\label{eq:multiBerk}
\end{equation}
with $\bm{d/dt}$ is the coderivation corresponding to $d/dt$ and $\bm{\delta}$ is the coderivation corresponding to $\delta$. It is important to note that $\del$ is nilponent:
\begin{equation}\del^2 = 0.\end{equation}
Therefore $\del$ defines an $A_\infty$ algebra. We refer to $\del$ as the {\it multiform $A_\infty$ structure}. The zero-form part of $\del$ must define an $A_\infty$ algebra on its own. This is the {\it constraint} $A_\infty$ structure:
\begin{equation}\C = \n-\m_2.\end{equation}
Meanwhile, the coderivation $\Q$ which appears in the equations of motion  
\begin{equation}
\left.\Q\frac{1}{1-A_\eta}\right|_{t=1}=0
\end{equation}
defines the {\it dynamical} $A_\infty$ structure:
\begin{equation}
\D = \Q.
\end{equation}
When using the degree grading it is natural to express the conformal field theory inner product as a symplectic form:
\begin{equation}\omega_L(A,B) \equiv (-1)^{\deg(A)}\langle A,B\rangle_L,\end{equation}
where the subscript $L$ indicates that the symplectic form is defined contracting states in the large Hilbert space. It is graded antisymmetric,
\begin{equation}\omega_L(A,B) = -(-1)^{\deg(A)\deg(B)}\omega_L(B,A),\end{equation}
and nondegenerate. The BRST operator, eta zero mode, and open string star product are {\it cyclic} with respect to $\omega_L$ in the sense that the following properties hold:
\begin{eqnarray}
0\lineup = \omega_L(A,QB) +(-1)^{\deg(A)}\omega_L(QA,B), \\
0\lineup = \omega_L(A,\eta B) +(-1)^{\deg(A)}\omega_L(\eta A,B), \\
0\lineup =\omega_L(A,m_2(B,C)) +(-1)^{\deg(A)}\omega_L(m_2(A,B),C).
\end{eqnarray}
We sometimes write the symplectic form as a ``double bra" state:
\begin{equation}\langle\omega_L| A\otimes B = \omega_L(A,B).\end{equation}
With this we can express cyclicity in an abbreviated form: 
\begin{eqnarray}
0\lineup = \langle \omega_L|\pi_2 \Q,\\
0\lineup = \langle\omega_L|\pi_2 \n,\\
0\lineup = \langle \omega_L|\pi_2\m_2.\label{eq:m2cyc}
\end{eqnarray}
In particular, this implies that dynamical and constraint $A_\infty$ structures of the Berkovits theory are cyclic with respect to $\omega_L$:
\begin{eqnarray}
0\lineup = \langle\omega_L|\pi_2\D,\\
0\lineup = \langle\omega_L|\pi_2\C.
\end{eqnarray}
The WZW-like action can be expressed: 
\begin{equation}
S = \int_0^1 dt\,\omega_L\left(A_t,\pi_1 \Q\frac{1}{1-A_\eta}\right).\label{eq:coBerk}
\end{equation}
All formulas in this paragraph work equally well if we replace $\Q$ and $\n-\m_2$ with more general $A_\infty$ structures $\D$ and $\C$, and we replace $\omega_L$ with any symplectic form $\omega$ with respect to which $\D$ and $\C$ are cyclic. This is the generalization of the WZW-like action we are after. 

\subsection{NS Open Superstring Field Theory in the Small Hilbert Space}
\label{subsec:liftsmall}

A more unusual example of a WZW-like action is given by starting with the action for open superstring field theory in the small Hilbert space,
\begin{equation}S=\frac{1}{2}\omega_S(\PsiN,Q\PsiN)+\frac{1}{3}\omega_S(\PsiN,M_2(\PsiN,\PsiN)) + \frac{1}{4}\omega_S(\PsiN,M_3(\PsiN,\PsiN,\PsiN))+...,\label{eq:smallaction}\end{equation}
and replacing $\PsiN$ with a dynamical string field $\PhiN$ in the large Hilbert space through the substitution \cite{WBlarge}
\begin{equation}\PsiN = \eta\PhiN.\end{equation}
Here $\PsiN$ is a degree even NS string field at ghost number $1$ and picture $-1$ in the small Hilbert space. The object $\omega_S$ is a symplectic form defined between states in the small Hilbert space. It is related to $\omega_L$ through 
\begin{equation}\omega_S(a,b)=\omega_L(A,b),\label{eq:wS}\end{equation}
where $a,b$ are in the small Hilbert space and $a=\eta A$. The small Hilbert space symplectic form is graded antisymmetric and nondegenerate. The products $M_{n+1}$ form a cyclic $A_\infty$ algebra, which means that the coderivation
\begin{equation}\M= \Q+\M_2+\M_3+...\end{equation}
satisfies
\begin{eqnarray}
\M^2\lineup =0,\\
\langle\omega_S|\pi_2\M\lineup = 0.
\end{eqnarray}
The products are also cyclic with respect to $\omega_L$ when embedded in the natural way into the large Hilbert space. An explicit construction of $\M$ is given in \cite{WittenSS,ClosedSS}, but the details are not important at the moment. One simple but important condition is that the products $M_{n+1}$ must multiply inside the small Hilbert space. This is implied~by 
\begin{equation}[\n,\M]=0.\end{equation}
Also note that
\begin{eqnarray}
\n^2\lineup =0,\\
\langle\omega_L|\pi_2\n\lineup = 0.
\end{eqnarray}
Therefore $\n$ and $\M$ define mutually commuting cyclic $A_\infty$ structures. It is natural to interpret $\n$ and $\M$ as constraint and dynamical $A_\infty$ structures of a WZW-like action. 

We can derive the WZW-like action as follows. Substituting $\PsiN=\eta\PhiN$ into the action gives
\begin{equation}S=\frac{1}{2}\omega_S(\eta\PhiN,Q\eta\PhiN)+\frac{1}{3}\omega_S(\eta\PhiN,M_2(\eta\PhiN,\eta\PhiN)) + \frac{1}{4}\omega_S(\eta\PhiN,M_3(\eta\PhiN,\eta\PhiN,\eta\PhiN))+...\ .\end{equation}
Next we introduce an interpolation $\PhiN(t)$,
\begin{equation}\PhiN(t) = f(t,\PhiN),\ \ \ \ \ \ \PhiN(0)=0,\ \ \ \PhiN(1)=\PhiN,\end{equation}
and write the action as the integral of a total derivative with respect to $t$:
\begin{eqnarray}
S \lineup = \int_0^1dt\,\frac{d}{dt}\left[\frac{1}{2}\omega_S(\eta\PhiN(t),Q\eta\PhiN(t))+\frac{1}{3}\omega_S(\eta\PhiN(t),M_2(\eta\PhiN(t),\eta\PhiN(t))) +...\right]\nonumber\\
\lineup = \int_0^1dt\,\Big[\omega_S(\eta\dot{\Phi}_\mathrm{NS}(t),Q\eta\PhiN(t)) + \omega_S(\eta\dot{\Phi}_\mathrm{NS}(t),M_2(\eta\PhiN(t),\eta\PhiN(t))) +...\Big]\nonumber\\
\lineup = \int_0^1 dt\, \omega_S\left(\eta\dot{\Phi}_\mathrm{NS}(t),\pi_1\M\frac{1}{1-\eta\PhiN(t)}\right).
\end{eqnarray}
The dot denotes the derivative with respect to $t$. In the second step, we used cyclicity of the products to place $\eta\dot{\Phi}_\mathrm{NS}(t)$ in the first entry of the symplectic form. Next we replace $\omega_S$ with $\omega_L$ using \eq{wS}: 
\begin{equation}S= \int_0^1 dt\, \omega_L\left(\dot{\Phi}_\mathrm{NS}(t),\pi_1\M\frac{1}{1-\eta\PhiN(t)}\right).\end{equation}
We define the potentials 
\begin{equation}A_\eta \equiv \eta\PhiN(t),\ \ \ A_t \equiv \dot{\Phi}_\mathrm{NS}(t).\label{eq:anat}\end{equation}
In particular
\begin{equation}A_\eta|_{t=1}=\PsiN\end{equation}
is the dynamical string field in the small Hilbert space. The action is expressed 
\begin{equation}S= \int_0^1 dt\, \omega_L\left(A_t,\pi_1\M\frac{1}{1-A_\eta}\right).\label{eq:WZWsmall}\end{equation}
This closely resembles the Berkovits action as written in \eq{coBerk}. The only difference is that $\Q$ has been replaced with $\M$, and $\n-\m_2$ has been replaced with $\n$. In fact,  \eq{WZWsmall} can be understood as WZW-like action, in a generalized sense. 

To make this statement meaningful we must formulate the flatness conditions appropriate for this action. Since the dynamical $A_\infty$ structure is $\M$ and the constraint $A_\infty$ structure is $\n$, by analogy to \eq{multiBerk} we should postulate a multiform $A_\infty$ structure
\begin{equation}\del = \n + dx^M \M + dt\bm{\frac{d}{dt}}+dx^\delta \bm{\delta}.\end{equation}
We have
\begin{equation}\del^2=0\end{equation}
by virtue of the fact that $\M$ and $\n$ are nilpotent and commute. We introduce a multiform potential
\begin{equation}A=A^{(0)}+A^{(1)}+A^{(2)}+...,\end{equation}
where
\begin{eqnarray}
A^{(0)} \lineup \equiv A_\eta,\\
A^{(1)}\lineup \equiv dx^M A_M + dt A_t + dx^\delta A_\delta,\\
\lineup\vdots\ .\nonumber
\end{eqnarray}
The 1-form component contains potentials $A_M,A_t$ and $A_\delta$. The flatness conditions are expressed
\begin{equation}
\del\frac{1}{1-A}=0.
\end{equation}
The 0-form component of the flatness conditions imply
\begin{equation}\eta A_\eta = 0.\end{equation}
At $t=1$, this is simply the statement that $\PsiN$ is in the small Hilbert space. The 1-form component of the flatness conditions imply
\begin{eqnarray}
\eta A_t \lineup = \frac{d}{dt}A_\eta,\nonumber\\
\eta A_\delta \lineup = \delta A_\eta,\nonumber\\
\eta A_M\lineup = \pi_1\M\frac{1}{1-A_\eta}.
\end{eqnarray}
We can solve these equations with 
\begin{equation}A_\eta = \eta\PhiN(t),\ \ \ A_t=\dot{\Phi}_\mathrm{NS}(t),\ \ \ A_\delta(t) = \delta\PhiN(t).\end{equation}
The first two potentials agree with \eq{anat}. The potential $A_M$ is more nontrivial. Following \cite{WBlarge} we can find a solution in the form
\begin{equation}A_M = \int_0^t dt'\,\left(\left.\pi_1\M\frac{1}{1-A_\eta}\otimes A_t\otimes \frac{1}{1-A_\eta}\right|_{t=t'}\right).\end{equation}
This expression appears to have little in common with the Maurer-Cartan elements that define the WZW-like action of the Berkovits theory. Also, the higher potentials $A^{(2)},A^{(3)},...$ will be nonvanishing.  A systematic procedure for determining them was given in \cite{WBlarge}. A more general version of this procedure is described in subsection~\ref{subsubsec:flat}. 

\subsection{WZW-like Action in the General Case}

We now describe the general formulation of the WZW-like action. We start with a state space $\mathcal{H}$ with a $\mathbb{Z}\times\mathbb{Z}$ grading, which we write
\begin{equation}
 (N_1,N_2) \in\mathbb{Z}\times\mathbb{Z}.
\end{equation}
The value of the lattice vector will be called {\it rank}. In open superstring field theory, the integer $N_1$ roughly corresponds to ghost number, and $N_2$ to picture number, but the precise relation will be described later. The state space also has a $\mathbb{Z}_2$ grading distinguishing even and odd elements, called {\it degree}. For our considerations, degree will always be related to rank by
\begin{equation}\deg = N_1+N_2\ \ \ \mathrm{mod}\ \mathbb{Z}_2.\end{equation}
The definition of the WZW-like action requires the following ingredients:
\begin{description}
\item{(1)} A degree odd dynamical field $\Phi\in \mathcal{H}$ with rank
\begin{equation}\mathrm{rank}(\Phi) = (0,1).\end{equation}
\item{(2)} A symplectic form $\omega:\H\otimes\H \to \mathbb{C}$
\begin{equation}\omega(A,B) = -(-1)^{\deg(A)\deg(B)}\omega(B,A).\end{equation}
The symplectic form is only nonvanishing on states whose ranks add to (1,1)
\item{(3)} A pair of degree odd coderivations $\D$ and $\C$ acting on the tensor algebra of $\H$, denoted $T\H$.  The coderivation $\D$ is called the {\it dynamical} $A_\infty$ structure, and $\C$ is called the {\it constraint} $A_\infty$ structure. The coderivations carry respective ranks
\begin{equation}\rank(\D) = (1,0),\ \ \ \rank(\C) = (0,-1),\end{equation}
are nilpotent and mutually commute
\begin{equation}
\D^2 = \C^2 = [\D,\C] = 0,
\end{equation}
and are cyclic with respect to the symplectic form
\begin{equation}\langle \omega|\pi_2\D = 0,\ \ \ \ \langle\omega|\pi_2\C = 0.\end{equation}
We also assume that $\C$ and $\D$ are independent of $\Phi$.
\end{description}
In this paper we are primarily interested in the open string, so we will discuss the WZW-like action from the perspective of $A_\infty$ algebras.  A very similar formulation based on $L_\infty$ algebras can be obtained by symmetrizing the tensor algebra, and would be appropriate for closed string field theories.

The WZW-like action is not expressed directly in terms of $\Phi$, but rather in terms of functionals of $\Phi$---the {\it potentials}---satisfying certain identities---the {\it flatness conditions}. The flatness conditions can be expressed 
\begin{equation}\del\frac{1}{1-A}=0,\end{equation}
where $\del$ is the {\it multiform $A_\infty$ structure} and $A$ is the {\it multiform potential}. Let us describe these ingredients in turn. The construction starts with three coderivations:
\begin{equation}\del_D\equiv \D,\ \ \ \ \del_t\equiv \bm{ \frac{d}{dt}},\ \ \ \ \del_\delta \equiv \bm{\delta}.\label{eq:d1}\end{equation}
The coderivation $\del_t=\bm{d/dt}$ corresponds to the derivative $d/dt$ with respect to an auxiliary variable $t\in[0,1]$. This is the trivial ``third dimension" of the Wess-Zumino-Witten action. The coderivation $\del_\delta=\bm{\delta}$ corresponds to the variation $\delta$ of a state with respect to the dynamical field $\Phi$. We introduce $\delta$ for the purposes of computing the variation of the action. Both $\bm{d/dt}$ and $\bm{\delta}$ are degree even and carry rank $(0,0)$. The three coderivations \eq{d1} will be labeled collectively as 
\begin{equation}\del_i\end{equation} 
with $i=D,t,\delta$. In fact, the index $i$ can be extended to label other coderivations, but for now \eq{d1} are the important ones. We assume that $\D$ and $\C$ are independent of $t$ and $\Phi$, so all four coderivations $\C,\D,\bm{d/dt},\bm{\delta}$ commute. In particular we have 
\begin{equation}\C^2 = 0,\ \ \ [\del_i,\C] = 0,\ \ \ [\del_i,\del_j] = 0,\label{eq:CDtdcom}\end{equation}
with $i,j=D,t,\delta$. For each $\del_i$ we introduce a basis 1-form 
\begin{equation} dx^i.\end{equation}
The basis 1-forms carry degree and rank
\begin{equation}
\mathrm{rank}(dx^i) \equiv (0,-1) -\mathrm{rank}(\del_i),\ \ \ \ \ \ \deg(dx^i) \equiv \deg(\del_i) +1 \ \ \mathrm{mod}\ \mathbb{Z}_2.\label{eq:rd1f}
\end{equation}
In the natural way, we allow the basis 1-forms $dx^i$ to multiply states and operators. The basis 1-forms commute through themselves and other objects with a sign given by degree. Next we introduce a sequence of coderivations of increasing form degree:
\begin{eqnarray}
\del^{(0)}\lineup \equiv \C,\\
\del^{(1)}\lineup \equiv dx^i \del_i ,\\
\del^{(2)}\lineup \equiv \frac{1}{2!}dx^i\wedge dx^j \, \del_{ij},\\
\lineup \vdots\nonumber\ .
\end{eqnarray}
The zero form coderivation $\del^{(0)}$ is precisely the constraint $A_\infty$ structure $\C$. The 1-form coderivation contains the dynamical $A_\infty$ structure $\D$ and the coderivations $\bm{d/dt}$ and $\bm{\delta}$ as listed in \eq{d1}. The higher form coderivations will be defined so that the sum
\begin{equation}\del = \del^{(0)}+\del^{(1)}+\del^{(2)} + ...\end{equation}
is nilpotent:
\begin{equation}\del^2 = 0.\end{equation}
This defines the {\it multiform $A_\infty$ structure}. The coderivation $\del$ is degree odd and carries rank $(0,-1)$. If we expand $\del^2=0$ in basis 1-forms, we obtain a hierarchy of identities 
\begin{eqnarray}
0 \lineup = \C^2,  \\ 
0\lineup = \[\C,\del_i\] , \\
0 \lineup = \[\C,\del_{ij}\] +(-1)^{(i+1)j} \[\del_i,\del_j\],  \\
\lineup\ \vdots\ ,\nonumber
\end{eqnarray}
where $i$ appearing in the exponent of $(-1)$ is shorthand for $\deg(\del_i)$. For the moment we are primarily concerned with the index $i$ taking values $D,t,\delta$. In this case, \eq{CDtdcom} implies that $\del_{ij},\del_{ijk},...$ can be chosen to vanish. We will proceed under this assumption. Later we will extend the index $i$ to include other coderivations which may not commute, and for these values of $i$ the higher form coderivations will need to be nonzero to ensure $\del^2=0$.  Next we introduce sequence of $\H$-valued differential forms of increasing form degree:
\begin{eqnarray}
A^{(0)}\lineup \equiv A_C,\\
A^{(1)}\lineup \equiv dx^i A_i ,\\
A^{(2)}\lineup \equiv \frac{1}{2!}dx^i\wedge dx^j \, A_{ij},\\
\lineup \vdots\ .\nonumber
\end{eqnarray}
We refer to $A^{(0)}$ as the {\it 0-potential}, $A^{(1)}$ as the {\it 1-potential}, and generally $A^{(n)}$ as the {\it n-potential}. The $0$-potential plays a special role and will be denoted $A_C$. Also important are $A_D,A_t$ and $A_\delta$ which appear as components of the 1-potential $A^{(1)}$. The sum 
\begin{equation}A = A^{(0)} +A^{(1)}+A^{(2)}+...\end{equation}
defines the {\it multiform potential}. The multiform potential is degree even and carries rank $(0,0)$. The potentials are generally functionals of the dynamical field $\Phi$. With these preparations, we can formulate the {\it flatness conditions}:
\begin{equation}
\del\frac{1}{1-A} = 0.\label{eq:flat}
\end{equation}
Expanding in basis 1-forms gives a hierarchy of identities: 
\begin{eqnarray}
0 \lineup =  \C \frac{1}{1-A_C},\phantom{\Bigg(}\label{eq:0f}\\
0 \lineup =  \left(\del_i \frac{1}{1-A_C}\right)  + (-1)^{i+1}\left(\C \frac{1}{1-A_C}\otimes A_i\otimes \frac{1}{1-A_C}\right)\phantom{\Bigg(},\label{eq:1f}\\
0 \lineup =   \left(\del_i \frac{1}{1-A_C}\otimes A_j\otimes \frac{1}{1-A_C}\right) + (-1)^{ij+1} \left(\del_j \frac{1}{1-A_C}\otimes A_i\otimes \frac{1}{1-A_C}\right)\nonumber\\
\lineup\ \ \  +  (-1)^{i+1}\left(\C \frac{1}{1-A_C}\otimes A_i\otimes \frac{1}{1-A_C}\otimes A_j\otimes \frac{1}{1-A_C}\right) \nonumber\\
\lineup\ \ \ + (-1)^{(i+1)j}\left(\C \frac{1}{1-A_C}\otimes A_j\otimes\frac{1}{1-A_C}\otimes A_i\otimes\frac{1}{1-A_C}\right)\ \ \ \ \ \nonumber\\
\lineup\ \ \ +(-1)^{(i+1)j}\left(\C \frac{1}{1-A_C}\otimes A_{ij}\otimes\frac{1}{1-A_C}\right)  +  (-1)^{i(j+1)}\left(\del_{ij}\frac{1}{1-A_C}\right),\label{eq:2f}\\
\lineup \vdots\ .
\end{eqnarray}
Further relations involving higher potentials exist but will play a limited role in our discussion. Not all solutions of the flatness conditions are interesting for the the purposes of defining a WZW-like action. For example, the solution $A=0$ would result in an action which vanishes identically. An acceptable solution should satisfy two additional properties: 
\begin{description}
\item{{\bf (i)}} The potential $A_\delta$ must vanish at $t=0$.
\item{{\bf (ii)}} For any degree odd state $\lambda$ or rank $(0,1)$, there is a unique choice of variation $\delta$ such that $A_\delta|_{t=1} = \lambda$.
\end{description}
Property {\bf (i)} is needed to derive the expected equations of motion from the action. Essentially, it implies that the dynamics of the theory are determined only by the the potentials evaluated at $t=1$. Property {\bf (ii)} implies that knowledge of $A_\delta$ at $t=1$ is equivalent to knowledge of the variation of the dynamical field $\Phi$.

Now we are ready to write the WZW-like action:
\begin{equation}
S = \int_0^1 dt\,\omega\left(A_t,\pi_1\D\frac{1}{1-A_C}\right).\label{eq:WZWaction}
\end{equation}
An interesting property of this action is that the equations of motion follow without explicit knowledge of how the potentials depend on $\Phi$. We leave the details to appendix \ref{app:EOM}, but the essential point is that the flatness conditions, together with cyclicity of $\C$ and $\D$, imply that the variation of the action takes the form
\begin{equation}\delta S = \int_0^1 dt\frac{d}{dt} \omega\left(A_\delta,\pi_1\D\frac{1}{1-A_C}\right).\end{equation}
Integrating the total derivative, the boundary term at $t=0$ vanishes on account of property {\bf (i)}. Thus we find 
\begin{equation}\delta S = \left.\omega\left(A_\delta,\pi_1\D\frac{1}{1-A_C}\right)\right|_{t=1}.\end{equation}
Next, note that property {\bf (ii)} implies that an arbitrary variation of $\Phi$ can produce an arbitrary $A_\delta$. Since the symplectic form is nondegenerate, setting $\delta S=0$ therefore implies 
\begin{equation}\left.\D\frac{1}{1-A_C}\right|_{t=1} = 0.\end{equation}
These are the equations of motion of the WZW-like action. Returning to the beginning, recall the equation of motion and constraint:
\begin{equation}\D\frac{1}{1-\varphi} = 0,\ \ \ \C\frac{1}{1-\varphi} =0.\end{equation}
If we identify
\begin{equation}\varphi = A_C|_{t=1},\end{equation}
then $\D\frac{1}{1-\varphi} = 0$ is the equation of motion derived from the WZW-like action, and $\C\frac{1}{1-\varphi} =0$ is solved identically by virtue of the fact that $A_C$ satisfies flatness conditions. 

One might notice that our discussion makes no reference to the interpolation $\Phi(t)$ of the dynamical field. In the conventional formulation of the Wess-Zumino-Witten action, it is important that the action is independent of the choice of $\Phi(t)$.  However, in the general context we see no reason to assume that the potentials depend on $t$ and $\Phi$ always through an interpolation $\Phi(t)$.  We prefer to view the interpolation as a feature of a particular kind of solution of the flatness conditions. The invariance under changes of $\Phi(t)$ is an example of a more general symmetry in the choice of solution to the flatness conditions, to be described in the next subsection. From this point of view it should be clear that we regard $\Phi$ as the fundamental dynamical variable, not $\Phi(t)$ which may or may not be a meaningful object. Changes in $\Phi(t)$ which do not effect $\Phi$ are therefore not gauge transformations, since they do not alter the dynamical variable.

\subsection{Symmetries}
\label{subsec:WZWsym}

The WZW-like action has a very large degree of symmetry. We interpret the symmetries as coming in three varieties:  
\begin{itemize}
\item Symmetries that transform the $A_\infty$ structures.
\item Symmetries that transform the potentials.
\item Symmetries that transform the dynamical field.
\end{itemize}
Of these three, only the last is a ``genuine" symmetry in the sense that it transforms the field variable. The first two symmetries relate different ways of writing the {\it same} action in WZW-like form. That is, if one substitutes an explicit formula for the potentials and integrates over $t$ to express the action directly in terms of $\Phi$, the first two symmetries act trivially. Sometimes symmetries can be interpreted in more than one sense, but we find the above categorization useful.

\subsubsection{Symmetries in the Choice of $A_\infty$ structures}
\label{subsubsec:U}

The WZW-like action is invariant under cyclic $A_\infty$ isomorphisms that simultaneously transform the dynamical and constraint $A_\infty$ structures. Such an isomorphism can be represented as an invertible cohomomorphism $\U$ of rank $(0,0)$ on $T\H$ which is cyclic with respect to the symplectic form:
\begin{equation}\langle\omega|\pi_2\U = \langle \omega|\pi_2.\end{equation}
We also assume that $\U$ is independent of $t$ and $\Phi$. We can use $\U$ to transform the dynamical and constraint $A_\infty$ structures:
\begin{eqnarray}
 \D' \lineup = \U\D\U^{-1},\\
 \C' \lineup = \U\C\U^{-1}.
\end{eqnarray}
More generally, we can transform the multiform $A_\infty$ structure:
\begin{equation}\del' = \U\del \U^{-1}.\end{equation}
If we also transform the multiform potential: 
\begin{equation}A' = \pi_1\U\frac{1}{1-A},\end{equation}
where $A$ satisfies the flatness conditions with respect $\del$, we obtain a solution to a new set of flatness conditions:
\begin{equation}
\del'\frac{1}{1-A'} = 0.
\end{equation}
It is easy to confirm that the new flatness conditions define a new form of the WZW-like action:
\begin{equation}S = \int_0^1dt\, \omega\left(A_t',\pi_1\D' \frac{1}{1-A_C'}\right).\end{equation}  
This action is actually the same as the original WZW-like action. To show this, note that cyclicity of $\U$ implies \cite{OkWB,WBlarge}
\begin{equation}
\omega(B_1,B_2) = \omega\left(\pi_1\U\frac{1}{1-A}\otimes B_1\otimes\frac{1}{1-A},\pi_1\U\frac{1}{1-A}\otimes B_2\otimes\frac{1}{1-A}\right).
\end{equation}
Therefore we can write
\begin{eqnarray}
S \lineup =  \int_0^1dt\, \omega\left(A_t',\pi_1\D' \frac{1}{1-A_C'}\right)\nonumber\\
\lineup = \int_0^1dt\, \omega\left(\pi_1\U\frac{1}{1-A_C}\otimes A_t\otimes\frac{1}{1-A_C},\pi_1\D' \U \frac{1}{1-A_C}\right) \nonumber\\
\lineup = \int_0^1dt\, \omega\left(\pi_1\U\frac{1}{1-A_C}\otimes A_t\otimes\frac{1}{1-A_C},\pi_1\U\D \frac{1}{1-A_C}\right) \nonumber\\
\lineup = \int_0^1dt\, \omega\left(\pi_1\U\frac{1}{1-A_C}\otimes A_t\otimes\frac{1}{1-A_C},\pi_1\U\frac{1}{1-A_C}\otimes\pi_1\left(\D \frac{1}{1-A_C}\right) \otimes \frac{1}{1-A_C}\right)\nonumber\\
\lineup = \int_0^1dt\, \omega\left(A_t,\pi_1\D \frac{1}{1-A_C}\right),
\end{eqnarray}
which is the original WZW-like action. In this sense the action is invariant under cyclic $A_\infty$ isomorphisms. This is a powerful advantage, since we are free to transform to an action where the dynamical and constraint $A_\infty$ structures take the simplest possible form. 

This symmetry is the main reason why the Berkovits theory is simpler than superstring field theories in the small Hilbert space. The dynamical and constraint $A_\infty$ structures of the Berkovits theory are
\begin{eqnarray}
\D \lineup = \Q,\\
\C \lineup = \n-\m_2.
\end{eqnarray}
To formulate superstring field theory in the small Hilbert space, however, the constraint $A_\infty$ structure must be $\n$ rather than $\n-\m_2$. Therefore we must construct a cyclic $A_\infty$ isomorphism $\g$ so that
\begin{eqnarray}
\D' \lineup = \g^{-1}\Q\g,\\
\C' \lineup = \g^{-1}(\n-\m_2)\g=\n .
\end{eqnarray} 
A construction of $\g$ is given in \cite{WittenSS} but is quite complicated.

\subsubsection{Symmetries in the Choice of Potentials}

Given a fixed choice of $A_\infty$ structures $\D$ and $\C$, there is still a symmetry in the action corresponding to the choice of potentials. Since $\del^2=0$, the flatness conditions are invariant under an infinitesimal ``gauge transformation:"
\begin{equation}
\delta_{\mathrm{potential}} A = \pi_1\del\frac{1}{1-A}\otimes \Lambda\otimes\frac{1}{1-A},\label{eq:pottrans}
\end{equation}
where $\Lambda$ is a degree odd multiform of rank $(0,1)$,
\begin{equation}
\Lambda \ \equiv\ \Lambda^{(0)} \ +\Lambda^{(1)} + \Lambda^{(2)}+...,
\end{equation}
with
\begin{eqnarray}
\Lambda^{(0)} \lineup \equiv \Lambda_C,\\
\Lambda^{(1)}\lineup \equiv  dx^i\Lambda_i,\\
\Lambda^{(2)}\lineup \equiv \frac{1}{2!}dx^i\wedge dx^j \Lambda_{ij} ,\\
\lineup \vdots \nonumber\ .
\end{eqnarray}
Generally, $\Lambda$ can be a functional of the dynamical field $\Phi$. This ``gauge transformation" is different from the gauge transformation of the WZW-like action, since it does not transform the dynamical field. Rather, it transforms how the potentials are chosen to depend on the dynamical field. The choice of $\Lambda$ is constrained by the requirement that $A_\delta$ must vanish at $t=0$. This implies
\begin{equation}\Lambda_C|_{t=0} = 0,\ \ \ \Lambda_\delta|_{t=0}=0.\label{eq:propiLambda}\end{equation}
Let us see how the action changes under a transformation of potentials. The $0$- and $1$-potentials transform explicitly~as
\begin{eqnarray}
\delta_\mathrm{potential} A_C\lineup = \pi_1\C\frac{1}{1-A_C}\otimes \Lambda_C\otimes\frac{1}{1-A_C},\label{eq:dpotAC}\\
\delta_\mathrm{potential} A_i\lineup =\pi_1\del_i\frac{1}{1-A_C}\otimes \Lambda_C\otimes\frac{1}{1-A_C}+(-1)^{i+1}\pi_1\C\frac{1}{1-A_C}\otimes\Lambda_i\otimes\frac{1}{1-A_C}\nonumber\\
\lineup\ \ \ +(-1)^{i+1}\pi_1\C\frac{1}{1-A_C}\!\otimes A_i\!\otimes\frac{1}{1-A_C}\!\otimes\! \Lambda_C\!\otimes\!\frac{1}{1-A_C}+\pi_1\C\frac{1}{1-A_C}\!\otimes\! \Lambda_C\!\otimes\!\frac{1}{1-A_C}\!\otimes\! A_i \!\otimes\!\frac{1}{1-A_C}.\nonumber\\\label{eq:dpotAi}
\end{eqnarray}
Observe that $\delta_\mathrm{potential}A_C$ and $\delta_\mathrm{potential}A_t$  take the same form as $\delta A_C$ and $\delta A_t$, respectively, with the replacement 
\begin{equation}
\delta_{\mathrm{potential}}\to\delta,\ \ \ \Lambda_C\to A_\delta,\ \ \ \Lambda_t\to - A_{t\delta}
\end{equation}
Therefore we can follow the calculation of appendix \ref{app:EOM} to find 
\begin{equation}
\delta_{\mathrm{potential}} S = \left.\omega\left(\Lambda_C,\pi_1\D\frac{1}{1- A_C}\right)\right|_{t=1}.
\end{equation}
Provided that
\begin{equation}\Lambda_C|_{t=1} = 0,\label{eq:invLambda}\end{equation}
the ``gauge transformation" of the potentials leaves the action invariant. This leaves a lot of freedom in the choice of solution of the flatness conditions. 

Let us explain the relation between this ``gauge transformation" and the invariance of the action under changes of interpolation $\Phi(t)$. We discuss this specifically in the context of the Berkovits theory, since in a more general context the interpolation need not be meaningful. The interpolation is defined by a function $f(t,\Phi)$ as in \eq{Berkint}, and we will write a variation of this function $\delta_f$. Assuming the standard Maurer-Cartan potentials \eq{Aeta}-\eq{Adelta}, one finds
\begin{eqnarray}
\delta_f A_\eta \lineup = \eta\Big((\delta_f e^{\Phi(t)})e^{-\Phi(t)}\Big) - \Big[A_\eta,(\delta_f e^{\Phi(t)})e^{-\Phi(t)}\Big],\\
\delta_f A_Q \lineup = Q\Big((\delta_f e^{\Phi(t)})e^{-\Phi(t)}\Big) - \Big[A_Q,(\delta_f e^{\Phi(t)})e^{-\Phi(t)}\Big],\\
\delta_f A_t \lineup = \frac{d}{dt}\Big((\delta_f e^{\Phi(t)})e^{-\Phi(t)}\Big) - \Big[A_t,(\delta_f e^{\Phi(t)})e^{-\Phi(t)}\Big],\\
\delta_f A_\delta \lineup = \delta\Big((\delta_f e^{\Phi(t)})e^{-\Phi(t)}\Big) - \Big[A_\delta,(\delta_f e^{\Phi(t)})e^{-\Phi(t)}\Big].
\end{eqnarray}
This is consistent with \eq{dpotAC} and \eq{dpotAi} if we identify
\begin{equation} \Lambda_C = (\delta_f e^{\Phi(t)})e^{-\Phi(t)}\end{equation}
and take all other higher form gauge parameters contained in $\Lambda$ to vanish. This is consistent with 
\begin{equation}\Lambda_C|_{t=0}=0,\ \ \ \Lambda_\delta|_{t=0}=0,\end{equation}
since the variation of $f$ must vanish at $t=0$ to preserve the boundary condition $f(0,\Phi)=0$, and $\Lambda_\delta$ is assumed to vanish identically. Moreover, we have
\begin{equation}\Lambda_C|_{t=1} = 0\end{equation}
since the variation of $f$ must vanish at $t=1$ to preserve the boundary condition $f(1,\Phi) = \Phi$. Therefore the action is independent of the form of $\Phi(t)$ inside the auxiliary ``third dimension" $t$. However, it is clear that this is a special example of a much more general symmetry of the action. 

It is interesting to note that if $\Lambda_C$ does not vanish at $t=1$, the action transforms in the same way as when taking the variation of the field. This means that the action after the transformation is identical to the action after making an infinitesimal field redefinition $\Phi\to\Phi+\delta\Phi$, where $\delta\Phi$ is determined by solving 
\begin{equation}A_\delta|_{t=1} = \Lambda_C|_{t=1}\end{equation}
Property {\bf (ii)} guarantees the existence of a solution to this equation. This has an important consequence: the structure of the WZW-like action is covariant under field redefinition. Specifically: two solutions of the flatness conditions correspond to WZW-like actions related (at most) by field redefinition. Conversely, field redefinition of a WZW-like action must produce a new WZW-like action characterized by a new solution of the flatness conditions.

\subsubsection{``True" Symmetries}
\label{subsubsec:sym}

Finally, let us consider the ``true" symmetries which transform the dynamical field. To motivate the structure, let us first consider superstring field theory in the small Hilbert space, characterized by a cyclic $A_\infty$ algebra $\M$. A symmetry transformation of the dynamical field $\PsiN$ takes the form
\begin{equation}\delta \PsiN = \pi_1{\bf v} \frac{1}{1-\PsiN},\end{equation}
where ${\bf v}$ is a degree even coderivation generating the symmetry transformation. For this to be a symmetry of the action, ${\bf v}$ must commute with $\M$, must be well defined in the small Hilbert space, and must be cyclic with respect to the symplectic form:
\begin{eqnarray}
[\M,{\bf v}] \lineup = 0,\\
\ [\n,{\bf v}] \lineup = 0,\\
\langle \omega_L|\pi_2{\bf v} \lineup = 0.
\end{eqnarray}
This structure translates to the WZW-like approach as follows: We replace $\PsiN$ with the 0-potential $A_C|_{t=1}$, $\M$ with the dynamical $A_\infty$ structure $\D$, and $\n$ with the constraint $A_\infty$ structure $\C$. Therefore we propose that a symmetry transformation $\delta\Phi$ of the dynamical field should result in a change of the 0-potential $A_C|_{t=1}$ of the form
\begin{equation}
\delta A_C|_{t=1} = \left.\pi_1 {\bf v} \frac{1}{1-A_C}\right|_{t=1},\label{eq:ACsym}
\end{equation}
where ${\bf v}$ is a degree even cyclic coderivation of rank $(0,0)$ which commutes with $\D$ and $\C$,
\begin{eqnarray}
\ [\D,{\bf v}] \lineup = 0,\\
\ [\C,{\bf v}] \lineup = 0,\\
\langle\omega|\pi_2{\bf v}\lineup = 0,
\end{eqnarray}
and is independent of $t$ and $\Phi$. We can check that this transformation preserves the 0-form part of the flatness conditions:
\begin{eqnarray}
\delta\left.\left(\C\frac{1}{1-A_C}\right)\right|_{t=1}\lineup = \left.\C\frac{1}{1-A_C}\otimes \delta A_C \otimes \frac{1}{1-A_C}\right|_{t=1}\nonumber\\
\lineup = \left.\C\frac{1}{1-A_C}\otimes \pi_1\left({\bf v}\frac{1}{1- A_C}\right) \otimes \frac{1}{1-A_C}\right|_{t=1}\nonumber\\
\lineup = \left.\C{\bf v} \frac{1}{1-A_C}\right|_{t=1}\nonumber\\
\lineup = \left.{\bf v} \C\frac{1}{1-A_C}\right|_{t=1}\nonumber\\
\lineup = 0.
\end{eqnarray}
A similar computation shows that it also preserves the equations of motion. Therefore any transformation of $\Phi$ which changes $A_C|_{t=1}$ according to \eq{ACsym} will define an on-shell symmetry of the theory. The question is how to generalize this to a symmetry of the action off-shell.

The key idea is to extend our list of coderivations $\del_i$ to include ${\bf v}$,
\begin{equation}
\del_v \equiv {\bf v},
\end{equation}
and introduce a basis 1-form $dx^v$:
\begin{equation}
\mathrm{rank}(dx^v) = (0,-1),\ \ \ \ \ \ \deg(dx^v) =\mathrm{odd}.
\end{equation}
We then assume that the multiform coderivation, multiform potential, and flatness conditions take the same form as before but that the index $i$ is extended to include $v$ in addition to $D,t,\delta$. This introduces a list of new potentials and coderivations
\begin{eqnarray}
\lineup A_v,\ \ \ A_{vi},\ \ \  A_{vij},\ \ \ A_{vijk},\ \ \ ... \ ,\\
\lineup \del_v,\ \ \ \,\del_{vi},\ \ \ \,\del_{vij},\ \ \ \,\del_{vijk},\ \ \ ...\ .
\end{eqnarray}
We assume that ${\bf v}$ is independent of $t$ and $\Phi$, so the higher coderivations $\del_{vi},\del_{vij},...$ can be assumed to vanish if $i,j = D,t,\delta$. However, it may be useful to further extend the flatness conditions to include coderivations representing other symmetry transformations. Generally such coderivations will form a nontrivial Lie algebra (corresponding to the symmetry group in question), and higher-form coderivations will be needed to ensure $\del^2=0$. This occurrence will not play a major role in our discussion. Once we have extended the flatness conditions to include the index $v$, we can write the symmetry transformation \eq{ACsym} of the 0-potential as:
\begin{equation}
\delta A_C|_{t=1} =\left.\pi_1 {\bf v} \frac{1}{1-A_C}\right|_{t=1} = \left.\pi_1\C\frac{1}{1-A_C}\otimes A_v\otimes\frac{1}{1-A_C}\right|_{t=1}.\label{eq:vflat}
\end{equation}
On the other hand, for a generic variation $\delta$ we also have  
\begin{equation}\delta A_C|_{t=1} =\left. \pi_1 \C\frac{1}{1-A_C}\otimes A_\delta\otimes\frac{1}{1-A_C}\right|_{t=1}.\end{equation}
Therefore the symmetry transformation of the dynamical field $\delta\Phi$ can be determined by equating
\begin{equation}
A_\delta|_{t=1} = A_v|_{t=1}.\label{eq:deltasym}
\end{equation}
From property {\bf (ii)} this equation uniquely determines $\delta\Phi$. It remains to be shown that this is a symmetry of the action. We give the details in appendix \ref{app:sym}.

\subsubsection{Gauge Symmetry}
\label{subsubsec:gauge}

The most important example of a ``true" symmetry is gauge symmetry. To see where gauge symmetries come from,  note we may trivially invent a symmetry transformation by positing a coderivation that it is simultaneously $\D$ and $\C$ exact,
\begin{equation}\del_g \equiv [\C,[\D,\llambda_D]],\label{eq:delg}\end{equation}
for some coderivation $\llambda_D$. Such a symmetry is in a sense trivial, since it is present simply by virtue of the fact that we have a WZW-like action, and is independent of the physics of the system that is realized by the action. In fact, we will interpret this as representing gauge symmetry. For present purposes, it is sufficient to view $\llambda_D$ as the coderivation derived from a string field $\lambda_D$ of rank $(-1,1)$ regarded as a 0-string product. The string field $\lambda_D$ is a gauge parameter. To complete the definition of the symmetry transformation, we must postulate an expression for the potential:
\begin{eqnarray}
A_g \lineup \equiv \pi_1\Big([\C,\llambda_C]+[\D,\llambda_D]\Big)\frac{1}{1-A_C}.\label{eq:Ag}
\end{eqnarray}
Here we introduce an additional coderivation $\llambda_C$ corresponding to a string field $\lambda_C$ of rank $(0,2)$. 
Therefore the transformation is defined by two gauge parameters $\lambda_D$ and $\lambda_C$. One may readily verify that
\begin{equation}
\del_g\frac{1}{1-A_C} = \C\frac{1}{1-A_C}\otimes A_g\otimes\frac{1}{1-A_C}.
\end{equation}
Therefore $\del_g$ and $A_g$ are consistent with flatness conditions to this order. The gauge transformation of $\Phi$ is determined by equating
\begin{equation}A_\delta|_{t=1} = A_g|_{t=1}.\label{eq:gtrans}\end{equation}
We will discuss higher order flatness conditions in a moment.

Let us confirm that this agrees with what we know about the gauge symmetry of Berkovits open superstring field theory. The gauge transformation of the Berkovits theory is often expressed
\begin{equation}
\delta e^\PhiN =  e^\PhiN \eta \omega + Q\Lambda e^\PhiN ,
\end{equation}
where the string fields $\omega$ and $\Lambda$ are gauge parameters. Multiplying this equation with $e^{-\PhiN}$ this can be rewritten
\begin{eqnarray}
A_\delta|_{t=1} \lineup =  e^\PhiN \eta \omega e^{-\PhiN}+ Q\Lambda\nonumber\\
\lineup = \eta\Big(e^\PhiN \omega e^{-\PhiN}\Big)-(\eta e^\PhiN) \omega e^{-\PhiN} + e^\PhiN \omega (\eta e^{-\PhiN})+ Q\Lambda\nonumber\\
\lineup = \left.\Big(\eta\big(e^\PhiN \omega e^{-\PhiN}\big)-[A_\eta, e^\PhiN \omega e^{-\PhiN}] + Q\Lambda\Big)\right|_{t=1}.
\end{eqnarray}
This should agree with the potential $A_g$ derived from \eq{gtrans} evaluated at $t=1$. Computing we find
\begin{eqnarray}
A_g \lineup = \pi_1\Big([\n-\m_2,\llambda_\eta] + [\Q,\llambda_Q]\Big)\frac{1}{1-A_\eta}\nonumber\\
\lineup =  \eta\lambda_\eta-m_2(A_\eta,\lambda_\eta) -m_2(\lambda_\eta,A_\eta) + Q\lambda_Q\nonumber\\
\lineup = \eta\lambda_\eta -[A_\eta,\lambda_\eta] + Q\lambda_Q.
\end{eqnarray}
Therefore we can identify
\begin{equation}\lambda_\eta = e^\PhiN \omega e^{-\PhiN},\ \ \ \lambda_Q = \Lambda,\end{equation}
and the gauge symmetry of the Berkovits theory is seen to be a special case of the general WZW-like gauge transformation \eq{gtrans}. It is interesting to mention that $A_g$ is perhaps the most elementary example of a potential that is not derived from a derivation of the open string star product. Instead, $A_g$ corresponds to the coderivation $\del_g$. We may define the action of $\del_g$ on a degree even string field $A$ through
\begin{eqnarray}
\pi_1\del_g\frac{1}{1-A} = \eta\big(Q\lambda_D\big) -[Q\lambda_D,A].
\end{eqnarray}
This operation is inhomogeneous in the string field $A$, and is clearly not a derivation of the star product. Also note that $A_g$ leads to a nonvanishing 2-potential:
\begin{equation}A_{gt} = [\lambda_\eta,A_t]\end{equation}
From this we see that a proper account of gauge symmetry requires the full formalism we have developed, even in the Berkovits theory based on a conventional WZW-like action.

As another example, let us see that the gauge transformation is consistent with the gauge symmetry of open superstring field theory in the small Hilbert space. Lifting the theory to the large Hilbert space following section \ref{subsec:liftsmall} the gauge transformation should be defined by the potential
\begin{eqnarray}a_g \lineup = \pi_1\Big([\n,\llambda_\eta]+[\M,\llambda_M]\Big)\frac{1}{1-a_\eta}\nonumber\\
\lineup = \eta\lambda_\eta + \pi_1\M\frac{1}{1-a_\eta}\otimes\lambda_M\otimes\frac{1}{1-a_\eta}.\end{eqnarray}
Since $a_\delta|_{t=1} = \delta\PhiN$, the gauge parameter $\lambda_\eta$ implies that the action is independent of the components of the string field $\PhiN$ which live in the small Hilbert space. This implies that the action can be expressed as a function of $\PsiN=\eta\PhiN$, which leads to the theory  formulated in the small Hilbert space. In terms of the small Hilbert space string field, the gauge transformation amounts to
\begin{equation}
\delta\PsiN = \pi_1\M\frac{1}{1-\PsiN}\otimes(-\eta\lambda_M)\otimes\frac{1}{1-\PsiN}.
\end{equation}
This is the standard $A_\infty$ gauge transformation defined by the gauge parameter $-\eta\lambda_M$ in the small Hilbert space.

Since gauge symmetry is present for all WZW-like actions, we would like to find a universal way to incorporate it into a complete solution of the flatness conditions. Suppose we have a multiform coderivation $\del'$ and multiform potential $A'$ describing all flatness conditions of interest {\it except} those associated with gauge symmetry. In particular,  $\del'$ and $A'$ do not include the basis 1-form $dx^g$. The goal is to find an ``improvement" of $\del'$ and $A'$ which includes the flatness conditions associated with the gauge symmetry. The improved multiform coderivation and potential will take the form
\begin{eqnarray}
\del \lineup = \del' + \del_\mathrm{gauge},\\
A \lineup = A'+A_\mathrm{gauge},
\end{eqnarray}
where $\del_\mathrm{gauge}$ and $A_\mathrm{gauge}$ are proportional to $dx^g$:
\begin{eqnarray}
\del_\mathrm{gauge} \lineup = dx^g \del_g+ dx^g \wedge dx^i \del_{gi} + \frac{1}{2!}dx^g \wedge dx^i\wedge dx^j \del_{gij} +...,\\
A_\mathrm{gauge} \lineup = dx^g A_g+ dx^g \wedge dx^i A_{gi} + \frac{1}{2!}dx^g \wedge dx^i\wedge dx^j A_{gij} +... \ .
\end{eqnarray}
The coderivation $\del_g$ and potential $A_g$ are already given in \eq{delg} and \eq{Ag}. We propose that this can be generalized to higher orders by 
\begin{eqnarray}
\del_\mathrm{gauge} \lineup = dx^g [\del',[\D,\llambda_D]],\label{eq:delgcomp}\\
A_\mathrm{gauge}\lineup = dx^g \pi_1\Big([\del',\llambda_C]+[\D,\llambda_D]\Big)\frac{1}{1-A'}.\label{eq:Agcomp}
\end{eqnarray}
This is related to $\del_g$ and $A_g$ by the replacement $\C\to\del'$ and $A_C\to A'$. To show that this is consistent, note that the improved multiform coderivation $\del$ is nilpotent:
\begin{equation}
\del^2 = (\del')^2 +[\del',\del_\mathrm{gauge}] + \del_\mathrm{gauge}^2 = 0.
\end{equation}
The first term vanishes since $\del'$ is nilpotent (by assumption); the second term vanishes since $\del_\mathrm{gauge}$ takes the form of a commutator with $\del'$; finally, the third term vanishes identically since $dx^g\wedge dx^g = 0$. Next we confirm the flatness conditions:
\begin{eqnarray}
\del \frac{1}{1-A}\lineup = (\del' +\del_\mathrm{gauge})\frac{1}{1-A'-A_\mathrm{gauge}}\nonumber\\
\lineup = (\del' +\del_\mathrm{gauge})\left(\frac{1}{1-A'}+\frac{1}{1-A'}\otimes A_\mathrm{gauge}\otimes\frac{1}{1-A'}\right)\nonumber\\
\lineup = \del_\mathrm{gauge}\frac{1}{1-A'} + \del' \frac{1}{1-A'}\otimes A_\mathrm{gauge}\otimes\frac{1}{1-A'}.
\end{eqnarray}
Substituting \eq{delgcomp} and \eq{Agcomp},
\begin{eqnarray}
\del \frac{1}{1-A}\lineup = dx^g [\del',[\D,\llambda_D]]\frac{1}{1-A'} - dx^g \del' \frac{1}{1-A'}\otimes \left(\pi_1\Big([\del',\llambda_C]+[\D,\llambda_D]\Big)\frac{1}{1-A'}\right)\otimes\frac{1}{1-A'}\nonumber\\
\lineup = dx^g \left([\del',[\D,\llambda_D]]\frac{1}{1-A'} - \del\Big( [\del',\llambda_C]+[\D,\llambda_D]\Big)\frac{1}{1-A'}\right)\nonumber\\
\lineup = dx^g \left([\del',[\D,\llambda_D]]\frac{1}{1-A'} - [\del',[\del,\llambda_C]+[\D,\llambda_D]]\frac{1}{1-A'}\right)\nonumber\\
\lineup = dx^g \left([\del',[\D,\llambda_D]]\frac{1}{1-A'} - [\del',[\D,\llambda_D]]\frac{1}{1-A'}\right)\nonumber\\
\lineup = 0.
\end{eqnarray}
To prove gauge invariance of the action following appendix \ref{app:sym}, it is important to know that the higher form coderivations
\begin{equation}\del_{gi},\ \ \del_{gij},\ \ \del_{gijk},\ \ ...\end{equation}
vanish when $i,j,k = D,t,\delta$. We can confirm this:
\begin{equation}
\del_\mathrm{gauge} = dx^g\left[\C +dx^D\D + dt\bm{\frac{d}{dt}} + dx^\delta \bm{\delta},[\D,\llambda_D]\right] = dx^g [\C,[\D,\llambda_D]].
\end{equation}
The 2-form component vanishes since $\D$ is nilpotent and $[\D,\llambda_D]$ is independent of $t$ and $\Phi$. Higher components of $\del_\mathrm{gauge}$ can only appear if the index $i$ labels other coderivations corresponding to symmetries which do not commute with gauge transformations. This proves gauge invariance of the action.

\section{WZW-like Action in Open Superstring Field Theory}
\label{sec:KO}

In this section we apply the formalism just developed to construct an action for open superstring field theory in the large Hilbert space including the Ramond sector. We follow \cite{complete} in the formulation of the Ramond sector. A very similar formulation of the Ramond sector can be given using spurious free fields \cite{1PIR}. See \cite{SenBV,RWaction,loop} for other discussions of this alternative. We start with the free action of open superstring field theory in the small Hilbert space:
\begin{equation}S_{\mathrm{free}} = \frac{1}{2}\Omega_S(\Psi,Q\Psi).\end{equation}
The string field $\Psi$ is in the small Hilbert space and carries ghost number $1$. It contains an Neveu-Schwarz (NS) and Ramond (R) sector component:
\begin{equation}\Psi =\PsiN+\PsiR.\end{equation}
 The NS string field carries picture $-1$ and the Ramond string field carries picture $-1/2$ and is subject to the constraint
\begin{equation}XY\PsiR = \PsiR\label{eq:Rconst}\end{equation}
We introduce the picture raising operator $X$, a picture lowering operator $Y$, and an operator $\Xi$ defined in the large Hilbert space with the properties
\begin{eqnarray}
XYX \lineup = X\ \ \ \ \mathrm{acting\ on\ small\ Hilbert\ space\ at\ picture}\ -3/2,\label{eq:XYX}\\
YXY \lineup = Y\ \ \ \ \mathrm{acting\ on\ small\ Hilbert\ space\ at\ picture}\ -1/2,\\
\ [\eta,X] \lineup = [\eta,Y] = 0,\\
\ [Q,X]\lineup = 0,\\
\ [Q,\Xi]\lineup = X,\\
\ [\eta,\Xi] \lineup = 1,\\
\Xi^2 \lineup= 0,\\
X,Y,\ \lineup\mathrm{and}\ \Xi \ \mathrm{are\ BPZ\ even}.
\end{eqnarray}
For our purposes these formal properties are all we need. See \cite{RWaction} for a concrete construction, and also \cite{complete,OhOk,supermod} for discussion of the required distributional operators in the $\beta\gamma$ system. The object $\Omega_S$ is called the {\it restricted} symplectic form, and is defined:
\begin{equation}
\Omega_S(a,b) \equiv \omega_S(\mathcal{G}^{-1} a,b),\label{eq:OmS}
\end{equation}
where
\begin{equation}
\mathcal{G}^{-1} \equiv \left\{\begin{matrix}\ \ \ \,\mathbb{I}\ \ \ \, \mathrm{acting\ on\ NS\ states}\Vspace \\ \ \ Y\ \ \ \mathrm{acting\ on\ R\ states}\Vspace\end{matrix}\right. .
\end{equation}
In the Ramond sector, $\Omega_S$ is assumed to operate between states in the small Hilbert space at picture $-1/2$ satisfying the constraint \eq{Rconst}. The restricted symplectic form is graded antisymmetric and nondegenerate, and the BRST operator is cyclic. 

We now reformulate the free theory using a WZW-like action. We replace $\Psi$ with a dynamical string field $\Phi$ in the large Hilbert space by making the substitution
\begin{equation}\Psi = \eta\Phi.\end{equation}
$\Phi$ has an NS and an R component:
\begin{equation}\Phi = \PhiN+\PhiR.\end{equation}
The NS component carries ghost and picture number $0$, while the Ramond component carries ghost number $0$ and picture $+1/2$ and is subject to the constraint
\begin{equation}XY\eta\PhiR = \eta\PhiR.\end{equation}
Following the steps of section \ref{subsec:liftsmall}, we arrive at a WZW-like action:
\begin{equation}
S_{\mathrm{free}} = \int_0^1 dt\, \Omega_L\big(A_t,QA_\eta\big),
\end{equation}
where $\Omega_L$ is the restricted symplectic form defined in the large Hilbert space: 
\begin{equation}\Omega_L(A,B)\equiv\omega_L(\mathcal{G}^{-1} A,B).\end{equation}
The dynamical $A_\infty$ structure is $\Q$ and the constraint $A_\infty$ structure is $\n$, and both are cyclic with respect to the restricted symplectic form. The multiform $A_\infty$ structure and multiform potential are given by
\begin{eqnarray}
\del \lineup = \n + dx^Q\Q+ dt\bm{\frac{d}{dt}} + dx^\delta \bm{\delta},\\
A \lineup = A_\eta + dx^Q A_Q + dt A_t + dx^\delta A_\delta,
\end{eqnarray}
with 
\begin{eqnarray}
A_\eta \lineup = \eta\Phi(t),\\
A_t \lineup = \dot{\Phi}(t),\\
A_\delta \lineup = \delta\Phi(t),\\
A_Q\lineup = Q\Phi(t),
\end{eqnarray}
where $\Phi(t)$ is an interpolation of $\Phi$. The flatness conditions are obeyed, 
\begin{equation}\del\frac{1}{1-A} = 0,\end{equation}
and the higher form potentials can be taken to vanish. 

Let us explain the relation between ghost and picture number and the concept of {\it rank} introduced earlier. Generally we may wish to describe the rank of an operator $\mathcal{O}$ on the tensor algebra, in which case it corresponds to a pair of eigenvalues $(N_1,N_2)$ under commutation with coderivations $\N_1,\N_2$: 
\begin{eqnarray}
[\N_1,\mathcal{O}] \lineup = N_1\mathcal{O},\\
\ [\N_2,\mathcal{O}]\lineup = N_2\mathcal{O},
\end{eqnarray}
given by 
\begin{eqnarray}
\N_1\lineup = \bm{\mathrm{gh}} + \bm{\mathrm{pic}} -\frac{1}{2}\bm{\mathrm{R}}, \label{eq:N1}\\
\N_2 \lineup = \bm{1}+\bm{\mathrm{pic}}-\frac{1}{2}\bm{\mathrm{R}}.
\end{eqnarray}
Here $\bm{1}$ is the coderivation corresponding to the identity operator $\mathbb{I}$ on the state space; ${\bf R}$ is the coderivation corresponding to an operator which acts as the identity on Ramond states and as zero on NS states; $\bm{\mathrm{gh}}$ is the coderivation corresponding to the operator which counts ghost number, and ${\bf pic}$ is the coderivation corresponding to the operator which counts picture. One may readily verify with this definition that
\begin{eqnarray}
\rank(\Phi) = (0,1),\ \ \ \rank(\Q) = (1,0),\ \ \ \rank(\n) = (0,-1),
\end{eqnarray}
in accordance with the grading assignments of the previous section.

Let us make an important remark about the restricted symplectic form $\Omega_L$. In the Ramond sector, we assume that $\Omega_L$ is defined when contracting a state $A$ at picture $+1/2$ in the large Hilbert space with a state $b$ at picture $-1/2$ in the small Hilbert space. In this case we have the equality
\begin{equation}\Omega_L(A,b) = \Omega_S(a,b),\end{equation}
where $a=\eta A$ and both $a$ and $b$ satisfy the constraint $XY=1$. We will not need a more general definition of $\Omega_L$, for example contracting two Ramond states in the large Hilbert space or at pictures other than $\pm 1/2$. This means that in the Ramond sector $\Omega_L$ and $\Omega_S$ are equivalent. This is related to the fact that the Ramond sector will effectively be formulated in the small Hilbert space, as was done by Kunitomo and Okawa \cite{complete}. The reason to describe the Ramond sector in the large Hilbert space is to allow a more unified treatment of the NS and R sectors in terms of a WZW-like action. 

\subsection{Dynamical and Constraint $A_\infty$ structures}
\label{subsec:DCAinf}

The next step is to find a nonlinear generalization of the dynamical and constraint $A_\infty$ structures. In fact, the requisite structures are already contained in the field equation \eq{field_equation}
\begin{eqnarray}(Q-\eta)\varphi +\varphi^2 = 0,\label{eq:field_equation2}\end{eqnarray}
provided that $\varphi$ is now assumed to contain both NS and R components,
\begin{equation}\varphi = \varphi_\mathrm{NS} + \varphi_\mathrm{R},\end{equation}
where $\varphi_\mathrm{NS}$ carries picture $-1$ and $\varphi_\mathrm{R}$ carries picture $-1/2$. We can expand the field equation according to picture:
\begin{eqnarray}
0\lineup = Q\varphi_\mathrm{R},\\
0\lineup = Q\varphi_\mathrm{NS} + \varphi_\mathrm{R}^2, \\
0\lineup = \eta\varphi_\mathrm{R} - [\varphi_\mathrm{R},\varphi_\mathrm{NS}], \\
0\lineup = \eta\varphi_\mathrm{NS}-\varphi_\mathrm{NS}^2.
\end{eqnarray}
The first two of equations naturally define a dynamical $A_\infty$ structure, while the second two define a constraint $A_\infty$ structure. Note that in the first two equations the star product only multiplies Ramond states. We account for this by writing the dynamical $A_\infty$ structure in the form
\begin{equation}\Q +\m_2|_2.\label{eq:noncycD}\end{equation}
The subscript after the vertical slash denotes {\it Ramond number} \cite{Ramond,susyS}. Ramond number counts the number of Ramond inputs minus the number of Ramond outputs required for the product to be nonzero. Alternatively, it is the eigenvalue under commutation with the coderivation ${\bf R}$ introduced in \eq{N1}:
\begin{equation} [{\bf R},{\bf b}_n|_r] = -r {\bf b}_n|_r. \end{equation}
In \eq{noncycD} the star product is restricted to Ramond number 2, which effectively means it is nonzero only when multiplying a pair of  Ramond states. Similarly the constraint $A_\infty$ structure can be written
\begin{equation}\n - \m_2|_0,\label{eq:noncycC}\end{equation}
where the star product is nonzero only when it contains the same number of Ramond inputs as outputs. We may then write the dynamical field equation and constraint in coalgebra form
\begin{eqnarray}
(\Q +\m_2|_2)\frac{1}{1-\varphi} \lineup = 0,\\
(\n-\m_2|_0)\frac{1}{1-\varphi} \lineup = 0.
\end{eqnarray}
This generalizes \eq{Bdyn} and \eq{Bconst} to include Ramond states. One may verify that\footnote{In proving this it is convenient to note that Ramond number is conserved when multiplying coderivations. For example we have the relation
\begin{equation}[\b,\c]\big|_{r} = \sum_{s=-1}^{r+1}[\b|_s,\c|_{r-s}].\end{equation}
$\Q$ and $\n$ carry Ramond number $0$, though we do not write it explicitly. See \cite{susyS} for more discussion of Ramond number.}
\begin{equation}
(\Q+\m_2|_2)^2 = (\n-\m_2|_0)^2 = [\Q+\m_2|_2,\n-\m_2|_0] = 0.
\end{equation}
Therefore $\Q+\m_2|_2$ and $\n-\m_2|_0$ define mutually commuting $A_\infty$ structures.

\begin{figure}
\begin{center}
\resizebox{4in}{1.6in}{\includegraphics{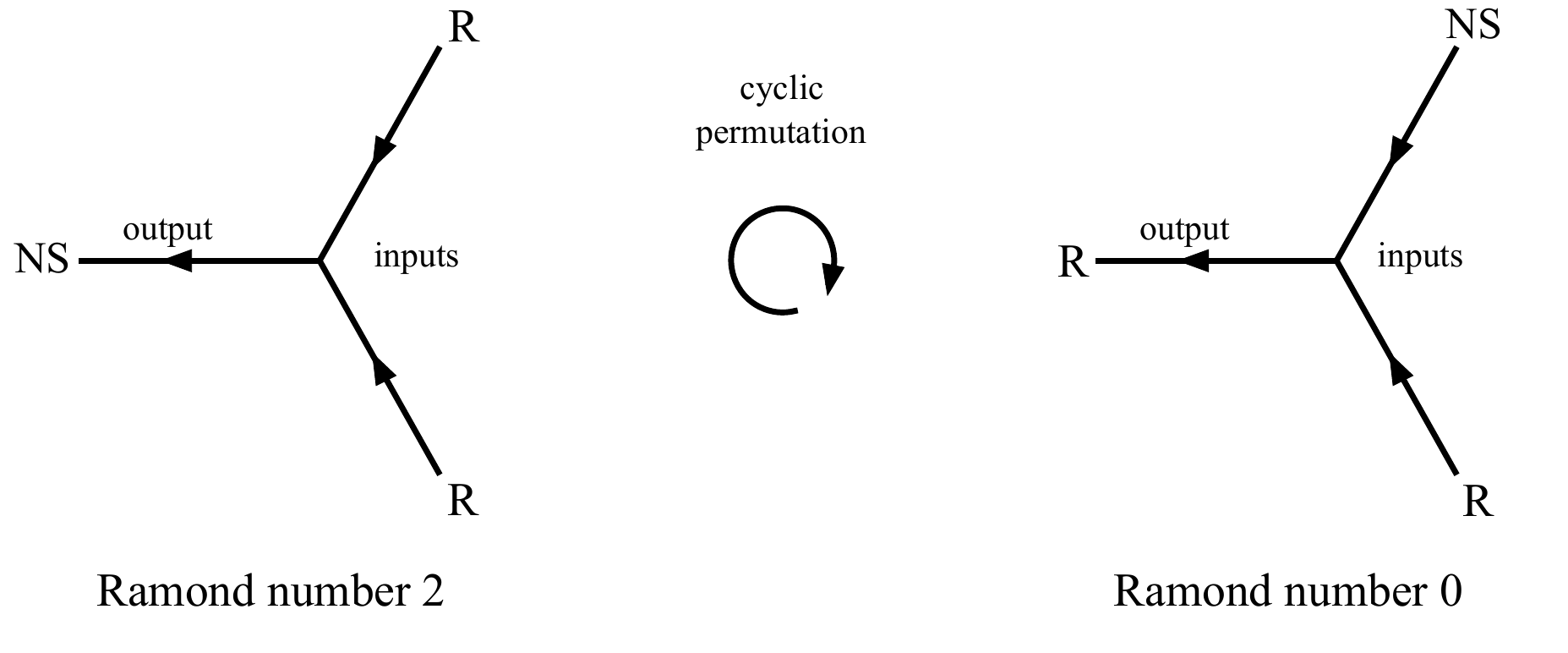}}
\end{center}
\caption{\label{fig:Rcyc} Cyclic permutation of a vertex changes the definition of the output and inputs of a product. Since Ramond number is not symmetric between inputs and outputs, products of fixed Ramond number are generally not cyclic.} 
\end{figure}

Unfortunately, these two $A_\infty$ structures do not define a WZW-like action. The problem is cyclicity: The star product restricted to a specific Ramond number is not invariant under cyclic permutations of the vertex. This is illustrated in figure \ref{fig:Rcyc}. Therefore we will need to modify the $A_\infty$ structures by making a similarity transformation
\begin{eqnarray}
\D \lineup = \F^{-1}(\Q+\m_2|_2)\F,\label{eq:openD}\\
\C \lineup = \F^{-1}(\n -\m_2|_0)\F,
\end{eqnarray}
where $\F$ is an invertible cohomomorphism defined in such a way that $\D$ and $\C$ are cyclic. This transformation is equivalent to a field redefinition
\begin{equation}\varphi\ \ \to\ \ \varphi' = \pi_1\F\frac{1}{1-\varphi},\end{equation}
and so should not alter the physical content of the field equations. 

Let us consider the constraint $A_\infty$ structure first. The problem is that $m_2|_0$ has a part that can multiply one Ramond state, but there is nothing to multiply two Ramond states as would be required by cyclicity. There are two possible ways to fix this:
\begin{description}
\item{(1)} Define $\F$ so as to add a piece to the constraint $A_\infty$ structure which multiplies two Ramond states. 
\item{(2)} Define $\F$ so as to remove the part of the constraint $A_\infty$ structure that multiplies one Ramond state. 
\end{description}
The first alternative requires that the product of two Ramond states contains a picture lowering operation, which seems unnatural. Therefore, we will follow the second alternative. Since we would like to reproduce the Berkovits theory in the NS sector, we propose that the constraint $A_\infty$ structure should take the form
\begin{equation}\F^{-1}(\n -\m_2|_0)\F = \n-\m_2|^0.\label{eq:openC}\end{equation}
The superscript after the vertical slash indicates {\it cyclic Ramond number} \cite{susyS}. A product of cyclic Ramond number $n$ is nonvanishing only if the number of Ramond inputs {\it plus} the number of Ramond outputs is equal to $n$. It is easily seen that cyclic Ramond number is invariant under cyclic permutations of a product. The star product $m_2|^0$ is nonvanishing only when the number of Ramond inputs plus Ramond outputs is equal to zero, which effectively means that $m_2|^0$ only operates on NS states. Therefore \eq{openC} reproduces the constraint $A_\infty$ structure $\n-\m_2$ of the Berkovits theory when acting on NS states, and when acting on Ramond states it reduces simply to $\n$. This leads to the fact that the Ramond sector can be reformulated in the small Hilbert space, as anticipated above. 

Let us now construct the cohomomorphism $\F$. Actually, it is easier to construct $\F^{-1}$ and derive $\F$ by computing $(\F^{-1})^{-1}$. Since $\F^{-1}$ is a cohomomorphism, it is completely specified by its output on the 1-string component of the tensor algebra:
\begin{equation}
\pi_1\F^{-1} = \sum_{n=0}^\infty F^{\mathrm{inv}}_n\pi_n,
\end{equation}
where $F^{\mathrm{inv}}_n$ are a sequence of multi-string products $\H^{\otimes n} \to\H$. Since we do not wish to modify the $\n$ contribution to the constraint $A_\infty$ structure, we may assume
\begin{equation}F^{\mathrm{inv}}_0 = 0,\ \ \ F^{\mathrm{inv}}_1 = \mathbb{I}.\end{equation}
Let us concentrate on the 2-string product $F^{\mathrm{inv}}_2$.  Multiplying \eq{openC} by $\F^{-1}$ from the right gives
\begin{equation}\F^{-1}(\n - \m_2|_0) = (\n-\m_2|^0)\F^{-1}.\label{eq:CFinv}\end{equation}
Projecting on the left with $\pi_1$ and on the right with $\pi_2$ we conclude that
\begin{equation}F^{\mathrm{inv}}_2(\eta\otimes\mathbb{I}+\mathbb{I}\otimes\eta)-m_2|_0 = \eta F^{\mathrm{inv}}_2-m_2|^0.\end{equation}
This is equivalent to
\begin{equation}[\n,{\bf F}^{\mathrm{inv}}_2] = -\m_2|_0^2,\label{eq:nF2}\end{equation}
where $m_2|_0^2$ is the star product restricted to Ramond number $0$ and cyclic Ramond number $2$. This essentially means that $m_2|_0^2$ is nonzero only when multiplying one Ramond state. 

This gives  $F^{\mathrm{inv}}_2$ up to an $\n$-closed term. To fix the $\n$-closed term we require that the dynamical $A_\infty$ structure $\D$ is also cyclic. This issue with $\D$ is that $m_2|_2$ can multiply two Ramond states, but there is no contribution that can multiply one Ramond state. Transformation by $\F$ cannot remove $m_2|_2$ while assuming \eq{nF2}, since $m_2|_2$ is nontrivial in the BRST cohomology in the small Hilbert space. Therefore, $\F$ must add a product of one Ramond state to accompany $m_2|_2$. Importantly, this product requires a picture changing insertion. To understand the nature of the picture changing insertion, we note that the restricted symplectic form contracts Ramond states in the small Hilbert space at picture $-1/2$ satisfying the constraint $XY=1$. Therefore the Ramond output of $\D$ must also satisfy $XY=1$. This will be true provided the Ramond output is  proportional to the picture changing operator $X$ by virtue of \eq{XYX}. Therefore we propose that the dynamical $A_\infty$ structure should take the form
\begin{equation}
\F^{-1}(\Q+\m_2|_2)\F = \Q +\m_2|_2 + X\m_2|_0^2 + ...,\label{eq:openDapprox1}
\end{equation}
where $...$ denote coderivations representing higher products. A comment about notation: We use $X\m_2|_0^2$ to denote the coderivation corresponding to the product $Xm_2|_0^2$. More generally, given an operator $\mathcal{O}:~\H\to\H$ and a coderivation $\b$ containing products $b_n:\H^{\otimes n}\to\H$,
\begin{equation}\pi_1\b = \sum_{n=1}^\infty b_n\pi_n,\end{equation}
we define the coderivation $\mathcal{O}\b$:
\begin{equation}\pi_1\mathcal{O}\b \equiv\sum_{n=0}^\infty \mathcal{O}b_n\pi_n.\end{equation}
It is helpful to write \eq{openDapprox} in the form
\begin{equation}
\F^{-1}(\Q+\m_2|_2)\F = \Q +\mathcal{G}\m_2|^2 + ...,\label{eq:openDapprox}
\end{equation}
where the operator $\mathcal{G}$ is defined \cite{1PIR}
\begin{equation}\mathcal{G} \equiv \mathbb{I}|^0 + X|^2,\end{equation}
and the star product $\m_2|^2$ is restricted to cyclic Ramond number 2. The utility of $\mathcal{G}$ is that it cancels the operator $\mathcal{G}^{-1}$ in $\Omega_L$. Then, cyclicity of a product $\mathcal{G}b_n$ with respect to the restricted symplectic form amounts to cyclicity of $b_n$ with respect to the ordinary symplectic form $\omega_L$ in the large Hilbert space. In our example, cyclicity of $\D$ is apparent since $\m_2|^2$ is cyclic with respect to $\omega_L$. One thing to be careful about, however, is that the operator $\mathcal{G}$ only cancels when $\Omega_L$ contracts suitably restricted states in the Ramond sector. Therefore the property 
\begin{equation}
\langle\Omega_L|\pi_2\D=0
\end{equation}
holds only if $\Omega_L$ pairs allowed Ramond states as described at the end of the previous section. In the WZW-like action, this is ensured by the fact that $\D$ always acts on suitable combinations of potentials satisfying flatness conditions.

Therefore we require that $\F^{-1}$ satisfies
\begin{equation}
\F^{-1}(\Q + \m_2|_2) = (\Q+\mathcal{G}\m_2|^2 +...)\F^{-1}.
\end{equation}
Projecting on the left with $\pi_1$ and on the right with $\pi_2$ we conclude that
\begin{equation}
F^{\mathrm{inv}}_2(Q\otimes\mathbb{I} +\mathbb{I}\otimes Q)+m_2|_2 = Q F^{\mathrm{inv}}_2+\mathcal{G}m_2|^2.
\end{equation}
This is equivalent to 
\begin{equation}[\Q,{\bf F}^{\mathrm{inv}}_2] = -X\m_2|_0^2.\end{equation}
Together with \eq{nF2}, this implies that $F^{\mathrm{inv}}_2$ can be written
\begin{equation}F^{\mathrm{inv}}_2 = -\Xi m_2|_0^2,\label{eq:F2inv}\end{equation}
up to a term which is both $Q$ and $\eta$ exact, which we will set to zero. 

We may now go on to compute the 3-string product $F^{\mathrm{inv}}_3$. Projecting \eq{CFinv} on the left with $\pi_1$ and on the right with $\pi_3$ implies
\begin{equation}
F^{\mathrm{inv}}_3(\eta\otimes\mathbb{I}\otimes\mathbb{I} +\mathbb{I}\otimes\eta\otimes\mathbb{I}+\mathbb{I}\otimes\mathbb{I}\otimes\eta)-F^{\mathrm{inv}}_2(m_2|_0\otimes\mathbb{I}+\mathbb{I}\otimes m_2|_0) = \eta F^{\mathrm{inv}}_3 -m_2|^0(F^{\mathrm{inv}}_2\otimes\mathbb{I}+\mathbb{I}\otimes F^{\mathrm{inv}}_2).
\end{equation}
The second term on the left hand side vanishes by associativity of the star product, while the second term on the right hand side vanishes since $F^{\mathrm{inv}}_2$ necessarily produces a Ramond state, while $m_2|^0$ vanishes when multiplying Ramond states. We therefore conclude that
\begin{equation}[\n,{\bf F}^{\mathrm{inv}}_3] = 0.\end{equation}
By a similar argument we can show that the higher products are also $\eta$-closed. We claim that it is consistent with cyclicity of $\D$ to take $F_n^{\mathrm{inv}}$ to vanish identically for $n\geq 3$. Therefore $\F^{-1}$ will be given by
\begin{eqnarray}
F^{\mathrm{inv}}_0 \lineup = 0,\\
F^{\mathrm{inv}}_1\lineup = \mathbb{I},\\
F^{\mathrm{inv}}_2\lineup = -\Xi m_2|_0^2,\\
F^{\mathrm{inv}}_n\lineup = 0,\ \ \ \ \ \ \ \ \ \ n\geq 3.
\end{eqnarray}
We will prove cyclicity of $\D$ in the next section.

To complete the story, let us derive explicit expressions for the products defining $\F$ and $\D$. The cohomomorphism 
$\F$ is completely specified by its output on the 1-string component of the tensor algebra:
\begin{equation}\pi_1\F = \sum_{n=0}^\infty F_n \pi_n.\end{equation}
To determine the products $F_n$, we use the formula
\begin{equation}\F^{-1}\F = \mathbb{I}_{T\mathcal{H}}\end{equation}
and project with $\pi_1$ on the left and $\pi_n$ on the right. This gives 
\begin{equation}F_n +F_2^{\mathrm{inv}}\sum_{k=0}^n F_k\otimes F_{n-k}=0.\label{eq:Frec}\end{equation}
From this we find
\begin{eqnarray}
F_0\lineup = 0,\\
F_1\lineup = \mathbb{I},\\
F_2\lineup = \Xi m_2|_0^2, \\\
F_n \lineup = \Xi m_2|_0^2(F_{n-1}\otimes\mathbb{I}+\mathbb{I}\otimes F_{n-1}),\ \ \ \ n\geq 3.
\end{eqnarray}
Note that the sum in \eq{Frec} simplifies since the formula itself implies that $F_n$ always produce Ramond states for 
$n\geq 2$, but the star product $m_2|_0^2$ can multiply at most one Ramond state. The dynamical $A_\infty$ structure is defined by its output on the 1-string component of the tensor algebra:
\begin{equation}\pi_1\D = \sum_{n=1}^{\infty} D_n\pi_n.\end{equation}
Projecting \eq{openD} on the left with $\pi_1$ and the right with $\pi_n$, we find
\begin{eqnarray}
D_1 \lineup = Q,\Vspace\label{eq:D1}\\
D_2 \lineup = \mathcal{G} m_2|^2,\Vspace\\
D_n\lineup = \mathcal{G} m_2|^2 \sum_{k=1}^{n-1}F_k\otimes F_{n-k},\ \ \ \ n\geq 3.\label{eq:Dn}
\end{eqnarray}
In this case the sum does not simplify since $m_2|^2$ can accept two Ramond inputs. This completes the definition of the dynamical and constraint $A_\infty$ structures. 

\subsection{Ramond Vertices and Feynman Diagrams}
\label{subsec:Feynman}

The construction of the previous section has an interesting interpretation in terms of the decomposition model of an $A_\infty$ algebra \cite{Kontsevich1,Kajiura}. Consider an $A_\infty$ algebra of the form
\begin{equation}\M = \Q +\delta\M,\end{equation}
where $\Q$ plays the role of the the BRST operator and $\delta\M$ represents the products $M_n$ for $n\geq 2$. Suppose the state space $\H$ takes the form of a direct sum
\begin{equation}\H = \H_p\oplus\H',\end{equation}
where the subspace $\H_p$ is preserved by the action of $Q$ and contains representatives of all elements of the cohomology of $Q$. In particular, $Q$ has no cohomology on the complementary subspace $\H'$. We introduce a projection operator $\Pi$ onto the subspace $\H_p$ which commutes with $Q$:
\begin{equation}[Q,\Pi] = 0.\end{equation}
The {\it decomposition theorem} \cite{Kajiura} says that $\M$ is $A_\infty$-isomorphic to a direct sum of an $A_\infty$ algebra which acts as $\Q$ on $\H'$ and an $A_\infty$ algebra called the {\it minimal model} which acts on $\H_p$, denoted $\M_\mathrm{min}$. Specifically, there exists an invertible cohomomorphism $\J$ satisfying
\begin{equation}
\M_\mathrm{decomp} = \J^{-1}\M\J,
\end{equation}
where the {\it decomposition model} $\M_\mathrm{decomp}$ takes the form
\begin{equation}\M_\mathrm{decomp} = (\mathbb{I}-\Pi)\Q +\M_\mathrm{min},\end{equation}
and the products $M_{\mathrm{min},n+1}$ of the minimal model have the property that they vanish when multiplying states in $\H'$. Often in the mathematics literature $\H_p$ is assumed to be isomorphic to the cohomology of $Q$. For physical applications in string field theory it is more useful to identify $\H_p$ with a slightly larger space given by states satisfying the mass-shell condition $L_0=0$ \cite{Konopka}. In this context, the products of the minimal model compute tree-level $S$-matrix elements when contracted with BRST invariant string states. Therefore, the decomposition theorem makes the remarkable claim that a string field theory action characterized by a cyclic $A_\infty$ algebra $\M$ can be related by field redefinition to an action which has the standard kinetic term,
\begin{equation}\frac{1}{2}\omega(\Psi,Q\Psi),\end{equation}
but whose interaction vertices only couple states satisfying $L_0=0$. When these states are BRST invariant, the vertices coincide with the on-shell tree-level $S$-matrix elements of string theory. The minimal model is unique up to an $A_\infty$ isomorphism on $\H_p$. There are various possible realizations of the minimal model. However, for present purposes we are interested in a specific and canonical construction characterized by Feynman graphs~\cite{Kontsevich2}. 

Let us describe the $A_\infty$ isomorphism $\J$ in this approach. For this we note that because $Q$ has no cohomology in $\H'$, we should be able to find a degree odd contracting homotopy $Q^+$ satisfying
\begin{equation}[Q,Q^+] = 1-\Pi.\label{eq:prop}\end{equation}
The operator $Q^+$ is called the {\it propagator}. The construction assumes that the propagator defines a Hodge decomposition, which requires the additional properties
\begin{equation}(Q^+)^2 = 0,\ \ \ \ Q^+\Pi = \Pi Q^+ = 0.\end{equation}
For example, open string field theory amplitudes computed in Siegel gauge are characterized by the projection operator 
\begin{equation}\Pi = e^{-\infty L_0},\end{equation}
and $Q^+$ is the Siegel gauge propagator
\begin{equation}Q^+ = \frac{b_0}{L_0}.\end{equation}
The cohomomorphism $\J^{-1}$ can be defined through its products
\begin{equation}\pi_1\J^{-1} = \sum_{n=0}^{\infty}J_n^\mathrm{inv}\pi_n.\end{equation}
The decomposition model $\M^\mathrm{decomp}$ shares the same 1-product (namely $Q$) with $\M$, so we assume $J_0^\mathrm{inv}=0$ and $J_1^\mathrm{inv}=\mathbb{I}$. The product $J_2^\mathrm{inv}$ is defined by
\begin{equation}
J_2^\mathrm{inv} = Q^+M_2 -\Pi M_2 Q^+\otimes\mathbb{I} -\Pi M_2 \Pi\otimes Q^+,
\end{equation}
where $M_2$ is the 2-product of $\M$. We now determine the 2-product of the minimal model by equating
\begin{equation}
\J^{-1}(\Q+\M_2+...) = (\Q+\M_{\mathrm{min},2})\J^{-1}.
\end{equation}
Projecting on the left with $\pi_1$ and on the right with $\pi_2$ we find
\begin{equation}
\M_{2,\mathrm{min}} = \M_2 - [\Q,\J_2^\mathrm{inv}].
\end{equation}
Computing the right hand side with \eq{prop} gives
\begin{eqnarray}
M_{2,\mathrm{min}}\lineup = M_2 - (\mathbb{I}-\Pi)M_2-\Pi M_2(\mathbb{I}-\Pi)\otimes\mathbb{I}-\Pi M_2\Pi\otimes(\mathbb{I}-\Pi)\nonumber\\
\lineup = \Pi M_2\Pi\otimes \Pi.
\end{eqnarray}
Therefore $M_{2,\mathrm{min}}$ multiplies states in $\H'$ to zero, as desired. We may continue to construct the higher products of $\J^{-1}$, but we will not need them.

\begin{figure}
\begin{center}
\resizebox{6.5in}{3.5in}{\includegraphics{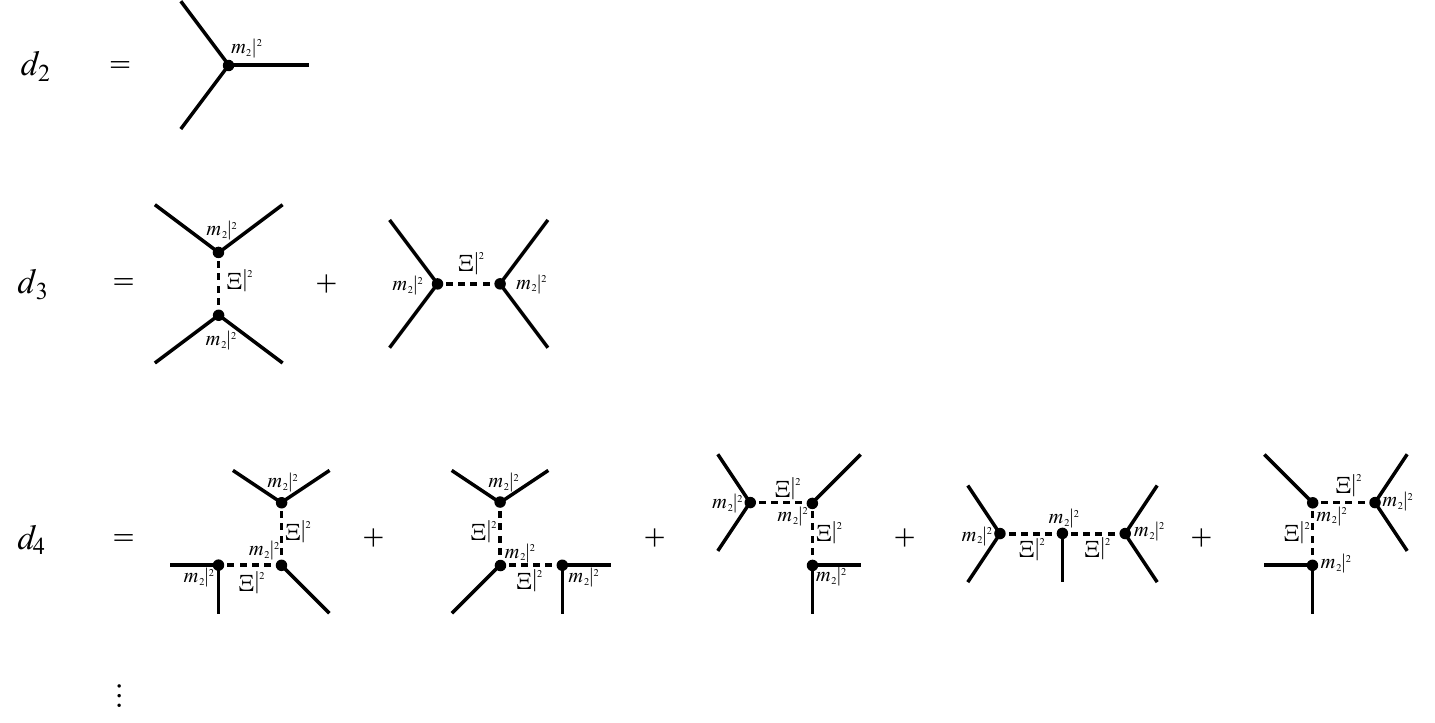}}
\end{center}
\caption{\label{fig:Dgraph} Feynman diagram expansion of the products of ${\bf d}$. We can choose output of the product $d_n$ to correspond to be the external leg in the left hand corner, and the $1$st through $n$th inputs to correspond to the other external legs proceeding counterclockwise. In this case, the diagrams correspond in order to the terms written in 
\eq{d2}-\eq{d4}.} 
\end{figure}

One might notice that the construction of the decomposition model addresses the same general problem that we encountered in defining the constraint $A_\infty$ structure for Ramond states. Namely, we wish to ``squeeze" all nonlinearities of the constraint $A_\infty$ structure into a subspace $\H_p$ (consisting of NS states), so that on the complementary subspace $\H'$ (consisting of Ramond states) the $A_\infty$ structure acts only as $\n$. We can make the identifications
\begin{eqnarray}
\H_p\ \ \lineup \rightarrow\ \ \mathrm{NS\ states}, \\
\H'\ \ \lineup \rightarrow\ \ \mathrm{R\ states}, \\
\Pi\ \ \lineup \rightarrow\ \ \mathbb{I}|^0,\\
\Q\ \ \lineup\rightarrow\ \ \n,\\
\M\ \ \lineup \rightarrow\ \ \n-\m_2|_0,\\
\M_\mathrm{decomp}\ \ \lineup \rightarrow\ \ \n-\m_2|^0,\\
\J\ \ \lineup \rightarrow\ \ \F.
\end{eqnarray}
In this context the ``propagator" is defined by
\begin{equation}Q^+\ \ \rightarrow\ \ \Xi|^2,\end{equation}
which acts as $\Xi$ on Ramond states and as $0$ on NS states. Following the previous paragraph, we may write down an expression for the 2-string product of $\F^{-1}$:
\begin{eqnarray}
F_2^{\mathrm{inv}} \lineup = -\Xi|^2 m_2|_0+\mathbb{I}|^0\,m_2|_0(\Xi|^2\otimes\mathbb{I})+\mathbb{I}|^0\,m_2|_0(\mathbb{I}|^0\otimes\Xi|^2)\nonumber\\
\lineup =-\Xi m_2|_0^2.
\end{eqnarray}
The second two terms vanish acting on any combination of NS and R states, and the first term is equivalent to \eq{F2inv}. The connection to Feynman diagrams, however, suggests \eq{F2inv} may be more naturally expressed as 
\begin{equation}
F_2^{\mathrm{inv}} = -\Xi|^2 m_2|^2,
\end{equation}
where $\Xi|^2$ is the ``propagator" and $m_2|^2$ defines a cyclic cubic vertex. We have seen that the higher products in $\F^{-1}$ can be chosen to vanish. In this way we obtain the decomposition model 
\begin{equation}\F^{-1}(\n-\m_2|_0)\F = \mathbb{I}|^2\n+\M_\mathrm{min},\end{equation}
where the products of the minimal model are defined by
\begin{eqnarray}
M_{\mathrm{min},1} \lineup =  \mathbb{I}|^0\eta,\\
M_{\mathrm{min},2} \lineup = -\mathbb{I}|^0 m_2|_0(\mathbb{I}|^0\otimes\mathbb{I}|^0).
\end{eqnarray}
This can be written more simply as 
\begin{equation}\F^{-1}(\n-\m_2|_0)\F = \n-\m_2|^0.\end{equation}
The decomposition model usually contains higher products given by a Feynman graph expansion with the vertices of $\M$ connected by propagators. In this example the higher products drop out since the products of the minimal model only operate on NS states, but the propagator is only nonzero acting on Ramond states. 

The Feynman graphs do however appear in the dynamical $A_\infty$ structure $\D$. From 
\eq{D1}-\eq{Dn} it is clear that $\D$ can be expressed in the form 
\begin{equation}\D = \Q+\mathcal{G}{\bf d},\label{eq:bfD}\end{equation}
where the coderivation ${\bf d}$ is defined to satisfy
\begin{equation}
\pi_1{\bf d} = \pi_1\m_2|^2 \F.\label{eq:bfd}
\end{equation}
Moreover, $\F$ can be defined 
\begin{equation}
\pi_1\F = \pi_1(\mathbb{I}_{T\mathcal{H}} + \Xi|^2 {\bf d}).\label{eq:bfFd}
\end{equation}
Together this gives a recursive definition of ${\bf d}$. For illustrative purposes, let us write out the first few products explicitly: 
\begin{eqnarray}
d_1\lineup = 0,\\
d_2\lineup = m_2|^2\label{eq:d2}\\
d_3\lineup = m_2|^2(\Xi|^2 m_2|^2\otimes \mathbb{I}) + m_2|^2(\mathbb{I}\otimes \Xi|^2 m_2|^2) ,\\
d_4 \lineup = m_2|^2(\Xi|^2 m_2|^2\otimes \mathbb{I})(\Xi|^2 m_2|^2\otimes \mathbb{I}\otimes\mathbb{I})+m_2|^2(\Xi|^2 m_2|^2\otimes \mathbb{I})(\mathbb{I}\otimes \Xi|^2 m_2|^2\otimes \mathbb{I})\nonumber\\
\lineup \ \ \ +m_2|^2(\mathbb{I}\otimes\Xi|^2 m_2|^2)(\mathbb{I}\otimes \Xi|^2 m_2|^2\otimes \mathbb{I})+m_2|^2(\mathbb{I}\otimes\Xi|^2 m_2|^2)(\mathbb{I}\otimes\mathbb{I}\otimes \Xi|^2 m_2|^2)\nonumber\\
\lineup\ \ \ +m_2|^2(\Xi|^2 m_2|^2\otimes \Xi|^2 m_2|^2),\label{eq:d4}\\
\lineup\vdots\nonumber\ .
\end{eqnarray}
For each term in each product we can draw a corresponding diagram as follows. We represent the output of the product by a line which meets a cubic vertex corresponding $m_2|
^2$. The line then splits into two lines. Each line either reaches an input of the product, or it carries an insertion of $\Xi|^2$ and meets another cubic vertex corresponding to $m_2|^2$, at which point the line again splits in two. Applying this process iteratively, it is clear that the products can be described by a sum of color-ordered Feynman graphs, with cubic vertices representing $m_2|^2$ and propagators representing $\Xi|^2$. For the products $d_2,d_3,d_4$, this is illustrated in figure \ref{fig:Dgraph}.

\begin{figure}
\begin{center}
\resizebox{3.6in}{1in}{\includegraphics{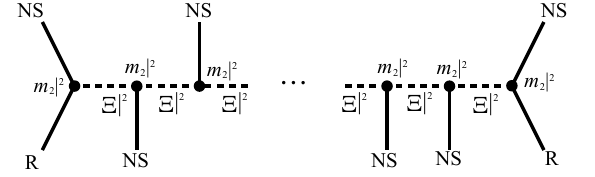}}
\end{center}
\caption{\label{fig:gend} All nonvanishing diagrams take the form of a line of propagators connecting a vertex with an NS and R external state to a vertex with an NS and R external state. The propagators are connected with vertices containing one NS external state.} 
\end{figure}

Let us make some observations based on this Feynman graph picture. First note that diagrams with three propagators meeting at a vertex will always vanish, since there is no cubic coupling of three Ramond states. This implies that all nonvanishing diagrams are characterized by a single line of connected propagators which start on a vertex containing two external states and terminate on a vertex containing two external states. The propagators are connected in sequence by vertices containing one external state. We also know that the line of propagators must start with a cubic vertex containing one NS and one R external state, and must end with a cubic vertex containing one NS and one R external state. All other external states must be NS, or the diagram vanishes. This is shown in figure \ref{fig:gend}. From this it is clear that the products $d_n$ carry cyclic Ramond number 2.

This representation makes it manifest that the dynamical $A_\infty$ structure is cyclic, since color-ordered amplitudes constructed by Feynman diagrams are cyclic. Let us give the proof. Since $Q$ is cyclic with respect to the restricted symplectic form, and the factor of $\mathcal{G}$ cancels when contracting appropriate states, the cyclicity of $\D$ effectively requires 
\begin{equation}\langle\omega_L|\pi_2{\bf d} = 0.\end{equation}
To show this, we represent the projection $\pi_2$ as 
\begin{equation}\pi_2 = \inverttriangle(\pi_1\otimes'\pi_1)\triangle,\end{equation}
where $\otimes'$ denotes the tensor product of tensor algebras, $\triangle$ is the coproduct, and $\inverttriangle$ is the product. See appendix A of \cite{WB} for explanation of this formalism. Acting the coproduct on ${\bf d}$ gives the expression 
\begin{equation}
\langle\omega_L|\pi_2{\bf d} = \langle\omega_L|\inverttriangle\Big(\pi_1{\bf d}\otimes'\pi_1+\pi_1\otimes'\pi_1{\bf d}\Big)\triangle.
\end{equation}
Next using \eq{bfFd} we can write 
\begin{equation}\pi_1 = \pi_1\F -\pi_1\Xi|^2{\bf d},\end{equation}
which gives 
\begin{equation}
\langle\omega_L|\pi_2{\bf d} = \langle\omega_L|\inverttriangle\Big(\pi_1{\bf d}\otimes'\pi_1\F+\pi_1\F\otimes'\pi_1{\bf d}-\pi_1{\bf d}\otimes'\pi_1\Xi|^2{\bf d}-\pi_1\Xi|^2{\bf d}\otimes'\pi_1{\bf d}\Big)\triangle.
\end{equation}
The last two terms amount to
\begin{equation}\langle\omega_L|(\mathbb{I}\otimes\Xi|^2-\Xi|^2\otimes\mathbb{I})\inverttriangle\big(\pi_1{\bf d}\otimes'\pi_1{\bf d}\big)\triangle,\end{equation}
which vanishes due to the BPZ even property of $\Xi$. Meanwhile, using \eq{bfd} we can write the first two terms
\begin{equation}
\langle\omega_L|\pi_2{\bf d} = \langle\omega_L|\inverttriangle\Big(\pi_1\m_2|^2\F\otimes'\pi_1\F+\pi_1\F\otimes'\pi_1\m_2|^2\F\Big)\triangle.
\end{equation}
In this form we can pull the coproduct back to the left, giving 
\begin{equation}
\langle\omega_L|\pi_2{\bf d} = \langle\omega_L|\pi_2\m_2|^2\F,
\end{equation}
which vanishes by cyclicity of the open string star product, \eq{m2cyc}. This demonstration can serve as a substitute for the proof of cyclicity given in \cite{RWaction}, and is much simpler. On the other hand, the Feynman diagram picture does not make it obvious that $\D$ is nilpotent and commutes with $\C$. We demonstrate this  in appendix \ref{app:DC}.

\subsection{Solution of Flatness Conditions}

To complete the construction of the WZW-like action, we must find a solution of the flatness conditions. There are many possible solutions, giving actions which differ (at most) by field redefinition. Since we are interested in coupling Berkovits' NS open superstring field theory to the Ramond sector, we seek an expression for the potentials which simplifies to the standard Maurer-Cartan elements \eq{Aeta}-\eq{Adelta} when the Ramond string field is set to zero. 

\subsubsection{A General Procedure}
\label{subsubsec:flat}

To find a unique solution of the flatness conditions we will ``fix a gauge" characterized by the assumption that all higher form potentials with an index $t$ vanish. Note that this is a gauge fixing from the perspective of the symmetry \eq{pottrans} of the flatness conditions, not from the perspective of the gauge symmetry of the action.  A special property of this gauge is that it leads to a first order differential equation in $t$ which can be integrated to give an explicit solution for the potentials. For this procedure to work we require the following assumptions: 
\begin{description}
\item{(1)} We are given (or have already found) potentials $A_C$ and $A_t$ satisfying 
\begin{eqnarray}
0\lineup =\pi_1\C\frac{1}{1-A_C},\label{eq:1flatC}\\
\dot{A}_C \lineup = \pi_1\C\frac{1}{1-A_C}\otimes A_t\otimes \frac{1}{1-A_C}.\label{eq:1flatt}
\end{eqnarray}
We also require that $A_C$ vanishes at $t=0$:
\begin{eqnarray}A_C|_{t=0} = 0.\end{eqnarray}
It is always possible to arrange this (see footnote \ref{foot:AC}).
\item{(2)} The multiform coderivation can be expanded
\begin{equation}\del = \C +dt\del_t+\del',\end{equation}
where $\del'$ contains no components proportional to the basis 1-form $dt$. As it happens, $\del'$ would only require components proportional to $dt$ if some coderivations inside $\del$ depended explicitly on $t$. This does not occur for the applications that interest us. 
\item{(3)} The multiform potential can be expanded as 
\begin{equation}A=A_C + dt A_t + A',\end{equation}
where $A'$ contains no components proportional to $dt$. In particular, all higher form potentials with an index $t$ must vanish. This condition can be viewed as a gauge fixing from the perspective of the symmetry \eq{pottrans} of the choice of potentials. Consistency will also require 
\begin{equation}A'|_{t=0} = 0.\end{equation}
Therefore all potentials besides $A_t$ are assumed to vanish at $t=0$.
\end{description}
With these assumptions, the component of the flatness conditions proportional to $dt$ can be written as: 
\begin{eqnarray}
\dot{A}' \lineup = \pi_1\del'\frac{1}{1-A_C-A'}\otimes A_t\otimes\frac{1}{1-A_C-A'}\nonumber\\
\lineup\ \ \  + \pi_1\C\frac{1}{1-A_C}\otimes
A'\otimes\frac{1}{1-A_C-A'}\otimes A_t\otimes\frac{1}{1-A_C-A'}\otimes A'\otimes \frac{1}{1-A_C}\nonumber\\
\lineup \ \ \ + \pi_1\C\left[\frac{1}{1-A_C}\otimes A'\otimes\frac{1}{1-A_C-A'}\otimes A_t\otimes\frac{1}{1-A_C}+\frac{1}{1-A_C}\otimes A_t\otimes\frac{1}{1-A_C-A'}\otimes A'\otimes\frac{1}{1-A_C}\right].\ \ \ \ \ \ \ \ \ \ \label{eq:flatdiff}
\end{eqnarray}
We can solve this recursively as follows. The 1-form component of \eq{flatdiff} gives a linear inhomogeneous differential equation which can be integrated to find the 1-potential. The 2-form component of \eq{flatdiff} is a linear inhomogeneous differential equation which can be integrated to find the 2-potential {\it provided} the 1-potential has been determined in the previous step. Continuing, it is clear that the $n$-form component of \eq{flatdiff} is a linear inhomogeneous differential equation which may be integrated to find the $n$-potential provided the $k$-potentials for $k<n$ have been obtained in previous steps. With each iteration of this procedure, we must solve a linear inhomogeneous differential equation of the general form:
\begin{equation}
\dot{a}(t) = b(t) + \mathcal{O}(t) a(t),\label{eq:linflatdiff}
\end{equation}
where $a(t), b(t)$ are elements of a vector space and $\mathcal{O}(t)$ is a linear operator on the vector space. In the present situation $a(t)$ corresponds to the $n$-potential, $b(t)$ arises from the first two terms in \eq{flatdiff}, and the linear operator $\mathcal{O}(t)$ arises from the last term in \eq{flatdiff}. The expressions for $b(t)$ and $\mathcal{O}(t)$ depend on potentials obtained in previous iterations. The desired solution of \eq{linflatdiff} is
\begin{equation}
a(t) = \int_0^t dt_1 \overleftarrow{\mathcal{P}}e^{\int_{t_1}^t dt_2 \mathcal{O}(t_2)}b(t_1),\label{eq:intlinflatdiff}
\end{equation}
where the path-ordering operator $\overleftarrow{\mathcal{P}}$ places $\mathcal{O}(t_2)$ from left to right in sequence of decreasing $t_2$. Note that $a(t)$ vanishes at $t=0$, which in particular implies that $A'$ vanishes at $t=0$ as required by our assumptions. In this way we may derive an explicit expression for the potentials to all orders.

Note that \eq{flatdiff} comes from the component of the flatness conditions proportional to $dt$. However, we must verify that other components of the flatness conditions are also satisfied. This requires 
\begin{equation}(\C+\del')\frac{1}{1-A_C-A'} = 0.\end{equation}
To see that this is satisfied, we take the derivative of the left hand side with respect to $t$ and use \eq{flatdiff}. This gives
\begin{eqnarray}
\lineup\frac{d}{dt}\pi_1(\C+\del')\frac{1}{1-A_C-A'}\nonumber\\
\lineup   = -\pi_1(\C+\del')\frac{1}{1-A_C-A'}\otimes\left(\pi_1(\C+\del')\frac{1}{1-A_C-A'}\right)\otimes\frac{1}{1-A_C-A'}\otimes A_t\otimes\frac{1}{1-A_C-A'}\nonumber\\
\lineup\ \ \ -\pi_1(\C+\del')\frac{1}{1-A_C-A'}\otimes A_t\otimes\frac{1}{1-A_C-A'}\otimes \left(\pi_1(\C+\del')\frac{1}{1-A_C-A'}\right)\otimes\frac{1}{1-A_C-A'}\ \ \ \ \ \ 
\end{eqnarray}
This is a first order homogeneous differential equation in the string field $\pi_1(\C+\del')\frac{1}{1-A_C-A'}$. Since this vanishes at  $t=0$ (because $A_C$ and $A'$ vanish at $t=0$), this implies that $\pi_1(\C+\del')\frac{1}{1-A_C-A'}$ must vanish for all $t$. Therefore all flatness conditions are obeyed provided we have a solution to \eq{flatdiff}.

One thing we need to verify is that the potentials derived by the above procedure reduce to the Maurer-Cartan elements of the Berkovits theory once the Ramond string field is set to zero. In this connection, note that in the conventional formulation of the WZW-like action, the higher form potentials vanish. In particular, the higher potentials with an index $t$ vanish. Therefore, the above procedure applied to the Berkovits theory will necessarily express the nonvanishing potentials as Maurer-Cartan elements, provided $A_\eta$ and $A_t$ are assumed to be Maurer-Cartan elements. Once we include the Ramond sector, the potentials will be more complicated, but they should still reduce to standard Maurer-Cartan elements if the Ramond string field is set to zero. 

\subsubsection{Expressions for the Potentials}
\label{subsubsec:Berkpot}

We now apply the above procedure to find concrete expressions for the potentials $A_C,A_t,A_\delta$ and $A_D$ for open superstring field theory including the Ramond sector. It is useful to decompose the potentials into NS and R components,
\begin{equation}A_C = N_C+R_C,\ \ \ \ A_i = N_i+R_i,\end{equation}
where $N$ and $R$ indicate NS and R sector fields, respectively. To start the recursion we must postulate suitable expressions for $A_C$ and $A_t$: 
\begin{eqnarray}
A_C \lineup = (\eta e^{\PhiN(t)})e^{-\PhiN(t)} + \eta\PhiR(t),\label{eq:RNSAC}\\ 
A_t\lineup =\left(\frac{d}{dt}e^{\PhiN(t)}\right)e^{-\PhiN(t)}+\dot{\Phi}_\mathrm{R}(t).\label{eq:RNSAt}
\end{eqnarray} 
where $\PhiN(t)$ and $\PhiR(t)$ are interpolations of the dynamical fields $\PhiN$ and $\PhiR$ and multiplication is performed with the open string star product. We will often switch back and forth between degree and Grassmann gradings to make contact with more familiar expressions. Note that $N_C$ and $N_t$ are what we were calling $A_\eta$ and $A_t$ in subsection \ref{subsec:BerkWZW}. In particular, $A_C$ and $A_t$ reduce to the standard Maurer-Cartan elements \eq{Aeta} and \eq{At} if we ignore the Ramond sector. We have the relations
\begin{eqnarray}
0\lineup =\eta N_C - N_C^2, \\
0 \lineup = \eta R_C,\\
0\lineup = \eta N_t -\dot{N}_C - [N_C,N_t],\\
0\lineup = \eta R_t -\dot{R}_C.
\end{eqnarray}
These equations are equivalent to \eq{1flatC} and \eq{1flatt} with $\C = \n - \m_2|^0$. Note also that $A_C$ vanishes at $t=0$.

From this starting point we can use \eq{intlinflatdiff} to derive expressions for $A_\delta$ and $A_D$. Let us start with $A_\delta$ since this is simpler. Extracting the 1-form component of \eq{flatdiff} proportional to $dx^\delta$ gives the equation
\begin{eqnarray}
\dot{A}_\delta \lineup= \delta A_t
-\pi_1\C\frac{1}{1-A_C}\otimes A_\delta\otimes \frac{1}{1-A_C}\otimes A_t\otimes\frac{1}{1-A_C}\nonumber\\
\lineup\ \ \ \ \ \ \ \ \ +\pi_1\C\frac{1}{1-A_C}\otimes A_t\otimes \frac{1}{1-A_C}\otimes A_\delta\otimes\frac{1}{1-A_C}\nonumber\\
\lineup = \delta A_t +m_2|^0(A_\delta\otimes A_t - A_t\otimes A_\delta),
\end{eqnarray}
where in the last step we substituted $\C = \n-\m_2|^0$. This can be equivalently expressed
\begin{eqnarray}
\dot{N}_\delta \lineup = \delta N_t+[N_t,N_\delta],\\
\dot{R}_\delta \lineup = \delta R_t.
\end{eqnarray}
Comparing to \eq{linflatdiff}, we see that for the NS equation the role of $b(t)$ is played by $\delta N_t$, and the role of $\mathcal{O}(t)$ is played by
\begin{equation}
[N_t,\cdot] = \mathrm{ad}_{N_t}.
\end{equation}
Meanwhile, in the Ramond equation the role of $b(t)$ is played by $\delta R_t$ and $\mathcal{O}(t)$ vanishes. Substituting into \eq{intlinflatdiff} we therefore obtain
\begin{eqnarray}
N_\delta \lineup = \int_0^t dt_1\overleftarrow{\mathcal{P}}e^{\int_{t_1}^t dt_2\mathrm{ad}_{N_t(t_2)}}\delta N_t(t_1),\\
R_\delta \lineup = \int_0^t dt_1 \delta R_t(t_1).
\end{eqnarray}
To simplify the expression for $N_\delta$, we note the equality
\begin{equation}
\overleftarrow{\mathcal{P}}e^{\int_{t_1}^t dt_2 \mathrm{ad}_{N_t(t_2)}} = e^{\PhiN(t)}e^{-\PhiN(t_1)} (\,\cdot\,) e^{\PhiN(t_1)}e^{-\PhiN(t)}.
\end{equation}
Then we can rewrite $N_\delta$
\begin{eqnarray}
N_\delta \lineup = e^{\PhiN(t)}\left[\int_0^t dt_1 e^{-\PhiN(t_1)}\delta\left(\left(\frac{d}{dt_1}e^{\PhiN(t_1)}\right)e^{-\PhiN(t_1)}\right)e^{\PhiN(t_1)}\right]e^{-\PhiN(t)}.
\end{eqnarray}
With some manipulation the integrand can be written as a total derivative:
\begin{equation}
N_\delta = e^{\PhiN(t)}\left[\int_0^t dt_1\frac{d}{dt_1}\left(e^{-\PhiN(t_1)}\delta e^{\PhiN(t_1)}\right)\right]e^{-\PhiN(t)}.
\end{equation}
We may also perform the integral for $R_\delta$ to obtain
\begin{equation} 
A_\delta = (\delta e^{\PhiN(t)})e^{-\PhiN(t)} + \delta\PhiR(t).\label{eq:RNSAdelta}
\end{equation}
This is analogous to the expressions \eq{RNSAC} and \eq{RNSAt} for $A_C$ and $A_t$. 

The potential $A_D$ has a more nontrivial structure. To make contact with the formulas of \cite{complete}, it is helpful to introduce
\begin{equation}D_\eta \equiv \eta  - \mathrm{ad}_{N_C},\end{equation}
and
\begin{equation}
F \equiv \frac{1}{\mathbb{I}-\Xi\mathrm{ad}_{N_C}} ,
\end{equation}
where
\begin{equation}\mathrm{ad}_{N_C} = [N_C,\cdot].\end{equation}
Using the identities of appendix \ref{app:FF} one can show that the differential equation for the NS are R components of $A_D$ takes the form
\begin{eqnarray}
\dot{N}_D\lineup = QN_t+[F(R_C),F(R_t)]+[F(R_C),F(\Xi[F(R_C),N_t])]+[N_t,N_D],\Vspace\\
\dot{R}_D\lineup = Q R_t +X[F(R_C),N_t]+X[N_C,F(R_t)+F(\Xi[F(R_C),N_t])].\Vspace
\end{eqnarray}
Substituting into \eq{intlinflatdiff} gives the expressions
\begin{eqnarray}
N_D\lineup = e^{\PhiN(t)}\!\!\left[\int_0^t dt_1 e^{-\PhiN(t_1)}\!\!\!\left.\Big(\! QN_t\!+\![F(R_C),F(R_t)]\!+\![F(R_C),F(\Xi[F(R_C),N_t])]\!\Big)\!\right|_{t=t_1}\!\!\!
 e^{\PhiN(t_1)} \!\right]\!\!e^{-\PhiN(t)},\ \ \ \ \ \ \ \ \ \ \label{eq:ND}\\
R_D \lineup = \int_0^t dt_1 \left.\Big(Q R_t +X[F(R_C),N_t]+X[N_C,F(R_t)+F(\Xi[F(R_C),N_t])]\Big)\right|_{t=t_1}.\label{eq:RD}
\end{eqnarray}
In this case the integrand is not a total derivative. However, it is a total derivative up to ``irrelevant" terms; with some manipulation with the identities of \cite{complete} we can show that
\begin{eqnarray}
N_D\lineup = (Qe^{\PhiN(t)})e^{-\PhiN(t)}+\frac{1}{2}[F(R_C),F(\PhiR(t))]+D_\eta Z_1,\Vspace\\
R_D\lineup = Q\PhiR(t)-XF(R_C) + \eta Z_2,\Vspace
\end{eqnarray}
where the remainder terms are defined by string fields $Z_1$ and $Z_2$ which take the form
\begin{eqnarray}
Z_1\lineup=e^{\PhiN(t)}\left[\int_0^t dt_1 e^{-\PhiN(t_1)}\left.\left(\frac{1}{2}\left[F(\PhiR(t)),\frac{d}{dt}F(\PhiR(t))\right]+\frac{1}{2}[F(\PhiR(t),[F(\PhiR(t)),N_t]]\right.\right.\right.\nonumber\\
\lineup\ \ \ \ \ \ \ \ \ \ \ \ \ \ \ \ \ \ \ \ \ \ \ \ \ \ \ \ \ \ \ \ \ \ \ \ \ \ \ \ \ 
\left.\left.\left.\phantom{\frac{1}{2}}+[F(R_C),F(\Xi[F(\PhiR(t)),N_t])]\right)\right|_{t=t_1}e^{\PhiN(t_1)}\right]e^{-\PhiN(t)},\\
Z_2\lineup = X\int_0^t dt_1\left.\Big(F(R_t)+F(\Xi[F(R_C),N_t])\Big)\right|_{t=t_1}.
\end{eqnarray}
The remainders $D_\eta Z_1$ and $\eta Z_2$ are irrelevant in the sense that they can be absorbed by a ``gauge transformation" of the potentials \eq{pottrans}, at the cost of introducing a nonvanishing 2-potential $A_{Dt}$. The transformation is defined by 
\begin{equation}\Lambda = dx^D(Z_1+Z_2).\end{equation}
Once the irrelevant terms are removed, $A_D$ will depend on $t$ and $\Phi$ exclusively through the interpolation $\Phi(t)$. It would be interesting to find expressions for the higher potentials with this property. 

\subsubsection{Gauge Symmetry}

To describe the gauge invariance of open superstring field theory we must find a suitable expression for the potential~$A_g$. Though we can determine $A_g$ as above by demanding that higher potentials with an index $t$ vanish, this form of $A_g$ will not generate gauge transformations corresponding to $\eta$. Therefore we will take $A_g$ from the general expression \eq{Ag} given in subsection \ref{subsubsec:gauge}. We decompose the gauge parameters $\lambda_C$ and $\lambda_D$ into NS and R components
\begin{equation}
\lambda_C = \nu_C+\rho_C,\ \ \ \ \ \lambda_D = \nu_D+\rho_D,
\end{equation}
where $\nu$ and $\rho$ denote NS and R string fields. The gauge parameters carry ghost number $-1$ and picture
\begin{eqnarray}
\mathrm{picture}(\nu_C)\lineup =1,\ \ \ \ \ \ \mathrm{picture}(\nu_D) = 0,\nonumber\\
\mathrm{picture}(\rho_C)\lineup = 3/2,\ \ \ \mathrm{picture}(\rho_D)=1/2.
\end{eqnarray}
In addition $\rho_D$ must satisfy the constraint
\begin{equation}XY\eta\rho_D = \eta\rho_D,\end{equation}
to be compatible with the constraint on the Ramond string field. Since we have an explicit form for $\D$ and $\C$, we may evaluate \eq{Ag} using the formulas of appendix \ref{app:FF} to find the NS and R components of $A_g$: 
\begin{eqnarray}
N_g\lineup = D_\eta \nu_C +Q\nu_D +[F(R_C),F(\Xi[F(R_C),\nu_D])]+ [F(R_C),F(\rho_D)],\label{eq:gauge1}\\
R_g\lineup = \eta\Big(\rho_C+F(\rho_D)+F(\Xi[F(R_C),\nu_D])\Big)+Q\rho_D+X F(\eta\rho_D)-F(\Xi[F(R_C),D_\eta\nu_D]).\label{eq:gauge2}
\end{eqnarray}
The gauge transformation of $\Phi$ is given by equating $A_\delta$ and $A_g$ at $t=1$. From \eq{RNSAdelta} we therefore have 
\begin{equation}
(\delta e^{\PhiN})e^{-\PhiN} = N_g|_{t=1},\ \ \ \ \ \ \delta\PhiR = R_g|_{t=1}.
\end{equation}
Note that the gauge parameter $\rho_C$ implies that the action is independent of the component of $\PhiR$ in the small Hilbert space. In particular we can write $\PhiR$ in the form
\begin{equation}\PhiR = \Xi\eta\PhiR + \eta\Xi\PhiR.\end{equation}
Since the action does not depend on the second term, it can only depend on the Ramond string field $\PsiR=\eta\PhiR$ in the small Hilbert space. This is the formulation of open superstring field theory given by Kunitomo and Okawa. Let us compare our expression for the gauge symmetry to theirs. In \cite{complete} the gauge transformation takes the form:
\begin{eqnarray}
(\delta e^{\PhiN})e^{-\PhiN} \lineup = \left.\Big(D_\eta\Omega + Q\Lambda + [F(\PsiR),F(\Xi\lambda)] +[F(\PsiR),F(\Xi[F(\PsiR),\Lambda])]\Big)\right|_{t=1},\label{eq:KOgauge1}\\
\delta\Psi_R\lineup = \left.\Big(Q\lambda + X\eta F(\lambda)+X\eta F(\Xi D_\eta[F(\PsiR),\Lambda])\Big)\right|_{t=1}.\label{eq:KOgauge2}
\end{eqnarray}
If we equate 
\begin{equation}\PsiR = \eta\PhiR = R_C|_{t=1},\end{equation}
we see that \eq{KOgauge1} and \eq{KOgauge2} agrees with \eq{gauge1} and \eq{gauge2} if we identify
\begin{equation}
\Omega = \nu_C,\ \ \ \Lambda = \nu_D,\ \ \ -\Xi\lambda = \rho_D.
\end{equation}
The Kunitomo-Okawa formulation has no analogue of the gauge parameter $\rho_C$, since this gauge symmetry has been eliminated by formulating the Ramond sector in the small Hilbert space. Note that we do not {\it a priori} assume that $\rho_D$ takes the form $-\Xi\lambda$. It could have an additional contribution in the small Hilbert space. However, such a contribution can be absorbed into the definition of $\nu_C$, and so does not represent an independent gauge freedom. Therefore the gauge transformation we have derived is in agreement with the result of Kunitomo and Okawa.

\subsection{Relation to the Kunitomo-Okawa Action}
\label{subsec:complete}

In this section we prove the equality of the WZW-like form of the open superstring field theory action
\begin{equation}
S = \int_0^1 dt\,\Omega_L\left(A_t,\pi_1\D\frac{1}{1-A_C}\right),\label{eq:WZWlike}
\end{equation}
and the action of Kunitomo and Okawa\footnote{To compare with the expression written in \cite{complete}, note that what we call $N_C$ and $N_t$ here is called $A_\eta$ and $A_t$ there. We prefer to reserve the symbol $A$ for the complete potential that includes NS and R components. In addition, $\PhiN$ and $\PsiR$ here is written as $\Phi$ and $\Psi$ there. Also, the BPZ inner products in the small and large Hilbert space are written $\langle\cdot,\cdot\rangle_S$ and $\langle\cdot,\cdot\rangle_L$ here, which is equivalent to $-\langle\!\langle\cdot,\cdot\rangle\!\rangle$ and $\langle\cdot,\cdot\rangle$ there.}
\begin{equation}
S = \frac{1}{2}\big\langle Y\PsiR,Q\PsiR\big\rangle_S -\int_0^1 dt\, \big\langle N_t,QN_C+F(\PsiR)^2\big\rangle_L.
\end{equation}
Note that the Kunitomo-Okawa action does not utilize an interpolation for the Ramond string field. That is, $\PsiR$ above is independent of $t$, and is identified with $\eta\PhiR(t)$ evaluated at $t=1$. Another form of the action which is more similar to \eq{WZWlike} and includes an interpolation for $\PsiR$ was found in \cite{Matsunaga}.

The WZW-like form of the action includes a term with only NS string fields, and two terms with two Ramond string fields:
\begin{eqnarray}
S \lineup = \int_0^1 dt\,\Omega_L\left(N_t,\pi_1\D\frac{1}{1-N_C}\right) + \int_0^1 dt\,\Omega_L\left(R_t,\pi_1\D\frac{1}{1-N_C}\otimes R_C\otimes \frac{1}{1-N_C}\right)\nonumber\\
\lineup\ \ \ \ \ \ +\int_0^1 dt\, \Omega_L\left(N_t,\pi_1\D\frac{1}{1-N_C}\otimes R_C\otimes\frac{1}{1-N_C}\otimes R_C\otimes\frac{1}{1-N_C}\right).\label{eq:RNSWZWlike}
\end{eqnarray}
We know that the NS term reproduces the Berkovits action, so let us concentrate on the Ramond terms. The strategy is to write these terms as the integral of a total derivative, and in this way eliminate the interpolation $\PhiR(t)$ of the Ramond string field. Using cyclicity of $\D$ to apply the identity \eq{grpcyc}, we may rearrange the last term to the form
\begin{eqnarray}
\lineup \!\!\!\!\!\!\int_0^1 dt\, \Omega_L\left(N_t,\pi_1\D\frac{1}{1-N_C}\otimes R_C\otimes\frac{1}{1-N_C}\otimes R_C\otimes\frac{1}{1-N_C}\right)\nonumber\\
\lineup =-\frac{1}{2}\!\int_0^1\! dt\, \Omega_L\!\!\left(\!R_C,\pi_1\D\left(\!\frac{1}{1-N_C}\!\otimes\! N_t\!\otimes\!\frac{1}{1-N_C}\!\otimes\! R_C\!\otimes\frac{1}{1-N_C}\!+\!\frac{1}{1-N_C}\!\otimes\! R_C\!\otimes\!\frac{1}{1-N_C}\!\otimes\! N_t\!\otimes\!\frac{1}{1-N_C}\!\right)\right).\nonumber\\
\end{eqnarray}
In the first entry of $\Omega_L$ we use the fact that $R_C = \eta\PhiR(t)$, and move the $\eta$ to the second entry of
$\Omega_L$. This gives 
\begin{eqnarray}
\lineup\!\!\!\!\!\!\! \int_0^1 dt\, \Omega_L\left(N_t,\pi_1\D\frac{1}{1-N_C}\otimes R_C\otimes\frac{1}{1-N_C}\otimes R_C\otimes\frac{1}{1-N_C}\right)\nonumber\\
\lineup =\!-\!\frac{1}{2}\!\int_0^1\! dt\, \Omega_L\!\!\left(\!\PhiR(t),\pi_1\C\D\left(\!\frac{1}{1-N_C}\!\otimes\! N_t\!\otimes\!\frac{1}{1-N_C}\!\otimes\! R_C\!\otimes\frac{1}{1-N_C}\!+\!\frac{1}{1-N_C}\!\otimes\! R_C\!\otimes\!\frac{1}{1-N_C}\!\otimes\! N_t\!\otimes\!\frac{1}{1-N_C}\!\right)\right).\nonumber\\
\end{eqnarray}
Next we operate $\C$ on the second entry of the symplectic form. Using the rules for acting coderivations on group like elements, we obtain a sum of terms containing the string fields
\begin{eqnarray}
\pi_1\C\frac{1}{1-N_C}\lineup = 0,\\
\pi_1\C\frac{1}{1-N_C}\otimes R_C\otimes\frac{1}{1-N_C} \lineup = \eta R_C  = 0,\\
\pi_1\C\frac{1}{1-N_C}\otimes N_t\otimes \frac{1}{1-N_C}\lineup =D_\eta N_t = \dot{N}_C,\label{eq:KOstep1}\\
\pi_1\C\frac{1}{1-N_C}\otimes R_C\otimes\frac{1}{1-N_C}\otimes N_t\otimes\frac{1}{1-N_C} \lineup = m_2|^0R_C\otimes N_t = 0.
\end{eqnarray}
Only \eq{KOstep1} is contributes. This gives 
\begin{eqnarray}
\lineup\!\!\!\!\!\!\! \int_0^1 dt\, \Omega_L\left(N_t,\pi_1\D\frac{1}{1-N_C}\otimes R_C\otimes\frac{1}{1-N_C}\otimes R_C\otimes\frac{1}{1-N_C}\right)\nonumber\\
\lineup =\frac{1}{2}\!\int_0^1\! dt\, \Omega_L\!\!\left(\!\PhiR(t),\pi_1\D\left(\!\frac{1}{1-N_C}\!\otimes\! \dot{N}_C\!\otimes\!\frac{1}{1-N_C}\!\otimes\! R_C\!\otimes\frac{1}{1-N_C}\!+\!\frac{1}{1-N_C}\!\otimes\! R_C\!\otimes\!\frac{1}{1-N_C}\!\otimes\! \dot{N}_C\!\otimes\!\frac{1}{1-N_C}\!\right)\right)\nonumber\\
\lineup = \frac{1}{2}\int_0^1\! dt\, \Omega_L\!\!\left(\!\PhiR(t),\frac{d}{dt}\pi_1\D\!\frac{1}{1-N_C}\!\otimes\! R_C\!\otimes\frac{1}{1-N_C}-\pi_1\D\!\frac{1}{1-N_C}\!\otimes\! \dot{R}_C\!\otimes\!\frac{1}{1-N_C}\right).
\end{eqnarray}
Using cyclicity of $\D$ and then $\C$ we can write the second term
\begin{equation}
-\frac{1}{2}\int_0^1dt\,\Omega_L\left(\PhiR(t),\pi_1\D\frac{1}{1-N_C}\otimes \dot{R}_C \otimes\frac{1}{1-N_C}\right)=
-\frac{1}{2}\int_0^1dt\,\Omega_L\left(R_t,\pi_1\D\!\frac{1}{1-N_C}\!\otimes\! R_C\!\otimes\frac{1}{1-N_C}\right).
\end{equation}
This partially cancels the second term in \eq{RNSWZWlike}. Substituting the Berkovits action for the purely NS contribution, we can therefore write the action
\begin{eqnarray}
S\lineup  = -\int_0^1 dt\,\big\langle N_t,QN_C\big\rangle+\frac{1}{2}\int_0^1dt\Omega_L\left(\dot{\Phi}_\mathrm{R}(t),\pi_1\D\!\frac{1}{1-N_C}\!\otimes\! R_C\!\otimes\frac{1}{1-N_C}\right)\nonumber\\
\lineup\ \ \ \ \ \ \ \ \ \ \ \ \ \ \ \ \ \ \ \ \ \ \ \ \ \ \ \ \ \,
+\frac{1}{2}\int_0^1 dt\,\Omega_L\!\!\left(\!\PhiR(t),\frac{d}{dt}\pi_1\D\!\frac{1}{1-N_C}\!\otimes\! R_C\!\otimes\frac{1}{1-N_C}\right).
\end{eqnarray}
The second two terms combine into an integral of a total derivative, giving
\begin{equation}
S  = -\int_0^1 dt\,\big\langle N_t,QN_C\big\rangle_L+\frac{1}{2}\left.\Omega_L\left(\PhiR,\pi_1\D\!\frac{1}{1-N_C}\!\otimes\! \eta\PhiR\!\otimes\frac{1}{1-N_C}\right)\right|_{t=1}.
\end{equation}
In this way we have eliminated the interpolation of the Ramond string field.

However, now the interpolation of the NS string field is missing from the second term. To reintroduce the NS interpolation, we again write the Ramond contribution to the action as the integral of a total derivative, but this time assuming $\PhiR$ is independent of $t$. We have the identity
\begin{equation}
\frac{1}{2}\!\!\left.\Omega_L\left(\PhiR,\!\pi_1\D\!\frac{1}{1-N_C}\!\otimes\! \eta\PhiR\!\otimes\frac{1}{1-N_C}\right)\right|_{t=1}-\frac{1}{2}\Omega_L(\PhiR,Q\eta\PhiR) = \frac{1}{2}\int_0^1dt\frac{d}{dt}\Omega_L\left(\!\PhiR,\!\pi_1\D\!\frac{1}{1-N_C}\!\otimes\! \eta\PhiR\!\otimes\frac{1}{1-N_C}\!\right).
\end{equation}
The second term on the left hand side comes from the boundary contribution at $t=0$, and can be written
\begin{equation}
\frac{1}{2}\Omega_L(\PhiR,Q\eta\PhiR)=\frac{1}{2}\big\langle Y \PsiR,Q\PsiR\big\rangle_S,
\end{equation}
with $\PsiR=\eta\PhiR$ is the Ramond string field in the small Hilbert space. Then the action is written
\begin{eqnarray}
S  = \frac{1}{2}\big\langle Y \PsiR,Q\PsiR\big\rangle_S-\int_0^1 dt\,\big\langle N_t,QN_C\big\rangle_L+
 \frac{1}{2}\int_0^1dt\frac{d}{dt}\Omega_L\left(\!\PhiR,\!\pi_1\D\!\frac{1}{1-N_C}\!\otimes\! \eta\PhiR\!\otimes\frac{1}{1-N_C}\!\right).\label{eq:KOstep2}
\end{eqnarray}
Next we operate the derivative in the integrand 
\begin{eqnarray}
\lineup \!\!\!\!\! \frac{1}{2}\int_0^1dt\frac{d}{dt}\Omega_L\left(\PhiR, \pi_1\D\frac{1}{1-N_C}\otimes\eta\PhiR\otimes\frac{1}{1-N_C}\!\right)\nonumber\\
\lineup =  \frac{1}{2}\int_0^1dt\Omega_L\left(\!\PhiR,\!\pi_1\D\!\left(\!\frac{1}{1-N_C}\!\otimes\! \dot{N}_C\!\otimes\!\frac{1}{1-N_C}\!\otimes\! \eta\PhiR\!\otimes\!\frac{1}{1-N_C}\!+\!\frac{1}{1-N_C}\!\otimes\! \eta\PhiR \!\otimes\!\frac{1}{1-N_C}\!\otimes\! \dot{N}_C\!\otimes\!\frac{1}{1-N_C}\!\right)\!\right).\nonumber\\
\end{eqnarray}
Using \eq{KOstep1} we can factor a $\C$ out of the second entry of the restricted symplectic form. Since $\pi_1\C = \eta$ acting on Ramond states, we can then pull $\eta$ to act on the first entry of the restricted symplectic form. Noting $\eta\PhiR = \PsiR$, this gives
\begin{eqnarray}
\lineup \!\!\!\!\! \frac{1}{2}\int_0^1dt\frac{d}{dt}\Omega_L\left(\PhiR, \pi_1\D\frac{1}{1-N_C}\otimes\eta\PhiR\otimes\frac{1}{1-N_C}\!\right)\nonumber\\
\lineup =  -\frac{1}{2}\int_0^1dt\Omega_L\left(\!\PsiR,\!\pi_1\D\!\left(\!\frac{1}{1-N_C}\!\otimes\! N_t\!\otimes\!\frac{1}{1-N_C}\!\otimes\! \PsiR \!\otimes\!\frac{1}{1-N_C}\!+\!\frac{1}{1-N_C}\!\otimes\! \PsiR \!\otimes\!\frac{1}{1-N_C}\!\otimes\! N_t\!\otimes\!\frac{1}{1-N_C}\!\right)\!\right).\nonumber\\
\end{eqnarray}
We then use cyclicity of $\D$ to pull $N_t$ onto the first entry of the symplectic form in both terms. This cancels the factor of $1/2$, giving
\begin{equation}
\frac{1}{2}\int_0^1dt\frac{d}{dt}\Omega_L\left(\PhiR, \pi_1\D\frac{1}{1-N_C}\otimes\eta\PhiR\otimes\frac{1}{1-N_C}\!\right)= \int_0^1dt\Omega_L\left(N_t,\pi_1\D\frac{1}{1-N_C}\otimes\PsiR \otimes \frac{1}{1-N_C}\otimes \PsiR\otimes\frac{1}{1-N_C}\right).\label{eq:KOstep3}
\end{equation}
Using \eq{F2R} we can simplify
\begin{eqnarray}
\lineup\pi_1\D\frac{1}{1-N_C}\otimes\PsiR \otimes \frac{1}{1-N_C}\otimes \PsiR\otimes\frac{1}{1-N_C}\nonumber\\
\lineup \ \ \ \ \ \ \ \ \ \ = \pi_1(\Q+\mathcal{G}{\bf d}) \frac{1}{1-N_C}\otimes\PsiR \otimes \frac{1}{1-N_C}\otimes \PsiR\otimes\frac{1}{1-N_C}\nonumber\\
\lineup \ \ \ \ \ \ \ \ \ \ = \pi_1 {\bf d} \frac{1}{1-N_C}\otimes\PsiR \otimes \frac{1}{1-N_C}\otimes \PsiR\otimes\frac{1}{1-N_C}\nonumber\\
\lineup \ \ \ \ \ \ \ \ \ \ =F(\PsiR)^2.
\end{eqnarray}
Therefore \eq{KOstep3} simplifies to
\begin{equation}
\frac{1}{2}\int_0^1dt\frac{d}{dt}\Omega_L\left(\PhiR, \pi_1\D\frac{1}{1-N_C}\otimes\eta\PhiR\otimes\frac{1}{1-N_C}\!\right) = -\int_0^1 dt\big\langle N_t,F(\PsiR)^2\big\rangle_L,
\end{equation}
and plugging into \eq{KOstep2} gives
\begin{equation}
S = \frac{1}{2}\big\langle Y\PsiR,Q\PsiR\big\rangle_S -\int_0^1 dt\, \big\langle N_t,QN_C+F(\PsiR)^2\big\rangle_L.
\end{equation}
which is the action as expressed by Kunitomo and Okawa. For the sake of comparison, we also expand the WZW-like form of the action \eq{WZWlike} explicitly in terms of the NS and R components of the potentials and the operator~$F$:
\begin{equation}
S = -\int_0^1dt\bigg[\big\langle N_t,Q N_C + F(R_C)^2\big\rangle_L +\big\langle Y R_t,QR_C+X[N_C,F(R_C)]\big\rangle_L\bigg],\label{eq:actionDF}
\end{equation}
This has a similar form to the Berkovits action 
\begin{equation}S = -\int_0^1 dt\,\langle N_t,QN_C\rangle_L.\end{equation}
The difference is that $Q$ has been replaced with a more complicated operation, and the inner product of states includes a factor of $Y$ in the Ramond sector. One may also check that \eq{actionDF} is essentially the same as the WZW-like action proposed by Matsunaga \cite{Matsunaga}.

\subsection{Relation to the Small Hilbert Space Formulation}
\label{subsec:small}

In subsection \ref{subsec:DCAinf} we constructed a cohomomorphism $\F$ which linearizes the constraint $A_\infty$ structure in the Ramond sector by removing the contribution of the star product with one Ramond state: 
\begin{equation}\F^{-1}(\n-\m_2|_0)\F = \n - \m_2|^0.\end{equation}
In the NS sector, the constraint $A_\infty$ structure remains nonlinear. We may wish to linearize the $A_\infty$ structure in the NS sector as well. This requires constructing another  cohomomorphism $\g$ so that
\begin{equation}\g^{-1}\F^{-1}(\n-\m_2|_0)\F\g = \n.\end{equation}
This should naturally give a formulation of open superstring field theory in the small Hilbert space. 

A suitable construction of $\g$ was given in \cite{WittenSS}. It is defined by a path-ordered exponential 
\begin{equation}\g = \overrightarrow{\mathcal{P}}\exp\left[\int_0^1 dt\, \mmu|^0(t)\right],\end{equation}
where $\mmu|^0(t)$ is a coderivation
\begin{equation}
\mmu|^0(t) = \sum_{n=0}^\infty t^n\mmu_{n+2}|^0,
\end{equation}
defined by a sequence of degree even multi-string products $\mu_{n+2}|^0$ called {\it gauge products}. The path ordering operation $\overrightarrow{\mathcal{P}}$ places $\mmu|^0(t)$ from left to right in sequence of increasing $t$. In the present context, the gauge products carry cyclic Ramond number 0, so they only multiply NS states. Also important are a sequence of degree odd multi-string products $m_{n+2}|^0$ called {\it bare products}, which can be packaged into a coderivation
\begin{equation}\m|^0(t) = \sum_{n=0}^\infty t^n \m_{n+2}|^0.\end{equation}
The bare products also carry cyclic Ramond number zero, and $m_2|^0$ is the cyclic Ramond number 0 component of Witten's open string star product. The bare products and gauge products are defined by recursive solution to a pair of equations
\begin{eqnarray}
\frac{d}{dt}\m|^0(t)\lineup = \big[\m|^0(t),\mmu|^0(t)\big],\label{eq:WZWm0}\\
\mu_{n+2}|^0 \lineup = \frac{1}{n+3}\Big(\Xi m_{n+2}|^0-m_{n+2}|^0(\Xi\otimes \mathbb{I}^{\otimes n+1}+...+\mathbb{I}^{\otimes n+1}\otimes \Xi)\Big).\label{eq:WZWmu0}
\end{eqnarray}
The second relation says the $\mu_{n+2}|^0$ is given by acting $\Xi$ symmetrically once on the output and each input of $m_{n+2}|^0$. This operation is naturally compatible with cyclicity, so both $\mmu|^0(t)$ and $\m|^0(t)$ are cyclic with respect to the large Hilbert space symplectic form $\omega_L$. This also implies that $\g$ is cyclic:
\begin{equation}\langle \Omega_L|\pi_2\g = \langle \Omega_L|\pi_2.\end{equation}
Cyclicity of $\g$ is trivial when $\Omega_L$ pairs Ramond states, since when acting on Ramond states  $\g$ reduces to the identity operator. With this definition of $\g$ one can show \cite{OkWB}
\begin{eqnarray} \g^{-1}(\n-\m_2|^0)\g=\n,
\end{eqnarray}
so the constraint $A_\infty$ structure has been reduced to $\n$ as desired. The dynamical $A_\infty$ structure is transformed into 
\begin{eqnarray}
\M \lineup =\g^{-1}\D\g\nonumber\\
\lineup  = (\F\g)^{-1}(\Q+\m_2|_2)\F\g\Vspace.\label{eq:WZWM}
\end{eqnarray}
This is cyclic with respect to the restricted symplectic form since both $\D$ and $\g$ are cyclic with respect to the restricted symplectic form.

The $A_\infty$ structure $\M$ naturally defines an open superstring field theory in the small Hilbert space including both NS and R sectors. However, a different construction of $\M$ was given in \cite{RWaction,KSR} in terms of a cohomomorphism $\G$
\begin{equation}\M = \G^{-1}(\Q+\m_2|_2)\G,\label{eq:RWM}\end{equation}
which is expressed by a path ordered exponential
\begin{equation}
\G =\overrightarrow{\mathcal{P}}\exp\left[\int_0^1 dt \mmu|_0(t)\right],\label{eq:Gpath}
\end{equation}
where
\begin{eqnarray}
\mmu|_0(t) \lineup = \sum_{n=0}^\infty t^n\mmu_{n+2}|_0,\\
\m|_0(t)\lineup = \sum_{n=0}^\infty t^n \m_{n+2}|_0,
\end{eqnarray}
are coderivations containing gauge products $\mu_{n+2}|_0$ and bare products $m_{n+2}|_0$. The difference from before is that the gauge products and bare products carry Ramond number 0 (not {\it cyclic} Ramond number 0) and  therefore can multiply one Ramond state. The gauge products are bare products are defined by recursive solution to the equations
\begin{eqnarray}
\frac{d}{dt}\m|_0(t) \lineup = [\m|_0(t),\mmu|_0(t)],\label{eq:m0diff}\\
\mu_{n+2}|_0^0 \lineup = \frac{1}{n+3}\Big(\Xi m_{n+2}|_0^0-m_{n+2}|_0^0(\Xi\otimes \mathbb{I}^{\otimes n+1}+...+\mathbb{I}^{\otimes n+1}\otimes \Xi)\Big),\label{eq:mu00rec}\\
\mu_{n+2}|_0^2\lineup = \Xi m_{n+2}|_0^2.\label{eq:mu02rec}
\end{eqnarray}
The gauge product with one Ramond input is defined differently from the gauge product with only NS inputs. Comparing \eq{RWM} and \eq{WZWM} it is natural to conjecture that the cohomomorphisms $\G$ is in fact the same as the cohomomorphism $\F\g$: 
\begin{equation}\G = \F\g.\end{equation} 
We prove this remarkable factorization property in appendix \ref{app:factor}. 

Therefore we may construct a WZW-like action using 
\begin{equation}
\C =\n,\ \ \ \ \ \D =\M.
\end{equation}
The action takes the form 
\begin{equation}
S = \int_0^1 dt\,\Omega_L\left(A_t,\pi_1\M\frac{1}{1-A_C}\right),\label{eq:RNSsmallWZW}
\end{equation}
where we postulate the simplest form of the potentials
\begin{eqnarray}
A_C\lineup = \eta \Phi(t),\\
A_t\lineup = \dot{\Phi}(t).
\end{eqnarray}
Note that these potentials are not equivalent to the potentials $A_C,A_t$ defined in \eq{RNSAC} and \eq{RNSAt}. Therefore, the field $\Phi$ in this action will be related to $\Phi$ in previous sections by a nontrivial field redefinition. Following the steps of section \ref{subsec:liftsmall} in reverse, \eq{RNSsmallWZW} can be rewritten 
\begin{equation}
S=\frac{1}{2}\Omega_S(\Psi,Q\Psi)+\frac{1}{3}\Omega_S(\Psi,M_2(\Psi,\Psi))+\frac{1}{4}\Omega_S(\Psi,M_3(\Psi,\Psi,\Psi))+...,
\end{equation}
where $\Psi=\eta\Phi$. This is the action for open superstring field theory formulated in the small Hilbert space. As demonstrated in \cite{OkWB,WB,WBlarge,RWaction}, the string field $\Psi=\PsiN+\PsiR$ may be related to $\PhiN^{(\mathrm{KO})}$ and $\PsiR^{(\mathrm{KO})}$ of the Kunitomo-Okawa theory through
\begin{eqnarray}
(\eta e^{\PhiN^{(\mathrm{KO})}})e^{-\PhiN^{(\mathrm{KO})}}\lineup = \pi_1\g\frac{1}{1-\Psi}, \\
\PsiR^{(\mathrm{KO})} \lineup = \PsiR.
\end{eqnarray}
The Ramond fields in the two theories are equal.

\section{Supersymmetry}
\label{sec:susy}

In this section we discuss supersymmetry in the WZW-like formulation of open superstring field theory. Analysis of supersymmetry in the small Hilbert space formulation appears in \cite{susyS}, and in the Kunitomo-Okawa formulation in \cite{Kunitomo}. The WZW-like approach gives a different and useful perspective on these results. 

We work with open superstring field theory on a BPS D-brane with 16 unbroken supersymmetries, and both NS and R sectors of the state space are GSO($+$) projected. The supersymmetries are described by the zero mode of the fermion vertex in the $-1/2$ picture:
\begin{equation}
s_1 = \sqrt{2}\oint_{|z|=1} \frac{dz}{2\pi i} \Theta_ae^{-\phi/2}(z)\eps_a.
\end{equation}
We follow the conventions of \cite{susyS}. $\Theta_a$ is the spin field and $\eps_a$ a constant anticommuting spinor variable of positive chirality---the supersymmetry parameter. For simplicity we leave the dependence of $s_1$ on the supersymmetry parameter implicit. The operator $s_1$ is degree even and carries ghost number $0$ and picture $-1/2$. It is BRST invariant and well-defined in the small Hilbert space:
\begin{eqnarray}[\Q,\s_1] \lineup = 0,\\
\ [\n,\s_1]\lineup  = 0.
\end{eqnarray} 
Since $s_1$ is the zero mode of a weight 1 primary operator, it is a derivation of the open string star product and is cyclic with respect to the large Hilbert space symplectic form:
\begin{eqnarray}
[\s_1,\m_2] \lineup = 0,\\
\langle\omega_L|\pi_2\s_1 \lineup = 0.
\end{eqnarray}
These properties are all we need.

To describe the transformation of the string field we need the coderivation $\del_s$ describing supersymmetry. We will not attempt to rederive it from scratch. Rather, we note that it should be related to the coderivation ${\bf S}$ constructed in \cite{susyS} through the $A_\infty$ isomorphism $\g$ introduced in section \ref{subsec:small}:
\begin{equation}\del_s = \g{\bf S}\g^{-1}.\end{equation}
In this way we find that $\del_s$ is given by
\begin{equation}
\del_s = \mathcal{G}\Ssigma,
\end{equation}
where the coderivation $\Ssigma$ is defined by
\begin{equation}
\pi_1\Ssigma = \pi_1\Big(\s_1+\s_1\,\Xi|^2{\bf d} + {\bf d}\,\Xi|^2\s_1\Big),
\end{equation}
where ${\bf d}$ introduced in section \ref{subsec:Feynman}. $\Ssigma$ has a simple interpretation. It is given by a sum over all Feynman diagrams in ${\bf d}$ together with a sum over all ways of attaching a propagator $\Xi|^2$ followed by the operator $s_1$ to each external state of the diagrams in ${\bf d}$. It is clear that $\del_s$ will be cyclic with respect to the restricted symplectic form, but what is less obvious is that it will commute with $\C$ and $\D$:
\begin{eqnarray}
[\C,\del_s] \lineup = 0,\\
\ [\D,\del_s]\lineup = 0.
\end{eqnarray}
We will prove this in appendix \ref{app:susy}.

\begin{figure}
\begin{center}
\resizebox{3.9in}{1in}{\includegraphics{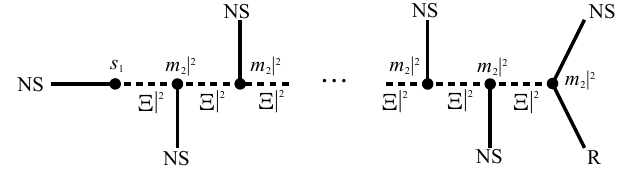}}
\end{center}
\caption{\label{fig:susydiag} All nonvanishing diagrams in the coderivation representing supersymmetry take the form of a single line of propagators connecting a cubic vertex with an NS and R external state to a quadratic vertex defined by $s_1$ with an NS external state. The line of propagators are connected by cubic vertices with an NS external state.} 
\end{figure}

Note that for a particular diagram contributing to ${\bf d}$ most ways of attaching $\Xi|^2$ followed by $s_1$ gives a vanishing result because there is no cubic coupling of Ramond states. Instead, the attached propagator must continue the line of propagators inside the diagram of ${\bf d}$. From this we can see that $\Ssigma$ is given by a sum over color-ordered diagrams with a single line of propagators; the line of propagators begins at a cubic vertex attached to two external states, and ends a quadratic vertex (given by $s_1$) attached to one external state. The line of propagators are connected by cubic vertices attached to one external state. We also know that the cubic vertex at the beginning of the line of propagators must contain one  Ramond external state, and all other external states must be NS to get a nonvanishing diagram. This is shown in figure \ref{fig:susydiag}. The products of $\del_s$ are nonvanishing only if there is a single Ramond state among the inputs and outputs, which is to say that $\del_s$ carries cyclic Ramond number $1$. It is interesting to mention that $\del_s$ is defined by the unique set of nonvanishing diagrams that can be built using $\Xi|^2,m_2|^2$ and $s_1$ with an odd number of Ramond external states. In this sense the realization of supersymmetry is very natural.

To complete the definition of the supersymmetry transformation we must find an expression for the potential $A_s$ representing supersymmetry. We expand $A_s$ into NS and R components:
\begin{equation}
A_s = N_s+R_s.
\end{equation}
There is some arbitrariness in the choice of $A_s$ due to the symmetry \eq{pottrans} in the solution of the flatness conditions. This ambiguity corresponds to the freedom to modify the supersymmetry transformation by a gauge transformation generated by the constraint $A_\infty$ structure. One way to fix this ambiguity is to construct $A_s$ so that higher potentials with an index $t$ vanish, following subsection \ref{subsubsec:flat}. This has the advantage of giving a canonical form for the higher potentials, but gives an expression for $A_s$ in terms of an integral analogous to \eq{ND}-\eq{RD} which is more complicated than it needs to be. Instead we will determine $A_s$ by computing the left hand side of the equation 
\begin{eqnarray}
\pi_1\del_s\frac{1}{1-A_C} \lineup = \pi_1\C \frac{1}{1-A_C}\otimes A_s\otimes\frac{1}{1-A_C},\nonumber\\
\lineup = D_\eta N_s +\eta R_s,\label{eq:flats}
\end{eqnarray}
and looking for a simple way to express it in the form given on the right hand side. Using the formulas of appendix \ref{app:FF} we find
\begin{eqnarray}
\pi_1\del_s\frac{1}{1-A_C} = s_1F(R_C)+[F(R_C),F(\Xi s_1 N_C)]+X\Big(s_1N_C+[N_C,F(\Xi s_1N_C)]\Big).\label{eq:susyA0}
\end{eqnarray}
To express this as the right hand side of \eq{flats}, we need factor $\eta$ out of this expression. Let us look at the NS part first. A factor of $\eta$ naturally appears in $R_C=\eta\PhiR(t)$. We note the identities \cite{complete}
\begin{eqnarray}
D_\eta^2 \lineup = 0,\\
F(\eta A)\lineup = D_\eta F(A),\\
F(\Xi D_\eta A)\lineup = A - D_\eta F(\Xi A).
\end{eqnarray}
Then we can write
\begin{eqnarray}
 s_1F(R_C)+[F(R_C),F(\Xi s_1 N_C)]\lineup =  s_1F(\eta \PhiR(t))+[F(\eta\PhiR(t)),F(\Xi s_1 N_C)]\nonumber\\
 \lineup = s_1D_\eta F(\PhiR(t))+[D_\eta F(\PhiR(t)),F(\Xi s_1 N_C)]\nonumber\\
 \lineup = D_\eta s_1 F(\PhiR(t))+D_\eta[F(\PhiR(t)),F(\Xi s_1 N_C)]\nonumber\\
 \lineup\ \ \ \ \ \ \ \ -[s_1 N_C,F(\PhiR(t))]-[F(\PhiR(t)),D_\eta F(\Xi s_1 N_C)]\nonumber\\
 \lineup =  D_\eta \Big(s_1 F(\PhiR(t))+[F(\PhiR(t)),F(\Xi s_1 N_C)]\Big)\nonumber\\
 \lineup\ \ \ \ \ \ \ \ -[s_1 N_C,F(\PhiR(t))]-[F(\PhiR(t)), s_1 N_C]+[F(\PhiR(t)),F(\Xi D_\eta s_1 N_C)]\nonumber\\
 \lineup =  D_\eta \Big(s_1 F(\PhiR(t))+[F(\PhiR(t)),F(\Xi s_1 N_C)]\Big)+[F(\PhiR(t)),F(\Xi D_\eta s_1 N_C)].\nonumber\\
 \label{eq:susyA1}
\end{eqnarray}
Note that
\begin{equation}s_1 N_C = D_\eta A_{s_1},\end{equation}
where $A_{s_1}$ is a Maurer-Cartan element defined
\begin{equation}A_{s_1}\equiv (s_1e^{\PhiN(t)})e^{-\PhiN(t)}.\end{equation}
Since $D_\eta$ is nilpotent, the last term in \eq{susyA1} drops out and we find
\begin{equation}
 s_1F(R_C)+[F(R_C),F(\Xi s_1 N_C)] =  D_\eta \Big(s_1 F(\PhiR(t))+[F(\PhiR(t)),F(\Xi s_1 N_C)]\Big).
\end{equation}
Now consider the Ramond component of \eq{susyA0}. Note the identity
\begin{eqnarray}
s_1 N_C \lineup = D_\eta A_{s_1}\nonumber\\
\lineup = D_\eta F(\Xi D_\eta A_{s_1}) +F(\Xi D_\eta^2 A_{s_1})\nonumber\\
\lineup = D_\eta F(\Xi s_1 N_C).
\end{eqnarray}
Therefore the Ramond component of \eq{susyA0} can be written
\begin{eqnarray}
X\Big(s_1N_C+[N_C,F(\Xi s_1N_C)]\Big)\lineup = X\Big(D_\eta F(\Xi s_1N_C)+[N_C,F(\Xi s_1N_C)]\Big),\nonumber\\
\lineup = \eta X F(\Xi s_1 N_C).
\end{eqnarray}
Therefore we propose
\begin{eqnarray}
N_s \lineup = s_1 F(\PhiR(t))+[F(\PhiR(t)),F(\Xi s_1 N_C)],\label{eq:Ns}\\
R_s\lineup = X F(\Xi s_1 N_C).\label{eq:Rs}
\end{eqnarray}
This seems fairly natural since the $N_s$ is identical to the NS part of \eq{susyA0} after replacing $R_C$ with $\PhiR(t)$. The supersymmetry transformation can be determined by equating
\begin{equation}(\delta e^\PhiN)e^{-\PhiN}  = N_s|_{t=1},\ \ \ \delta\PhiR = R_s|_{t=1}.\end{equation}
The general arguments of subsection \ref{subsubsec:sym} and appendix \ref{app:sym} then guarantee that we have a symmetry of the action.

A different expression for the supersymmetry transformation was obtained in recent work of Kunitomo \cite{Kunitomo}:\footnote{We thank H. Kunitomo for a discussion of the relation between these supersymmetry transformations.}
\begin{eqnarray}
(\delta_\mathrm{Kun} e^\PhiN)e^{-\PhiN}  \lineup = \left.\bigg(\,e^\PhiN\Big(s_1 \Xi \big(e^{-\PhiN}F(\PsiR) e^\PhiN\big)\Big)e^{-\PhiN}+[F(\Xi A_{s_1}),F(\PsiR)]\,\bigg)\right|_{t=1},\label{eq:KunNs}\\
\delta_\mathrm{Kun}\PsiR \lineup = X\eta F(\Xi s_1 N_C)|_{t=1}.\label{eq:KunRs}
\end{eqnarray}
This transformation uses the Ramond string field $\PsiR=\eta\PhiR$ in the small Hilbert space. The Ramond sector supersymmetry variations \eq{Rs} and \eq{KunRs} clearly agree, but the NS sector variations are rather different. To compare them, we will assume that $\PhiR$ takes the form $\Xi\PsiR$; if $\PhiR$ were to contain an additional $\eta$-exact piece, this would only modify the supersymmetry variation by a gauge transformation, as follows from a computation essentially identical to \eq{susyA1}. Assuming $\PhiR=\Xi\PsiR$, our expression for the supersymmetry transformation takes the form
\begin{equation}(\delta e^\PhiN)e^{-\PhiN} = \left.\bigg(s_1 F(\Xi\PsiR)+[F(\Xi\PsiR),F(\Xi s_1 N_C)]\bigg)\right|_{t=1}.\label{eq:NsSm}\end{equation}
With some algebra one can rewrite this as 
\begin{eqnarray}
(\delta e^\PhiN)e^{-\PhiN}  \lineup = \left.\bigg(\,e^\PhiN\Big(s_1 \Xi \big(e^{-\PhiN}F(\PsiR) e^\PhiN\big)\Big)e^{-\PhiN}+[F(\Xi A_{s_1}),F(\PsiR)]\,\bigg)\right|_{t=1}\nonumber\\ \lineup\ \ \ \ \ \ +\left.D_\eta\bigg(\,e^\PhiN\Big(s_1 \Xi \big(e^{-\PhiN}F(\Xi\PsiR) e^\PhiN\big)\Big)e^{-\PhiN}+[F(\Xi A_{s_1}),F(\Xi\PsiR)]\bigg)\right|_{t=1}.
\end{eqnarray}
The first line agrees with Kunitomo's expression, and the second line is an infinitesimal gauge transformation of $\PhiN$ generated by the constraint $A_\infty$ structure. Therefore the two supersymmetry transformations are physically equivalent. In more detail, we can expand the supersymmetry transformations \eq{KunNs} and \eq{NsSm} up to second order in the string field:
\begin{eqnarray}
\delta\PhiN\lineup = s_1\Xi\PsiR-\frac{1}{2}[\PhiN,s_1\Xi\PsiR]+s_1\Xi[\eta\PhiN,\Xi\PsiR]+[\Xi\PsiR, \Xi s_1\eta\PhiN]+...,\\
\delta_\mathrm{Kun}\PhiN \lineup = s_1\Xi\PsiR+\frac{1}{2}[\PhiN,s_1\Xi\PsiR]-s_1\Xi[\PhiN,\PsiR]+[\PsiR,\Xi s_1\PhiN]+... .
\end{eqnarray}
At the linearized level they agree, but they differ at the quadratic level and beyond.

As a physical application, we now demonstrate that transverse displacement of a BPS D-brane in open superstring field theory does not break any supersymmetries. An analogous discussion appears in \cite{Ramond}, but concerning a different supersymmetry transformation relevant for the equations of motion formulated in \cite{BerkRamond}. The displacement of a D-brane can be described analytically by a classical solution in Berkovits open superstring field theory \cite{marginal,Okmarg}:
\begin{equation}e^\PhiN = 1+ \sqrt{\Omega} X\frac{1}{1-B\frac{1-\Omega}{K}J}\sqrt{\Omega}.\label{eq:sol}\end{equation}
We express the solution in the formalism of \cite{Okawa} and following the conventions of \cite{simple}. The string field $\Omega=e^{-K}$ is the $SL(2,\mathbb{R})$ vacuum, $B$ satisfies $QB=K$. The string field $J$ corresponds to a picture $0$ marginal operator which generates a displacement of the D-brane, and $X$ satisfies $QX=J$:
\begin{equation}J = \lambda (i\sqrt{2}c\d X^+ +\gamma\psi^+),\ \ \ \ X = \lambda \xi e^{-\phi}c\psi^+.\end{equation}
$\lambda$ is a marginal parameter that measures the D-brane displacement, and the spatial component of the Lorentz index $+$ is transverse to the D-brane. The solution also turns on a timelike Wilson line whose strength matches the D-brane displacement, but the timelike Wilson line is physically trivial \cite{KOsing} and its only role is to regulate divergent OPEs between the marginal operators. For the purposes of supersymmetry, the relevant identity satisfied by this solution is
\begin{equation}
s_1(e^{-\PhiN}Qe^\PhiN) = s_1\left(\sqrt{\Omega}J\frac{1}{1-B\frac{1-\Omega}{K}J}\sqrt{\Omega}\right) = 0.\label{eq:susysol}
\end{equation}
Since $s_1$ is the zero mode of a weight 1 primary it acts as a derivation of the star product and annihilates the string field $K$. It also annihilates $B$ and $J$ since there is no pole in the OPE between the picture $-1/2$ fermion vertex and either the $b$-ghost or the picture $0$ marginal operator. To prove that the translated D-brane preserves supersymmetry, we must show that that the supersymmetry transformation of the classical solution is physically trivial. The transformation of the NS component vanishes since the Ramond component is identically zero. Therefore we must see how the (vanishing) Ramond component of the solution changes under supersymmetry:
\begin{equation}
\delta\PhiR= XF(\Xi s_1 N_C)|_{t=1}.
\end{equation}
The right hand side does not vanish, but we will show that it can be expressed as a gauge transformation. We write
\begin{eqnarray}
\delta\PhiR\lineup = XF(\Xi D_\eta A_{s_1})|_{t=1}\nonumber\\
\lineup = X A_{s_1}-X D_\eta F(\Xi A_{s_1})\nonumber\\
\lineup = X A_{s_1} - X F(\eta \Xi A_{s_1}).
\end{eqnarray}
Next we note that $A_{s_1}$ is BRST invariant,
\begin{equation}Q A_{s_1} = e^{\PhiN}s_1\Big(e^{-\PhiN}Qe^\PhiN\Big)e^{-\PhiN} = 0,\end{equation}
as follows from \eq{susysol}. This allows us to write 
\begin{equation}
\delta\PhiR = Q(\Xi A_{s_1}) -X F(\eta\Xi A_{s_1}).
\end{equation}
Comparing to \eq{gauge2}, we see that this is a gauge transformation defined by Ramond gauge parameters
\begin{eqnarray}
\rho_D \lineup = \Xi A_{s_1},\label{eq:rhoD}\\
\rho_C \lineup = -F(\Xi A_{s_1}).
\end{eqnarray}
with all other gauge parameters vanishing. Therefore displacement of the D-brane preserves supersymmetry.

One technical complication is that the gauge parameter $\rho_D$ does not satisfy the constraint
\begin{equation}XY\eta\rho_D = \eta\rho_D.\label{eq:strongrhoD}\end{equation}
Strictly speaking, the constraint on the Ramond string field only implies a weaker condition on the gauge parameter
\begin{equation}XYQ\eta\rho_D = Q\eta\rho_D,\end{equation}
which is satisfied by \eq{rhoD}. However, the stronger condition \eq{strongrhoD} is physically equivalent and natural since it corresponds to the assumed constraint on the Ramond gauge parameter in the small Hilbert space formulation. We can modify \eq{rhoD} to be consistent with the stronger constraint by noting that the gauge symmetry is reducible, so different gauge parameters may implement the same gauge transformation. In the background of the solution \eq{sol}, the potential $A_g$ generated by $\rho_D$ takes the form
\begin{equation}A_g = \pi_1\D\frac{1}{1-N_C}\otimes \rho_D\otimes\frac{1}{1-N_C}.\end{equation}
The potential takes the same form if we modify $\rho_D$ by
\begin{eqnarray}
\rho_D' \lineup = \rho_D + \pi_1\D\frac{1}{1-N_C}\otimes\mu\otimes\frac{1}{1-N_C},\nonumber\\
\lineup = \rho_D+Q\mu +X[N_C,F(\mu)],\label{eq:rhodp}
\end{eqnarray}
since $\D^2=0$. Since $Q$ has no cohomology in the large Hilbert space, we know that
\begin{equation}A_{s_1} = Q\alpha\end{equation}
for some $\alpha$. Therefore we write $\rho_D$ in \eq{rhoD} as 
\begin{equation}\rho_D = X\alpha - Q(\Xi\alpha).\end{equation}
The first term is consistent with the stronger constraint \eq{strongrhoD}, but the second term is not. However, we can remove the second term by choosing a different gauge parameter as in \eq{rhodp} with $\mu=\Xi\alpha$. This gives 
\begin{equation}\rho_D' = X\alpha +X[N_C,F(\Xi\alpha)].\end{equation}
This gauge parameter is consistent with the stronger constraint \eq{strongrhoD}.

\section{Concluding Remarks}

In this paper we developed a generalized notion of the Wess-Zumino-Witten action with the goal of giving a more conceptual understanding of the Ramond sector of open superstring field theory in the large Hilbert space. The main reason why a WZW-like formulation of the Ramond sector was unclear is that the required dynamical $A_\infty$ structure is necessarily nonlinear. In the conventional WZW action, the dynamical $A_\infty$ structure is simply an antiholomorphic exterior derivative, and higher products are not needed. The original form of heterotic string field theory \cite{heterotic} is a somewhat intermediate example. The constraint $A_\infty$ structure (actually $L_\infty$ structure) requires an infinite collection of higher closed string products, but the dynamical $A_\infty$ structure is still linear, given simply by the eta zero mode.\footnote{In our discussion of the open superstring, the constraint $A_\infty$ structure was taken as a nonlinear extension of $\eta$ while the dynamical $A_\infty$ structure a nonlinear extension of $Q$. However, in the heterotic string field theory of \cite{heterotic} the role of $Q$ and $\eta$ is reversed.} Therefore, the principal new possibility in our formulation is that the dynamical $A_\infty$ structure contains string products to all orders. 

There are a number of important questions we have not touched upon. One notable feature of the development, already discussed in \cite{heterotic}, is that the WZW-like action is generalized in a nonstandard form which does not exhibit manifest 3-dimensional covariance, and is asymmetric between dynamical and constraint $A_\infty$ structures. By contrast, the conventional WZW-like action is a sum of terms displaying manifest 2- and 3-dimensional covariance, and the holomorphic and antiholomorphic exterior derivatives appear symmetrically. This suggests that the generalized WZW-like action does not natrually reflect the structure of 2-dimensional field theory, and should be given a different kind of geometrical interpretation. Also interesting is whether the action has any connection to ($L_\infty$~generalizations of) Lie groups. Progress on these questions may shed light on the relation between the flatness conditions and the peculiar form of the generalized WZW-like action. Another interesting question is whether there is a WZW-like analogue of the minimal model theorem~\cite{Kajiura}. This could be useful for understanding the structure of perturbation theory in the large Hilbert space \cite{INOT,fermscat}. 

Our results have potentially useful implications for the construction of Ramond sector vertices in heterotic and type II string theories. Partial results for the Ramond sector of the heterotic string were found in \cite{Kunitomohet}. For the open superstring in the small Hilbert space \cite{RWaction}, the Ramond vertices were found following the construction of the classical equations of motion \cite{Ramond}, which closely parallels the NS sector construction. This setup, however, appears inadequate for finding heterotic and type II actions. For the heterotic string the procedure breaks down at 5-points, where there is no elementary solution of the recursion consistent with cyclicity of the vertex. On the other hand, Feynman graphs give an entirely different description which seems to more efficiently capture the structure of Ramond vertices. It would be interesting to see if this could be generalized to heterotic and type II string field theories. 

One motivation behind our analysis is the hope that a better understanding of the classical action may give a hint as to what structures to look for in the quantum theory. Batalin-Vilkovisky quantization in the large Hilbert space is now understood at the free level \cite{BerkBV,Torii1,Torii2}, but the nonlinear master action has proven to be very difficult to find.\footnote{See \cite{MatsunagaBV} for recent work in this direction.} At first it might seem more practical to ignore the large Hilbert space and quantize superstring field theory directly in the small Hilbert space, where there is a ready-made procedure for constructing the master action. However, spurious poles in loop amplitudes will likely complicate the quantization of the open superstring field theory in the small Hilbert space if the vertices are constructed using Witten's associative star product.\footnote{In Sen's discussions of superstring field theories, it is proposed that string vertices should come with long stubs so that the majority of the contribution to a given amplitude is given by the corresponding fundamental vertex in the string field theory action. The PCOs in this vertex may then be distributed so as to avoid divergences from spurious poles in integration over the moduli space. The Witten vertex is the opposite extreme from this setup. The entire contribution to the integration over moduli space is given by propagators connected to tree-level vertices. The PCO locations have already been fixed by the tree level vertices, and there is no possibility to alter them as needed to avoid spurious poles in loops. An explicit analysis of what happens in this situation, however, has not been carried out.} By contrast, a quantum superstring field theory in the large Hilbert space would likely be quite novel from the perspective of the conventional understanding of superstring perturbation theory. For example, genus $g$ correlators of the $\eta,\xi$ system contain $g-1$ more $\eta$ insertions than $\xi$ insertions, while $\beta\gamma$ correlators expressed in bosonized form \cite{Verlinde} always contain one fewer $\eta$ than $\xi$ insertion \cite{Watamura}. Meanwhile, the results of \cite{BerkBV} imply that states of many pictures will propagate through loops in large Hilbert space amplitudes. In the small Hilbert space, intermediate states always carry a fixed picture. It would be interesting to see the consequences of these differences for spurious poles.

\vspace{.5cm}

\noindent{\bf Acknowledgments}

\vspace{.25cm}

\noindent The author thanks S. Konopka, H. Matsunaga, and I. Sachs for conversations. This work is supported in part by the DFG Transregional Collaborative Research Centre TRR 33 and the DFG cluster of excellence Origin and Structure of the Universe and by ERDF and M\v{S}MT (Project CoGraDS -CZ.02.1.01/0.0/0.0/15\_ 003/0000437) . 

\begin{appendix}

\section{Equations of Motion}
\label{app:EOM}

In this appendix we derive the equations of motion from the WZW-like action:
\begin{equation}
S = \int_0^1 dt\, \omega\left(A_t,\pi_1\D\frac{1}{1-A_C}\right).
\end{equation} 
Consider the variation of the integrand:
\begin{equation}
\delta\omega\left(A_t,\pi_1\D\frac{1}{1-A_C}\right) = \omega\left(\delta A_t,\pi_1\D\frac{1}{1-A_C}\right) +  \omega\left(A_t,\pi_1\D\frac{1}{1-A_C}\otimes\delta A_C \otimes\frac{1}{1-A_C}\right).\label{eq:WZWvar1}
\end{equation}
The flatness conditions \eq{1f} and \eq{2f} imply
\begin{eqnarray}
\delta A_C \lineup = \pi_1\C \frac{1}{1-A_C}\otimes A_\delta\otimes\frac{1}{1-A_C},\\
\delta A_t \lineup = \dot{A}_\delta-\pi_1\C \frac{1}{1-A_C}\otimes A_t\otimes \frac{1}{1-A_C}\otimes A_\delta\otimes \frac{1}{1-A-C} + \pi_1\C\frac{1}{1-A_C}\otimes A_\delta\otimes \frac{1}{1-A_C}\otimes A_t\otimes \frac{1}{1-A_C} \nonumber\\
\lineup \ \ \ \ \ \  \ + \pi_1\C\frac{1}{1-A_C}\otimes A_{ t\delta }\otimes \frac{1}{1-A_C},
\end{eqnarray}
where we use the dot to denote differentiation with respect to $t$. Plugging into \eq{WZWvar1} gives the terms
\begin{eqnarray}
\delta\omega\left(A_t,\pi_1\D\frac{1}{1-A_C}\right)\lineup = 
\omega\left(\dot{A}_\delta,\pi_1\D\frac{1}{1-A_C}\right) - \omega\left(\pi_1\C \frac{1}{1-A_C}\otimes A_t\otimes \frac{1}{1-A_C}\otimes A_\delta\otimes \frac{1}{1-A_C} ,\pi_1\D\frac{1}{1-A_C}\right) \nonumber\\
\lineup\ \ \ + \omega\left(\pi_1\C\frac{1}{1-A_C}\otimes A_\delta\otimes \frac{1}{1-A_C}\otimes A_t\otimes \frac{1}{1-A_C},\pi_1\D\frac{1}{1-A_C}\right) \nonumber\\
\lineup\ \ \ +\omega\left(\pi_1\C\frac{1}{1-A_C}\otimes A_{ t\delta }\otimes \frac{1}{1-A_C},\pi_1\D\frac{1}{1-A_C}\right) \nonumber\\
\lineup  \ \ \ + \omega\left(\pi_1\D\C \frac{1}{1-A_C}\otimes A_\delta\otimes\frac{1}{1-A_C},A_t\right) .\label{eq:WZWvar2}
\end{eqnarray}
To proceed we use the fact that a coderivation ${\bf b}$ which is cyclic with respect to a symplectic form $\omega$ satisfies the identity:
 \begin{eqnarray}\lineup \omega\left(\pi_1{\bf b}\frac{1}{1-A}\otimes B_1\otimes \frac{1}{1-A}\otimes... \otimes \frac{1}{1-A}\otimes B_{n+1}\otimes \frac{1}{1-A},B_{n+2}\right) \nonumber\\
\lineup = -(-1)^{\deg(\D)\deg(B_1)}\omega\left(B_1,\pi_1{\bf b}\frac{1}{1-A}\otimes B_2\otimes \frac{1}{1-A}\otimes... \otimes \frac{1}{1-A}\otimes B_{n+2}\otimes \frac{1}{1-A}\right).\label{eq:grpcyc}\end{eqnarray}
In particular, using cyclicity of $\C$ we can rewrite the second and third terms of \eq{WZWvar2} so that $A_t$ appears on the second entry of the symplectic form. This gives:
\begin{eqnarray}
\delta\omega\left(A_t,\pi_1\D\frac{1}{1-A_C}\right)\lineup = 
\omega\left(\dot{A}_\delta,\pi_1\D\frac{1}{1-A_C}\right) - \omega\left(\pi_1\C \frac{1}{1-A_C}\otimes A_\delta\otimes \frac{1}{1-A_C}\otimes \left(\pi_1\D\frac{1}{1-A_C}\right)\otimes \frac{1}{1-A_C} ,A_t\right) \nonumber\\
\lineup\ \ \ + \omega\left(\pi_1\C\frac{1}{1-A_C}\otimes \left(\pi_1\D\frac{1}{1-A_C}\right)\otimes \frac{1}{1-A_C}\otimes A_\delta\otimes \frac{1}{1-A_C},A_t\right) \nonumber\\
\lineup\ \ \ +\omega\left(\pi_1\C\frac{1}{1-A_C}\otimes A_{ t\delta }\otimes \frac{1}{1-A_C},\pi_1\D\frac{1}{1-A_C}\right) \nonumber\\
\lineup\ \ \ - \omega\left(\pi_1\C\D \frac{1}{1-A_C}\otimes A_\delta\otimes\frac{1}{1-A_C},A_t\right).\ \ \ \ \ \ \ 
\end{eqnarray}
The second, third and last terms can be combined, and the equation simplifies to 
\begin{eqnarray}
\delta\omega\left(A_t,\pi_1\D\frac{1}{1-A_C}\right)\lineup = 
\omega\left(\dot{A}_\delta,\pi_1\D\frac{1}{1-A_C}\right) +\omega\left(\pi_1\C\frac{1}{1-A_C}\otimes A_{ t\delta }\otimes \frac{1}{1-A_C},\pi_1\D\frac{1}{1-A_C}\right) \nonumber\\
\lineup\ \ \ - \omega\left(\pi_1\C\frac{1}{1-A_C}\otimes\left(\pi_1\D \frac{1}{1-A_C}\otimes A_\delta\otimes\frac{1}{1-A_C}\right)\otimes\frac{1}{1-A_C},A_t\right).
\end{eqnarray}
Using cyclicity of $\C$ again on the second two terms,
\begin{eqnarray}
\delta\omega\left(A_t,\pi_1\D\frac{1}{1-A_C}\right)\lineup = 
\omega\left(\dot{A}_\delta,\pi_1\D\frac{1}{1-A_C}\right) -\omega\left(A_{ t\delta },\C\frac{1}{1-A_C}\otimes\left(\pi_1\D\frac{1}{1-A_C}\right)\otimes \frac{1}{1-A_C}\right) \nonumber\\
\lineup\ \ \ + \omega\left(\pi_1\D \frac{1}{1-A_C}\otimes A_\delta\otimes\frac{1}{1-A_C},\pi_1\C\frac{1}{1-A_C}\otimes A_t\otimes\frac{1}{1-A_C}\right)\nonumber\\
\lineup = 
\omega\left(\dot{A}_\delta,\pi_1\D\frac{1}{1-A_C}\right) +\omega\left(A_{ t\delta },\D\C\frac{1}{1-A_C}\right) \nonumber\\
\lineup\ \ \ + \omega\left(\pi_1\D \frac{1}{1-A_C}\otimes A_\delta\otimes\frac{1}{1-A_C},\pi_1\C\frac{1}{1-A_C}\otimes A_t\otimes\frac{1}{1-A_C}\right).
\end{eqnarray}
The second term drops out by the flatness condition \eq{0f}. The third term can be simplified with the flatness condition \eq{1f},
\begin{equation}
\dot{A}_C = \pi_1\C\frac{1}{1-A_C}\otimes A_t\otimes\frac{1}{1-A_C},
\end{equation}
so we find 
\begin{equation}
\delta\omega\left(A_t,\pi_1\D\frac{1}{1-A_C}\right) = 
\omega\left(\dot{A}_\delta,\pi_1\D\frac{1}{1-A_C}\right)+ \omega\left(\pi_1\D \frac{1}{1-A_C}\otimes A_\delta\otimes\frac{1}{1-A_C},\dot{A}_C\right).
\end{equation}
Using cyclicity we can pull $\D$ to the last entry of the symplectic form:
\begin{eqnarray}
\delta\omega\left(A_t,\pi_1\D\frac{1}{1-A_C}\right)\lineup = 
\omega\left(\dot{A}_\delta,\pi_1\D\frac{1}{1-A_C}\right)+ \omega\left( A_\delta,\pi_1\D \frac{1}{1-A_C}\otimes\dot{A_C}\otimes\frac{1}{1-A_C}\right)\nonumber\\
\lineup =\frac{d}{dt}\omega\left(A_\delta,\pi_1\D\frac{1}{1-A_C}\right).
\end{eqnarray}
Integrating the total derivative, the boundary term at $t=0$ vanishes by property {\bf (i)}. Therefore we obtain
\begin{equation}
\delta S = \left.\omega\left(A_\delta,\pi_1\D\frac{1}{1-A_C}\right)\right|_{t=1}.
\end{equation}
Property {\bf (ii)} together with nondegeneracy of the symplectic form then implies the equations of motion. 

\section{Symmetry of the Action}
\label{app:sym}

In this appendix we demonstrate that the symmetry transformation proposed in subsection \ref{subsubsec:sym} leaves the WZW-like action invariant. As shown in appendix \ref{app:EOM}, the variation of the action is given by 
\begin{equation}
\delta S = \left.\omega\left(A_\delta,\pi_1\D\frac{1}{1-A_C}\right)\right|_{t=1}.
\end{equation}
A symmetry transformation of the dynamical field is characterized by the potential $A_v$ as described in subsection \ref{subsubsec:sym},
\begin{equation}
A_\delta|_{t=1} = A_v|_{t=1},
\end{equation}
so the variation of the action is  
\begin{equation}
\delta S = \left.\omega\left(A_v,\pi_1\D\frac{1}{1-A_C}\right)\right|_{t=1}.
\end{equation}
This can be written as the integral of a total derivative with respect to $t$:
\begin{eqnarray}
\delta S \lineup = \int_0^1 dt\, \frac{d}{dt}\omega\left(A_v,\pi_1\D\frac{1}{1-A_C}\right)\nonumber\\
\lineup = \int_0^1 dt\, \left[\omega\left(\dot{A}_v,\pi_1\D\frac{1}{1-A_C}\right)+ \omega\left(A_v,\pi_1\D\frac{1}{1-A_C}\otimes\dot{A}_C\otimes \frac{1}{1-A_C}\right)\right].
\end{eqnarray}
To simplify the argument we assume that $A_C|_{t=0}$ vanishes, so that the boundary term at $t=0$ can be ignored. This can always be arranged by adding (at most) a constant to the action.\footnote{\label{foot:AC} Suppose $A_C|_{t=0}$ does not vanish. Then we can add a constant to the action in the form 
\begin{equation}\mathrm{constant} = \int_{-1}^0 dt\, \omega\left(A_t,\pi_1\D\frac{1}{1-A_C}\right),\label{eq:footsym}\end{equation}
where we assume that the potentials have been extended to satisfy flatness conditions in the entire range $t\in[-1,1]$, and for $t\in[0,1]$ they agree with the potentials we started with. For $t\in[-1,0]$ the potentials will be defined so that $A_\delta$ and $A_C$ vanish at $t=-1$. In particular, since $A_\delta$ vanishes at $t=-1$ and $t=0$, the constant we are adding is indeed independent of $\Phi$, as follows from the computation of appendix \ref{app:EOM}. By a reparameterization $t' = (t+1)/2$ we may then write the action
\begin{equation}
S = \mathrm{constant}+ \int_{0}^1 dt\, \omega\left(A_t,\pi_1\D\frac{1}{1-A_C}\right) = \int_{0}^1 dt'\, \omega\left(A_t,\pi_1\D\frac{1}{1-A_C}\right).
\end{equation}
In this way we have reformulated the WZW-like action so that $A_C$ vanishes at $t'=0$ by construction.} Now we substitute the flatness condition for $\dot{A}_C$ and $\dot{A}_v$,
\begin{eqnarray}
\dot{A}_C \lineup = \pi_1\C\frac{1}{1-A_C}\otimes A_t\otimes\frac{1}{1-A_C},\\
\dot{A}_v \lineup= \pi_1{\bf v}\frac{1}{1-A_C}\otimes A_t\otimes\frac{1}{1-A_C} - \pi_1\C \frac{1}{1-A_C}\otimes A_v\otimes \frac{1}{1-A_C}\otimes A_t\otimes \frac{1}{1-A_C}\nonumber\\
\lineup\ \ \  + \pi_1\C\frac{1}{1-A_C}\otimes A_t\otimes \frac{1}{1-A_C}\otimes A_v\otimes \frac{1}{1-A_C} + \pi_1\C\frac{1}{1-A_C}\otimes A_{ v t }\otimes \frac{1}{1-A_C},
\end{eqnarray}
to find
\begin{eqnarray}
\delta S \lineup = \int_0^1 dt\left[\omega\left(\pi_1{\bf v}\frac{1}{1-A_C}\otimes A_t\otimes\frac{1}{1-A_C},\pi_1\D\frac{1}{1-A_C}\right)\right.
\nonumber\\
\lineup\ \ \ \ \ \ \ \ -\omega\left(\pi_1\C \frac{1}{1-A_C}\otimes A_v\otimes \frac{1}{1-A_C}\otimes A_t\otimes \frac{1}{1-A_C},\pi_1\D\frac{1}{1-A_C}\right)\nonumber\\
\lineup\ \ \ \ \ \ \ \ +\omega\left(\pi_1\C\frac{1}{1-A_C}\otimes A_t\otimes \frac{1}{1-A_C}\otimes A_v\otimes \frac{1}{1-A_C},\pi_1\D\frac{1}{1-A_C}\right)\nonumber\\
\lineup\ \ \ \ \ \ \ \ +\omega\left(\pi_1\C\frac{1}{1-A_C}\otimes A_{ v t }\otimes \frac{1}{1-A_C},\pi_1\D\frac{1}{1-A_C}\right)\nonumber\\
\lineup\ \ \ \ \ \ \ \ \left.+\omega\left(A_v,\pi_1\D\C\frac{1}{1-A_C}\otimes A_t\otimes \frac{1}{1-A_C}\right)\right].
\end{eqnarray}
The term with $A_{vt}$ drops out using cyclicity of $\C$ and the flatness condition \eq{0f}. Next we use cyclicity of the coderivations to place $A_t$ on the first entry of the symplectic form in the first term, and $A_v$ on the first entry of the symplectic form in the second two terms:
\begin{eqnarray}
\delta S \lineup = \int_0^1 dt\left[-\omega\left(A_t ,\pi_1{\bf v}\frac{1}{1-A_C}\otimes\left(\pi_1\D\frac{1}{1-A_C}\right)\otimes\frac{1}{1-A_C}\right)\right.
\nonumber\\
\lineup\ \ \ \ \ \ \ \ \ 
+\omega\left(A_v, \pi_1\C\frac{1}{1-A_C}\otimes A_t\otimes \frac{1}{1-A_C}\otimes \left(\pi_1\D\frac{1}{1-A_C}\right)\otimes \frac{1}{1-A_C}\right)\nonumber\\
\lineup\ \ \ \ \ \ \ \ 
-\omega\left(A_v,\pi_1\C \frac{1}{1-A_C}\otimes \left(\pi_1\D\frac{1}{1-A_C}\right)\otimes \frac{1}{1-A_C}\otimes A_t\otimes \frac{1}{1-A_C}\right)\nonumber\\
\lineup\ \ \ \ \ \ \ \ \left.
-\omega\left(A_v,\pi_1\C\D\frac{1}{1-A_C}\otimes A_t\otimes \frac{1}{1-A_C}\right)\right].
\end{eqnarray}
The last three terms combine giving
\begin{eqnarray}
\delta S \lineup = \int_0^1 dt\left[-\omega\left(A_t ,\pi_1{\bf v}\frac{1}{1-A_C}\otimes\left(\pi_1\D\frac{1}{1-A_C}\right)\otimes\frac{1}{1-A_C}\right)\right.\nonumber\\
\lineup\ \ \ \ \ \ \ \ \left.
-\omega\left(A_v,\pi_1\C\frac{1}{1-A_C}\otimes\left(\pi_1\D\frac{1}{1-A_C}\otimes A_t\otimes \frac{1}{1-A_C}\right)\otimes\frac{1}{1-A_C}\right)\right].
\end{eqnarray}
In the second term we can use cyclicity of the coderivations to place $A_t$ on the first entry of the symplectic form:
\begin{eqnarray}
\delta S \lineup = \int_0^1 dt\left[-\omega\left(A_t ,\pi_1{\bf v}\frac{1}{1-A_C}\otimes\left(\pi_1\D\frac{1}{1-A_C}\right)\otimes\frac{1}{1-A_C}\right)\right.\nonumber\\
\lineup\ \ \ \ \ \ \ \ \left.
+\omega\left(A_t,\pi_1\D\frac{1}{1-A_C}\otimes\left(\pi_1\C\frac{1}{1-A_C}\otimes A_v\otimes \frac{1}{1-A_C}\right)\otimes\frac{1}{1-A_C}\right)\right].
\end{eqnarray}
Using the flatness condition \eq{vflat} we obtain
\begin{eqnarray}
\delta S  \lineup = \int_0^1 dt\left[-\omega\left(A_t ,\pi_1{\bf v}\frac{1}{1-A_C}\otimes\left(\pi_1\D\frac{1}{1-A_C}\right)\otimes\frac{1}{1-A_C}\right)\right.\nonumber\\
\lineup\ \ \ \ \ \ \ \ \left.
+\omega\left(A_t,\pi_1\D\frac{1}{1-A_C}\otimes\left(\pi_1{\bf v}\frac{1}{1-A_C}\right)\otimes\frac{1}{1-A_C}\right)\right]\nonumber\\
\lineup = \int_0^1 dt\, \omega\left(A_t,\pi_1\Big(\D{\bf v} -{\bf v}\D\Big)\frac{1}{1-A_C}\right).\nonumber\\
\lineup = 0
\end{eqnarray}
This vanishes on the assumption that ${\bf v}$ and $\D$ commute. This completes the proof that the proposed symmetry transformation leaves the WZW-like action invariant. 

\section{Proof of $A_\infty$ Relations} 
\label{app:DC}

In this appendix we prove that $\D$ and $\C$ are commuting $A_\infty$ structures:
\begin{equation}\D^2=\C^2 = [\C,\D]=0,\label{eq:appDC}\end{equation}
where in open superstring field theory the dynamical and constraint $A_\infty$ structures are defined
\begin{eqnarray}
\C \lineup = \n-\m_2|^0,\label{eq:appC}\\
\D \lineup = \Q +\mathcal{G}{\bf d},\label{eq:appD}
\end{eqnarray}
and the coderivation ${\bf d}$ and cohomomorphism $\F$ are defined recursively by:
\begin{eqnarray}
\pi_1 {\bf d} \lineup = \pi_1 \m_2|^2 \F, \label{eq:appd}\\
\pi_1\F \lineup = \pi_1\big(\mathbb{I}_{T\H} + \Xi|^2 {\bf d}\big).\label{eq:appF}
\end{eqnarray}
This recursion generates the Feynman graph expansion described in section \ref{subsec:Feynman}. The simplest way to see that $\D$ and $\C$ are nilpotent and commute is to transform to the simpler $A_\infty$ structures of the field equation \eq{field_equation2}. However, it is interesting to understand this more directly from the definition of $\C$ and $\D$ given in  \eq{appC}-\eq{appF}.

Computing the square of \eq{appD}, the condition $\D^2=0$ is equivalent to 
\begin{equation}\pi_1\Big([\Q,{\bf d}]+{\bf d}\mathcal{G}{\bf d}\Big) = 0.\label{eq:app2}\end{equation}
Acting this equation on an $n$-string state gives an identity involving the $n$th product contained in ${\bf d}$. The first nontrivial instance is given by acting on a 2-string state, which implies
\begin{equation}[\Q,\m_2|^2] = 0.\end{equation}
This is true by virtue of the fact that $Q$ is a derivation of the star product. To prove the higher identities we use an inductive argument. We start by computing 
\begin{equation}
\pi_1[\Q,{\bf d}] = -m_2|^2\pi_2[\Q,\F].
\end{equation}
We used \eq{appd} and $\pi_1\m_2|^2 = m_2|^2\pi_2$. As in section \ref{subsec:Feynman} and \cite{WB}, we represent $\pi_2$ in terms of the product and coproduct,
\begin{equation}\pi_2=\inverttriangle(\pi_1\otimes'\pi_1)\triangle,\label{eq:pcp}\end{equation}
and pull the coproduct towards the right. This gives
\begin{equation}
\pi_1[\Q,{\bf d}] = -m_2|^2\inverttriangle\Big(\pi_1[\Q,\F]\otimes'\pi_1\F + \pi_1\F\otimes'\pi_1[\Q,\F]\Big)\triangle.\label{eq:app3}
\end{equation}
In computing $\pi_1[\Q,\F]$, we allow ourselves to assume \eq{app2}. The reason is that our argument is inductive. In particular, in \eq{app3} $\pi_1[\Q,\F]$ on the right hand side acts on fewer states than $\pi_1[\Q,{\bf d}]$ on the left hand side, since $\F$ does not contain a $0$-string product. The idea is to show that \eq{app2} holds acting on larger numbers of states if it holds acting on fewer states. With this we can compute
\begin{eqnarray}
\pi_1[\Q,\F] \lineup = \pi_1\big[\Q,\mathbb{I}_{T\mathcal{H}}+\Xi|^2{\bf d}\big]\nonumber\\
\lineup = \pi_1\Big(X|^2{\bf d} - \Xi|^2[\Q,{\bf d}]\Big)\nonumber\\
\lineup = \pi_1\Big(X|^2{\bf d} - \Xi|^2{\bf d}\mathcal{G}{\bf d}\Big)\nonumber\\
\lineup = \pi_1\Big(\F\mathcal{G}{\bf d} - \mathbb{I}|^0{\bf d}\Big)\nonumber\\
\lineup = \pi_1\Big(\F\mathcal{G}{\bf d} - \mathbb{I}|^0\m_2|^2\F\Big).\label{eq:app35}
\end{eqnarray}
In the first step we used the definition of $\F$ in \eq{appF}, in the second step we acted $\Q$, in the third step we assumed \eq{app2}, in the fourth step we used the definition of $\F$ in \eq{appF}, and in the final step we used the definition of ${\bf d}$ in \eq{appd}. Plugging in to \eq{app3} gives
\begin{eqnarray}
\pi_1[\Q,{\bf d}] \lineup = -m_2|^2\inverttriangle\Big(\pi_1\big(\F\mathcal{G}{\bf d} - \mathbb{I}|^0\m_2|^2\F\big)\otimes'\pi_1\F + \pi_1\F\otimes'\pi_1\big(\F\mathcal{G}{\bf d} - \mathbb{I}|^0\m_2|^2\F\big)\Big)\triangle\nonumber\\
\lineup = -m_2|^2\inverttriangle(\pi_1\otimes'\pi_1)\triangle\Big(\F\mathcal{G}{\bf d} - \mathbb{I}|^0\m_2|^2\F\Big)\nonumber\\
\lineup = -\pi_1\m_2|^2 \Big(\F\mathcal{G}{\bf d} - \mathbb{I}|^0\m_2|^2\F\Big)\nonumber\\
\lineup = -\pi_1{\bf d}\mathcal{G}{\bf d} +\pi_1\m_2|^2\mathbb{I}|^0\m_2|^2\F.\label{eq:app4}
\end{eqnarray}
Note that
\begin{equation}
\pi_1\m_2|^2\mathbb{I}|^0\m_2|^2=0.\label{eq:app45}
\end{equation}
This expression vanishes identically in all cases except when acting on three Ramond states, and then it vanishes by associativity of the star product. Therefore we can drop the second term in \eq{app4}. This establishes \eq{app2} and~$\D^2=0$.

The proof of $\C^2=0$ is elementary, so let us prove $[\C,\D]=0$. This is equivalent to
\begin{equation}[\n,{\bf d}] = [\m_2|^0,{\bf d}].\label{eq:app5}\end{equation}
When acting on a 2-string state this implies 
\begin{equation}[\n,\m_2|^2] = 0,\end{equation}
which is true by virtue of the fact that $\n$ is a derivation of the star product. As above, we use an inductive argument to show that \eq{app5} must then also hold acting on a larger number of string states. We start by computing
\begin{eqnarray}
\pi_1[\n,{\bf d}] \lineup = -m_2|^2\pi_2[\n,\F]\nonumber\\
\lineup = -m_2|^2\inverttriangle\Big(\pi_1[\n,\F]\otimes'\pi_1\F + \pi_1\F\otimes'\pi_1[\n,\F]\Big)\triangle,\label{eq:app6}
\end{eqnarray}
where we substituted \eq{pcp} and pulled the coproduct to the right. In computing $\pi_1[\n,\F]$ we allow ourselves to assume that \eq{app5} is satisfied. Again, $\pi_1[\n,\F]$ on the right hand side acts on fewer states than $\pi_1[\n,{\bf d}]$ on the left hand side, and the idea is to prove \eq{app5} holds acting on more states if it holds acting on fewer states.
With this we can compute
\begin{eqnarray}
\pi_1[\n,\F]\lineup = \pi_1\big[\n,\mathbb{I}_{T\H}+\Xi|^2{\bf d}]\nonumber\\
\lineup = \pi_1\Big(\mathbb{I}|^2{\bf d} -\Xi|^2[\n,{\bf d}]\Big)\nonumber\\
\lineup = \pi_1\Big(\mathbb{I}|^2{\bf d} -\Xi|^2[\m_2|^0,{\bf d}]\Big)\nonumber\\
\lineup = \pi_1\Big(\mathbb{I}|^2{\bf d} -\Xi|^2{\bf d}\m_2|^0\Big)\nonumber\\
\lineup = \pi_1\Big(\mathbb{I}|^2{\bf d} - \F\m_2|^0 + \m_2|^0\Big)\nonumber\\
\lineup = \pi_1\Big(\mathbb{I}|^2\m_2|^2\F +\m_2|^0\F -\F\m_2|^0\Big)\nonumber\\
\lineup = \pi_1\Big(\m_2|_0\F-\F\m_2|^0\Big).\label{eq:app65}
\end{eqnarray}
In the first step we substituted the definition of $\F$ in \eq{appF}, in the second step we acted with $\n$, in the third step we assumed \eq{app5}, in the fourth step we dropped a vanishing term from the commutator where $\Xi|^2$ acts on $m_2|^0$, in the fifth step we substituted the definition of $\F$ in \eq{appF}, in the sixth step we substituted the definition of ${\bf d}$ in \eq{appd} and noted that $\F$ acts as the identity on the input of $m_2|^0$. Finally, we used that the star product at Ramond number 0 is given by
\begin{equation}\m_2|_0 = \mathbb{I}|^2\m_2|^2+\m_2|^0\end{equation}
Substituting this result into \eq{app6} gives
\begin{eqnarray}
\pi_1[\n,{\bf d}] \lineup = -m_2|^2\inverttriangle\Big(\pi_1\Big(\m_2|_0\F-\F\m_2|^0\Big)\otimes'\pi_1\F + \pi_1\F\otimes'\pi_1\Big(\m_2|_0\F-\F\m_2|^0\Big)\Big)\triangle\nonumber\\
\lineup = -m_2|^2\inverttriangle(\pi_1\otimes'\pi_1)\triangle \Big(\m_2|_0\F-\F\m_2|^0\Big)\nonumber\\
\lineup = -\pi_1\m_2|^2\Big(\m_2|_0\F-\F\m_2|^0\Big)\nonumber\\
\lineup = \pi_1{\bf d}\m_2|^0 -\pi_1\m_2|^2\m_2|_0\F.\label{eq:app7}
\end{eqnarray}
Next we note that 
\begin{equation}\pi_1\m_2|^2\m_2|_0 = -\pi_1\m_2|^0\m_2|^2.\end{equation}
This can be shown as follows:
\begin{eqnarray}
\pi_1\m_2|^2\m_2|_0 \lineup =\pi_1 (\mathbb{I}|^2\m_2|_0+\mathbb{I}|^0\m_2|_2)\m_2|_0\nonumber\\
\lineup = \pi_1\mathbb{I}|^0\m_2|_2\m_2|_0\nonumber\\
\lineup = -\pi_1\mathbb{I}|^0\m_2|_0\m_2|_2\nonumber\\
\lineup = -\pi_1\mathbb{I}|^0\m_2|_0(\mathbb{I}|^0\m_2|_2+\mathbb{I}|^2\m_2|_0)\nonumber\\
\lineup = -\pi_1\mathbb{I}|^0\m_2|_0\m_2|^2\nonumber\\
\lineup = -\pi_1\m_2|^0 \m_2|^2.\label{eq:app8}
\end{eqnarray}
In the second and third step we used associativity of the star product in the form of the identities
\begin{equation}
(\m_2|_0)^2 = 0,\ \ \ [\m_2|_0,\m_2|_2] = 0,
\end{equation}
and in the fourth step we added a term which produces a Ramond state which is annihilated by the projection $\mathbb{I}|^0$ in front. Then \eq{app7} becomes
\begin{equation}
\pi_1[\n,{\bf d}] = \pi_1\m_2|^0\m_2|^2\F +\pi_1{\bf d}\m_2|^0.
\end{equation}
It is tempting to substitute ${\bf d}$ for $\m_2|^2\F$ on the right hand side. However, this equality only holds when both objects are projected onto a single string state. So we have a few additional steps:
\begin{eqnarray}
\pi_1\m_2|^0\m_2|^2\F\lineup = m_2|^0\pi_2\m_2|^2\F\nonumber\\
\lineup = m_2|^0\inverttriangle (\pi_1\otimes'\pi_1)\triangle \m_2|^2\F\nonumber\\
\lineup =  m_2|^0\inverttriangle \Big(\pi_1\m_2|^2\F\otimes'\pi_1\F+\pi_1\F\otimes'\pi_1\m_2|^2\F\Big)\triangle \nonumber\\
\lineup =  m_2|^0\inverttriangle \Big(\pi_1{\bf d}\otimes'\pi_1\F+\pi_1\F\otimes'\pi_1{\bf d}\Big)\triangle \nonumber\\
\lineup =  m_2|^0\inverttriangle \Big(\pi_1{\bf d}\otimes'\pi_1+\pi_1\otimes'\pi_1{\bf d}\Big)\triangle \nonumber\\
\lineup =  \pi_1\m_2|^0{\bf d}.\label{eq:app9}
\end{eqnarray}
The crucial point comes in the second to last step, where we dropped the cohomomorphism $\F$. This is allows since $m_2|^0$ only accepts NS inputs, and in this case $\F$ reduces to the identity operator. Therefore we find
\begin{equation}
\pi_1[\n,{\bf d}] =\pi_1\Big(\m_2|^0{\bf d}+{\bf d}\m_2|^0\Big).
\end{equation}
This establishes \eq{app3} and $[\C,\D] = 0$.

\section{$\F$ and the Kunitomo-Okawa operator $F$}
\label{app:FF}

In this appendix we list some identities relating the cohomomorphism $\F$, defined recursively in terms a coderivation ${\bf d}$ by
\begin{eqnarray}
\pi_1\F\lineup =\pi_1(\mathbb{I}_{T\mathcal{H}}+\Xi|^2{\bf d}),\label{eq:appbfFd}\\
\pi_1{\bf d}\lineup = \pi_1\m_2|^2\F,
\end{eqnarray}
and the operator $F$ introduced by Kunitomo and Okawa:
\begin{equation}
F \equiv \frac{1}{\mathbb{I}-\Xi\mathrm{ad}_{N_C}},
\end{equation}
where $\mathrm{ad}_{N_C}$ acts on a string field $A$ as 
\begin{equation}
\mathrm{ad}_{N_C} A \equiv [N_C,A],
\end{equation}
and $N_C = (\eta e^{\PhiN(t)})e^{-\PhiN(t)}$ is the NS part of the 0-potential in the Berkovits theory.

If $n$ and $r$ are generic NS and R string fields, respectively, we note the relations
\begin{eqnarray}
\pi_1\F\frac{1}{1-N_C} \lineup = N_C,\\
\pi_1\F\frac{1}{1-N_C}\otimes n\otimes\frac{1}{1-N_C} \lineup = n\\
\pi_1\F\frac{1}{1-N_C}\otimes r\otimes\frac{1}{1-N_C}\lineup = F(r),\label{eq:FF}\\
\pi_1\F\frac{1}{1-N_C}\otimes n\otimes\frac{1}{1-N_C}\otimes r\otimes\frac{1}{1-N_C} \lineup = (-1)^{\eps(n)+1}F(\Xi(nF(r))),\\
\pi_1\F\frac{1}{1-N_C}\otimes r\otimes\frac{1}{1-N_C}\otimes n\otimes\frac{1}{1-N_C} \lineup = (-1)^{\eps(r)+1}F(\Xi(F(r)n)),
\end{eqnarray}
where $\eps$ denotes Grassmann parity. We can write generalizations where $\F$ acts on group-like elements tensored with two or more NS and R states. These vanish in all cases except when there is a single Ramond state in the string of tensor products. Also useful are closely related identities involving the coderivation ${\bf d}$:
\begin{eqnarray}
\pi_1{\bf d}\frac{1}{1-N_C}\otimes r\otimes\frac{1}{1-N_C}\lineup =[N_C,F(r)],\\
\pi_1{\bf d}\frac{1}{1-N_C}\otimes n\otimes\frac{1}{1-N_C}\otimes r\frac{1}{1-N_C}\lineup = (-1)^{\eps(n)+1}\Big(nF(r)+[N_C,F(\Xi(nF(r))]\Big),\\
\pi_1{\bf d}\frac{1}{1-N_C}\otimes r\otimes\frac{1}{1-N_C}\otimes n\frac{1}{1-N_C}\lineup = (-1)^{\eps(r)+1}\Big(F(r)n+[N_C,F(\Xi(F(r)n)]\Big),\\
\pi_1{\bf d}\frac{1}{1-N_C}\otimes r_1\otimes\frac{1}{1-N_C}\otimes r_2\frac{1}{1-N_C}\lineup = (-1)^{\eps(r_1)+1}F(r_1)F(r_2) .\label{eq:F2R}
\end{eqnarray}
We can write generalizations where ${\bf d}$ acts on group-like elements tensored with additional NS and R states. These vanish in all cases except when one or two Ramond states appear in the string of tensor products. 

Let us demonstrate the simplest nontrivial relation \eq{FF}. The other identities follow from similar arguments. Using \eq{appbfFd} we have
\begin{equation}
\pi_1\F\frac{1}{1-N_C}\otimes r \otimes \frac{1}{1-N_C} = r +\Xi|^2 m_2|^2\pi_2\F\frac{1}{1-N_C}\otimes r\otimes \frac{1}{1-N_C}.
\end{equation}
Next we write $\pi_2=\inverttriangle(\pi_1\otimes'\pi_1)\triangle$ and note 
\begin{eqnarray}
\triangle \F\frac{1}{1-N_C}\otimes r\otimes \frac{1}{1-N_C} \lineup =\left( \F\frac{1}{1-N_C}\otimes r\otimes \frac{1}{1-N_C}\right)\otimes'\left( \F\frac{1}{1-N_C}\right) \nonumber\\
\lineup\ \ \ \ \ \ + \left( \F\frac{1}{1-N_C}\right)\otimes'\left( \F\frac{1}{1-N_C}\otimes r\otimes \frac{1}{1-N_C}\right)\nonumber\\
\lineup =\left( \F\frac{1}{1-N_C}\otimes r\otimes \frac{1}{1-N_C}\right)\otimes'\left(\frac{1}{1-N_C}\right) \nonumber\\
\lineup\ \ \ \ \ \ + \left(\frac{1}{1-N_C}\right)\otimes'\left( \F\frac{1}{1-N_C}\otimes r\otimes \frac{1}{1-N_C}\right),
\end{eqnarray}
In the last step we use the fact that $\F$ acts as the identity on NS states. Acting with $\pi_1\otimes'\pi_1$ and the product $\inverttriangle$ then gives 
\begin{equation}
\pi_2 \F\frac{1}{1-N_C}\otimes r\otimes \frac{1}{1-N_C} = \left( \pi_1\F\frac{1}{1-N_C}\otimes r\otimes \frac{1}{1-N_C}\right)\otimes N_C + N_C\otimes\left(\pi_1 \F\frac{1}{1-N_C}\otimes r\otimes \frac{1}{1-N_C}\right).
\end{equation}
We therefore find 
\begin{eqnarray}
\pi_1\F\frac{1}{1-N_C}\otimes r \otimes \frac{1}{1-N_C}\lineup = r +\Xi|^2 m_2|^2\left[\left( \F\frac{1}{1-N_C}\otimes r\otimes \frac{1}{1-N_C}\right)\otimes N_C + N_C\otimes\left( \F\frac{1}{1-N_C}\otimes r\otimes \frac{1}{1-N_C}\right)\right]\nonumber\\
\lineup =r+\Xi\mathrm{ad}_{N_C}\pi_1\F\frac{1}{1-N_C}\otimes r \otimes \frac{1}{1-N_C}.
\end{eqnarray}
This in particular implies 
\begin{equation}
(\mathbb{I}-\Xi\mathrm{ad}_{N_C})\pi_1\F \frac{1}{1-N_C}\otimes r \otimes \frac{1}{1-N_C} = r.
\end{equation}
Inverting the operator $\mathbb{I}-\Xi\mathrm{ad}_{N_C}$ we therefore find
\begin{equation}
\pi_1\F \frac{1}{1-N_C}\otimes r \otimes \frac{1}{1-N_C} = \frac{1}{\mathbb{I}-\Xi\mathrm{ad}_{N_C}} r = F(r).
\end{equation}
which reproduces \eq{FF}.

\section{The Factorization $\G=\F\g$}
\label{app:factor}

In this appendix we prove the equality
\begin{equation}
\G=\F\g,
\end{equation}
where the cohomomorphism $\G$ was constructed in \cite{RWaction} and is characterized by the solution to \eq{m0diff}-\eq{mu02rec}, $\F$ is defined by \eq{bfFd} and \eq{bfd}, and $\g$ is characterized by the solution to \eq{WZWm0}-\eq{WZWmu0}.

To do this we express $\F\g$ as the path ordered exponential of a coderivation $\mmu|_0(t)$. Then we can compare $\mmu|_0(t)$ to the expression defined by \eq{m0diff}-\eq{mu02rec}. We introduce the cohomomorphism
\begin{equation}t^{\bf 1},\end{equation}
where ${\bf 1}$ is the coderivation corresponding to the identity operator and $t$ is a parameter. Conjugating a coderivation with $t^{\bf 1}$ has the effect of multiplying products with a power of $t$, as follows:
\begin{equation}t^{-{\bf 1}}\b_{n+1}t^{\bf 1} = t^n \b_{n+1}.\end{equation}
We introduce $t$-dependent cohomomorphisms
\begin{eqnarray}
\F(t) \lineup \equiv t^{-{\bf 1}}\F t^{\bf 1},\\
\g(t) \lineup \equiv t^{-{\bf 1}}\g t^{\bf 1},
\end{eqnarray}
and define 
\begin{eqnarray}
\mmu|_0(t)\lineup \equiv \big(\F(t)\g(t)\big)^{-1}\frac{d}{dt}\big(\F(t)\g(t)\big)\nonumber\\
\lineup = \frac{1}{t}t^{-{\bf 1}}\Big((\F\g)^{-1}[{\bf 1},\F\g]\Big)t^{\bf 1}.\label{eq:mu0fac}
\end{eqnarray}
It follows from general arguments that the expression in parentheses is a coderivation. Since it appears conjugated with $t^{\bf 1}$, we know that $\mmu|_0(t)$ can be expanded
\begin{equation}\mmu|_0(t) = \sum_{n=0}^\infty t^n \mmu_{n+2}|_0,\end{equation}
where $\mmu_{n+2}|_0$ are coderivations corresponding to $n+2$-string products $\mu_{n+2}|_0$. These are the gauge products. We also know that $\mmu|_0(t)$ carries Ramond number zero since both $\F$ and $\g$ can be defined by products which carry Ramond number zero. We may extract the $n+2$nd gauge product by computing
\begin{equation}\mu_{n+2}|_0\pi_{n+2} = \pi_1\mmu|_0(t)\pi_{n+2}\big|_{t=1}.\end{equation}
This definition of $\mmu|_0(t)$ automatically implies 
\begin{equation}
\F\g = \overrightarrow{\mathcal{P}}\exp\left[\int_0^1 dt\,\mmu|_0(t)\right].
\end{equation}
Therefore to prove $\G=\F\g$, we must show that $\mu_{n+2}|_0$ as defined by \eq{mu0fac} agrees with $\mu_{n+2}|_0$ as defined by \eq{m0diff}-\eq{mu02rec}.

We introduce a coderivation for gauge products:
\begin{eqnarray}
\m|_0(t) \lineup \equiv \big(\F(t)\g(t)\big)^{-1}\m_2|_0\big(\F(t)\g(t)\big)\nonumber\\
\lineup = \frac{1}{t}t^{-{\bf 1}}\Big((\F\g)^{-1}\m_2|_0(\F\g)\Big) t^{\bf 1}.\label{eq:smallm0}
\end{eqnarray}
It follows from this expression that $\m|_0(t)$ is a coderivation with Ramond number zero that has an expansion in powers of $t$:
\begin{equation}\m|_0(t)=\sum_{n=0}^\infty t^n \m_{n+2}|_0,\end{equation}
where $\m_{n+2}|_0$ are coderivations corresponding to products $m_{n+2}|_0$. These are the bare products. The constant term in this expansion is Witten's open string star product restricted to Ramond number 0. The definition of the bare products implies the relation
\begin{equation}
\frac{d}{dt}\m|_0(t) = [\m|_0(t),\mmu|_0(t)],
\end{equation}
in agreement with \eq{m0diff}. We may extract the bare products using
\begin{equation}m_{n+2}|_0 \pi_{n+2} = \pi_1\m|_0(t)\pi_{n+2}\big|_{t=1}.\end{equation}
When $m_{n+2}|_0$ multiplies only NS states, it agrees with $m_{n+2}|^0$ as defined by \eq{WZWm0}. This can be shown as follows:
\begin{eqnarray}
m_{n+2}|_0 \pi_{n+2}^0 \lineup = \pi_1\big(\F\g\big)^{-1}\m_2|_0\big(\F\g\big)\pi_{n+2}^0\nonumber\\
\lineup = \pi_1\g^{-1}\m_2|^0\g\pi_{n+2}^0,\label{eq:ml0mu0}
\end{eqnarray}
where $\pi_m^r$ is the projection onto $m$ string states carrying $r$ Ramond factors. In the second step $\F$ drops out since $\F$ is equal to the identity unless it receives Ramond inputs, and $m_2|_0$ and $m_2|^0$ are equal when multiplying NS states.  The final expression \eq{ml0mu0} is equivalent to the definition of $m_{n+2}|^0$ implied by \eq{WZWm0}. Now we can compute the gauge products. Consider the part which multiplies NS states
\begin{eqnarray}\mu_{n+2}|_0^0\pi_{n+2}\lineup =\pi_1\mmu|_0(t)\pi_{n+2}^0|_{t=1}\nonumber\\
\lineup = \left.\pi_1 \big(\F(t)\g(t)\big)^{-1}\frac{d}{dt}\big(\F(t)\g(t)\big)\pi_{n+2}^0\right|_{t=1}.
\end{eqnarray}
$\F$ drops out since no Ramond states are present, and we get 
\begin{eqnarray}
\mu_{n+2}|_0^0\pi_{n+2}\lineup =\left.\pi_1 \g(t)^{-1}\frac{d}{dt}\g(t)\pi_{n+2}^0\right|_{t=1}\nonumber\\
\lineup = \pi_1\mmu|^0(t)\pi_{n+2}\big|_{t=1}\nonumber\\
\lineup = \frac{1}{n+3}\Big(\Xi m_{n+2}|^0-m_{n+2}|^0(\Xi\otimes \mathbb{I}^{\otimes n+1}+...+\mathbb{I}^{\otimes n+1}\otimes \Xi)\Big)\pi_{n+2}^0.
\end{eqnarray}
Finally, we have argued that $m_{n+2}|^0$ derived from \eq{WZWm0} is equivalent to $m_{n+2}|_0$ derived from \eq{smallm0} when multiplying NS states. Therefore, the gauge products derived from $\F\g$ satisfy the same relation \eq{mu00rec} as the gauge products derived from $\G$. Let us look at what happens when $\mu_{n+2}|_0$ multiplies a Ramond state:
\begin{eqnarray}\mu_{n+2}|_0^2\pi_{n+2}\lineup =\pi_1\mmu|_0(t)\pi_{n+2}^1|_{t=1}\nonumber\\
\lineup = \left.\pi_1 \g(t)^{-1}\F(t)^{-1}\frac{d}{dt}\big(\F(t)\g(t)\big)\pi_{n+2}^1\right|_{t=1}.
\end{eqnarray}
The factor $\g^{-1}(t)$ is equivalent to the identity when it produces a Ramond output, so we can further simplify
\begin{eqnarray}\mu_{n+2}|_0^2\pi_{n+2}\lineup = -\pi_1 \left.\left(\frac{d}{dt}\F(t)^{-1}\right)\right|_{t=1}\F\g\pi_{n+2}^1.
\end{eqnarray}
Now note that 
\begin{eqnarray}
\pi_1\F(t)^{-1}\lineup = \pi_1t^{-{\bf 1}}(\mathbb{I}_{T\mathcal{H}}-\Xi|^2\m_2|^2)t^{\bf 1}\nonumber\\
\lineup =\pi_1(\mathbb{I}_{T\mathcal{H}}-t\Xi|^2\m_2|^2),
\end{eqnarray}
and 
\begin{equation}\left.\frac{d}{dt}\pi_1\F(t)^{-1}\right|_{t=1}=-\Xi|^2\pi_1\m_2|^2.\end{equation}
Therefore 
\begin{eqnarray}\mu_{n+2}|_0^2\pi_{n+2}\lineup = \Xi\pi_1 \m_2|_0\F\g\pi_{n+2}^1.
\end{eqnarray}
When producing a Ramond output $m_2|^2$ is equivalent to $m_2|_0$. Now we can insert $\F^{-1}$ as follows
\begin{eqnarray}
\mu_{n+2}|_0^2\pi_{n+2}\lineup = \Xi\pi_1\big(\mathbb{I}-\Xi|^2\m_2|^2\big) \m_2|_0\F\g\pi_{n+2}^1\nonumber\\
\lineup = \Xi\pi_1\F^{-1} \m_2|_0\F\g\pi_{n+2}^1.
\end{eqnarray}
In this case $\F^{-1}$ is equivalent to the identity operator by associativity of the star product. We may also insert $\g^{-1}$
\begin{eqnarray}
\mu_{n+2}|_0^2\pi_{n+2}\lineup  = \Xi\pi_1\big(\F\g\big)^{-1} \m_2|_0\big(\F\g\big)\pi_{n+2}^1,
\end{eqnarray}
since $\g^{-1}$ is equivalent to the identity operator when it receives Ramond inputs. This implies
\begin{equation}\mu_{n+2}|_0^2 = \Xi m_{n+2}|_0^2.\end{equation}
Therefore, the gauge products defined by \eq{mu0fac} satisfies the same relation \eq{mu02rec} as the gauge products of $\G$. We conclude that the gauge products defined by the two constructions must be equal. This proves the factorization $\G=\F\g$.

\section{Proof of Supersymmetry}
\label{app:susy}

In this appendix we prove that the dynamical and constraint $A_\infty$ structures are invariant under supersymmetry:
\begin{eqnarray}
[\del_s,\D] \lineup = 0,\\
\ [\del_s,\C] \lineup = 0,
\end{eqnarray}
where $\del_s$ is defined by
\begin{equation}\del_s \equiv \mathcal{G}\Ssigma,\label{eq:appds}\end{equation}
and 
\begin{equation}
\pi_1\Ssigma \equiv \pi_1\big(\s_1 + \s_1\Xi|^2{\bf d} + {\bf d}\Xi|^2\s_1\big).\label{eq:appSigma}
\end{equation}
The demonstration follows the same strategy as appendix \ref{app:DC}. An equivalent but technically quite different proof of supersymmetry appears in \cite{susyS}.

Invariance of the dynamical $A_\infty$ structure implies that $\Ssigma$ should satisfy
\begin{equation}\pi_1\Big([\Q,\Ssigma] +{\bf d}\mathcal{G}\Ssigma - \Ssigma\mathcal{G}{\bf d} \Big)=0.\label{eq:appsusy0}\end{equation}
When acting on a 1-string state this implies
\begin{equation}[\Q,\s_1] = 0,\end{equation}
which is true by virtue of the fact that the zero mode of the fermion vertex is BRST invariant.  As in appendix \ref{app:DC}, we use an inductive argument to show that \eq{appsusy0} must then also hold when acting on a larger number of string states. We start by computing
\begin{eqnarray}
\pi_1[\Q,\Ssigma] \lineup = \pi_1\big[\Q,\s_1 + \s_1\Xi|^2{\bf d} + {\bf d}\Xi|^2\s_1\big]\nonumber\\
 \lineup = \pi_1\Big(\s_1X|^2{\bf d}-\s_1\Xi|^2[\Q,{\bf d}] +[\Q,{\bf d}\Xi|^2\s_1]\Big)\nonumber\\
 \lineup =  \pi_1\Big(\s_1X|^2{\bf d}+\s_1\Xi|^2{\bf d}\mathcal{G}{\bf d} +[\Q,{\bf d}\Xi|^2\s_1]\Big).\label{eq:appsusy025}
\end{eqnarray}
In the second step we used \eq{app2}. We can also use \eq{app2} to evaluate the last term, but this is not helpful. Instead we compute the last term as follows:
\begin{eqnarray}
[\Q,{\bf d}\Xi|^2\s_1]\lineup = -m_2|^2\pi_2[\Q,\F\Xi^2\s_1]\nonumber\\
\lineup = -m_2|^2\inverttriangle\Big(\pi_1[\Q,\F\Xi|^2\s_1]\otimes'\pi_1\F - \pi_1\F\Xi|^2\s_1\otimes'\pi_1[\Q,\F]\nonumber\\
\lineup\ \ \ \ \ \ \ \ \ \ \ \ \ \ \ \ + \pi_1[\Q,\F]\otimes'\pi_1\F\Xi|^2\s_1 + \pi_1\F\otimes'\pi_1[\Q,\F\Xi|^2\s_1]\Big)\triangle.\label{eq:appsusy05}
\end{eqnarray}
We substituted
\begin{equation}\pi_2=\inverttriangle(\pi_1\otimes'\pi_1)\triangle\end{equation}
and pulled the coproduct to the right. An expression for $\pi_1[\Q,\F]$ was found in appendix \ref{app:DC}, so the new object to compute is $\pi_1[\Q,\F\Xi|^2\s_1]$. In summary, we have found that computing $\pi_1[\Q,\Ssigma]$ requires computing $\pi_1[\Q,{\bf d}\Xi|^2\s_1]$ which in turn requires computing $\pi_1[\Q,\F\Xi|^2\s_1]$:
\begin{equation}\pi_1[\Q,\Ssigma]\ \to\ \pi_1[\Q,{\bf d}\Xi|^2\s_1]\ \to\ \pi_1[\Q,\F\Xi|^2\s_1].\label{eq:appsusy075}\end{equation}
In deriving $\pi_1[\Q,\F\Xi|^2\s_1]$ we allow ourselves to assume \eq{appsusy0}. Again, the idea is to assume that \eq{appsusy0} holds when acting on fewer states, and then show that \eq{appsusy0} must hold acting on more states. Note the relation
\begin{equation}\pi_1\F\Xi|^2\s_1 = \pi_1\Xi|^2\Ssigma,\label{eq:appsusy1}\end{equation}
and substitute:
\begin{eqnarray}
\pi_1[\Q,\F\Xi|^2\s_1] \lineup = \pi_1[\Q,\Xi|^2\Ssigma]\nonumber\\
\lineup = \pi_1\Big(X|^2\Ssigma -\Xi|^2[\Q,\Ssigma]\Big)\nonumber\\
\lineup =\pi_1\Big( X|^2\Ssigma +\Xi|^2({\bf d}\mathcal{G}\Ssigma-\Ssigma\mathcal{G}{\bf d})\Big)\nonumber\\
\lineup = \pi_1\Big(\F\mathcal{G}\Ssigma-\mathbb{I}|^0\Ssigma-\F\Xi|^2\s_1\mathcal{G}{\bf d}\Big).
\end{eqnarray}
In the second step we computed the BRST variation, in the third step we assumed \eq{appsusy0}, and in the fourth step we used \eq{appF} and \eq{appsusy1}. Note that 
\begin{eqnarray}
\pi_1\Ssigma \lineup = \pi_1\Big(\s_1\F+{\bf d}\Xi|^2\s_1\Big)\nonumber\\
\lineup = \pi_1\Big(\s_1\F+\m_2|^2\F\Xi|^2\s_1\Big).\label{eq:appsusy15}
\end{eqnarray}
Therefore
\begin{equation}
\pi_1[\Q,\F\Xi|^2\s_1]=\pi_1\Big(\F\mathcal{G}\Ssigma-\F\Xi|^2\s_1\mathcal{G}{\bf d}-\mathbb{I}|^0\s_1\F+\mathbb{I}|^0\m_2|^2\F\Xi|^2\s_1\Big).
\end{equation}
Returning to \eq{appsusy075}, we can go backward and derive $\pi_1[\Q,{\bf d}\Xi|^2\s_1]$. Noting \eq{app35},
\begin{equation}\pi_1[\Q,\F] = \pi_1\Big(\F\mathcal{G}{\bf d} - \mathbb{I}|^0\m_2|^2\F\Big),\end{equation}
we can plug into \eq{appsusy05} to obtain
\begin{eqnarray}
\pi_1[\Q,{\bf d}\Xi|^2\s_1]\lineup = -m_2|^2\inverttriangle\Big(\pi_1\big(\F\mathcal{G}\Ssigma-\F\Xi|^2\s_1\mathcal{G}{\bf d}-\mathbb{I}|^0\s_1\F+\mathbb{I}|^0\m_2|^2\F\Xi|^2\s_1\big)\otimes'\pi_1\F\nonumber\\
\lineup\ \ \ \ \ \ \ \ \ \ \ \ \ \ \ \ - \pi_1\F\Xi|^2\s_1\otimes'\pi_1\big(\F\mathcal{G}{\bf d} - \mathbb{I}|^0\m_2|^2\big) + \pi_1\big(\F\mathcal{G}{\bf d} - \mathbb{I}|^0\m_2|^2)\otimes'\pi_1\F\Xi|^2\s_1 \nonumber\\
\lineup\ \ \ \ \ \ \ \ \ \ \ \ \ \ \ \ + \pi_1\F\otimes'\pi_1\big(\F\mathcal{G}\Ssigma-\F\Xi|^2\s_1\mathcal{G}{\bf d}-\mathbb{I}|^0\s_1\F+\mathbb{I}|^0\m_2|^2\F\Xi|^2\s_1\big)\Big)\triangle\nonumber\\
\lineup = -m_2|^2\inverttriangle(\pi_1\otimes'\pi_1)\triangle\Big(\F\mathcal{G}\Ssigma-\F\Xi|^2\s_1\mathcal{G}{\bf d}-\mathbb{I}|^0\s_1\F+\mathbb{I}|^0\m_2|^2\F\Xi|^2\s_1\Big)\nonumber\\
\lineup = -\pi_1\m_2|^2\Big(\F\mathcal{G}\Ssigma-\F\Xi|^2\s_1\mathcal{G}{\bf d}-\mathbb{I}|^0\s_1\F+\mathbb{I}|^0\m_2|^2\F\Xi|^2\s_1\Big)\nonumber\\
\lineup = -\pi_1{\bf d}\mathcal{G}\Ssigma+\pi_1{\bf d}\Xi|^2\s_1\mathcal{G}{\bf d}+\pi_1\m_2|^2\mathbb{I}|^0\s_1\F+\pi_1\m_2|^2\mathbb{I}|^0\m_2|^2\F\Xi|^2\s_1\nonumber\\
\lineup = -\pi_1{\bf d}\mathcal{G}\Ssigma+\pi_1{\bf d}\Xi|^2\s_1\mathcal{G}{\bf d}+\pi_1\m_2|^2\mathbb{I}|^0\s_1\F.\label{eq:appsusy2}
\end{eqnarray}
In the last step we dropped the term containing $\pi_1\m_2|^2\mathbb{I}|^0\m_2|^2$ recalling \eq{app45}. Next observe that the combination $\pi_1\m_2|^2\mathbb{I}|^0\s_1$ is nonzero only when acting on two Ramond states, in which case we have 
\begin{equation}
\m_2|^2\mathbb{I}|^0\s_1(R_1\otimes R_2) = \m_2\s_1R_1\otimes R_2 = \s_1\m_2 (R_1\otimes R_2) = \s_1\mathbb{I}|^0\m_2|^2 (R_1\otimes R_2),
\end{equation}
where we used that $s_1$ is a derivation of the star product. The final expression on the right is also nonvanishing only when acting on two Ramond states. We can therefore substitute this into \eq{appsusy2} to find
\begin{equation}\pi_1[\Q,{\bf d}\Xi|^2\s_1]=\pi_1\Big(-{\bf d}\mathcal{G}\Ssigma+{\bf d}\Xi|^2\s_1\mathcal{G}{\bf d}+\s_1\mathbb{I}|^0{\bf d}\Big).\end{equation}
Returning to \eq{appsusy075}, we can go back further to derive $\pi_1[\Q,\Ssigma]$. From \eq{appsusy025} we have
\begin{eqnarray}
\pi_1[\Q,\Ssigma] \lineup = \pi_1\Big(\s_1X|^2{\bf d} + \s_1\Xi|^2{\bf d}\mathcal{G}{\bf d} -{\bf d}\mathcal{G}\Ssigma+{\bf d}\Xi|^2\s_1\mathcal{G}{\bf d}+\s_1\mathbb{I}|^0{\bf d}\Big)\nonumber\\
\lineup = \pi_1\Big(\big(\s_1+ \s_1\Xi|^2{\bf d}+{\bf d}\Xi|^2\s_1\big)\mathcal{G}{\bf d}  -{\bf d}\mathcal{G}\Ssigma\Big)\nonumber\\
\lineup = \pi_1\Big(\Ssigma\mathcal{G}{\bf d}-{\bf d}\mathcal{G}\Ssigma\Big).
\end{eqnarray}
This proves \eq{appsusy05} and invariance of the dynamical $A_\infty$ structure under supersymmetry.

Finally, we prove supersymmetry of the constraint $A_\infty$ structure. This requires that $\Ssigma$ should satisfy
\begin{equation}\pi_1[\n,\Ssigma] = \pi_1[\m_2|^0,\Ssigma].\label{eq:appsusy3}\end{equation}
When acting on a 1-string state this implies 
\begin{equation}[\n,\s_1] = 0,\end{equation}
which is true by virtue of the fact that the zero mode of the fermion vertex is in the small Hilbert space. As in earlier examples, we use an inductive argument to show that \eq{appsusy3} must then also hold when acting on a larger number of string states. We start by computing
\begin{eqnarray}
\pi_1[\n,\Ssigma] \lineup = \pi_1[\n,\s_1 + \s_1\Xi|^2{\bf d} + {\bf d}\Xi|^2\s_1]\nonumber\\
\lineup = \pi_1\Big(\s_1\mathbb{I}|^2{\bf d}-\s_1\Xi|^2[\m_2|^0,{\bf d}]+[\n,{\bf d}\Xi|^2\s_1]\Big)\nonumber\\
\lineup =\pi_1\Big(\s_1\mathbb{I}|^2{\bf d}-\s_1\Xi|^2{\bf d}\m_2|^0+[\n,{\bf d}\Xi|^2\s_1]\Big).\label{eq:appsusy325}
\end{eqnarray}
In the second step we used \eq{app3} and in the third step we dropped one term from the commutator since $m_2|^0$ only produces an NS output. We can also use \eq{app3} to evaluate the last term, but this is not helpful. Instead we compute the last term as follows:
\begin{eqnarray}
\pi_1[\n,{\bf d}\Xi|^2\s_1]\lineup = -m_2|^2\pi_2[\n,\F\Xi|^2\s_1]\nonumber\\
\lineup =  -m_2|^2\inverttriangle\Big(\pi_1[\n,\F\Xi|^2\s_1]\otimes'\pi_1\F +\pi_1[\n,\F]\otimes'\pi_1\F\Xi|^2\s_1\nonumber\\
\lineup\ \ \ \ \ \ \ \ \ \ \ \ \ \ \ \ -\pi_1\F\Xi|^2\s_1\otimes'\pi_1[\n,\F] + \pi_1\F\otimes'\pi_1[\n,\F\Xi|^2\s_1]\Big)\triangle.\label{eq:appsusy35}
\end{eqnarray}
We substituted
\begin{equation}\pi_2=\inverttriangle(\pi_1\otimes'\pi_1)\triangle\end{equation}
and pulled the coproduct to the right. An expression for $\pi_1[\n,\F]$ was found in appendix \ref{app:DC}, so the new object to compute is $\pi_1[\n,\F\Xi|^2\s_1]$. In summary, we find that computing $\pi_1[\n,\Ssigma]$ requires computing $\pi_1[\n,{\bf d}\Xi|^2\s_1]$ which in turn requires computing $\pi_1[\n,\F\Xi|^2\s_1]$:
\begin{equation}\pi_1[\n,\Ssigma]\ \to\ \pi_1[\n,{\bf d}\Xi|^2\s_1]\ \to\ \pi_1[\n,\F\Xi|^2\s_1].\label{eq:appsusy4}\end{equation}
In deriving $\pi_1[\n,\F\Xi|^2\s_1]$ we allow ourselves to assume \eq{appsusy3}. Again, the idea is to assume that \eq{appsusy3} holds when acting on fewer states, and then show that \eq{appsusy3} must hold acting on more states. Recalling \eq{appsusy1} we have
\begin{eqnarray}
\pi_1[\n,\F\Xi|^2\s_1]\lineup  = \pi_1[\n,\Xi|^2\Ssigma]\nonumber\\
\lineup = \pi_1\Big(\mathbb{I}|^2\Ssigma -\Xi|^2[\n,\Ssigma]\Big)\nonumber\\
\lineup = \pi_1\Big(\mathbb{I}|^2\Ssigma-\Xi|^2[\m_2|^0,\Ssigma]\Big)\nonumber\\
\lineup = \pi_1\Big(\mathbb{I}|^2\Ssigma+\Xi|^2\Ssigma\m_2|^0\Big)\nonumber\\
\lineup = \pi_1\Big(\mathbb{I}|^2\s_1\F +\mathbb{I}|^2\m_2|^2\F\Xi|^2\s_1+\F\Xi|^2\s_1\m_2|^0\Big)\nonumber\\
\lineup = \pi_1\Big(\mathbb{I}|^2\s_1\F +\m_2|_0\F\Xi|^2\s_1+\F\Xi|^2\s_1\m_2|^0\Big).
\end{eqnarray}
In the second step we computed the action of $\n$, in the third step we assumed \eq{appsusy3}, in the fourth step we dropped a term from the commutator since $m_2|^0$ only produces an NS output, in the fifth step we substituted \eq{appsusy1} and \eq{appsusy15}, and in the final step we noted that $\pi_1\mathbb{I}|^2\m_2|^2$ is equivalent to $\pi_1\m_2|_0$ since in this equation both receive at least one Ramond input. Returning to \eq{appsusy4}, we can go back and derive $\pi_1[\n,{\bf d}\Xi|^2\s_1]$. Recalling \eq{app65},
\begin{equation}\pi_1[\n,\F] =\pi_1(\m_2|_0\F-\F\m_2|^0),\end{equation}
we can plug into \eq{appsusy35} to obtain
\begin{eqnarray}
\pi_1[\n,{\bf d}\Xi|^2\s_1]\lineup =  -m_2|^2\inverttriangle\Big(\pi_1\big(\mathbb{I}|^2\s_1\F +\m_2|_0\F\Xi|^2\s_1+\F\Xi|^2\s_1\m_2|^0\big)\otimes'\pi_1\F \nonumber\\
\lineup\ \ \ \ \ \ \ \ \ \ \ \ \ \ \ \ +\pi_1\big(\m_2|_0\F-\F\m_2|^0\big)\otimes'\pi_1\F\Xi|^2\s_1 -\pi_1\F\Xi|^2\s_1\otimes'\pi_1\big(\m_2|_0\F-\F\m_2|^0\big)\nonumber\\ \lineup\ \ \ \ \ \ \ \ \ \ \ \ \ \ \ \ + \pi_1\F\otimes'\pi_1\big(\mathbb{I}|^2\s_1\F +\m_2|_0\F\Xi|^2\s_1+\F\Xi|^2\s_1\m_2|^0\big)\Big)\triangle\nonumber\\
\lineup = -m_2|^2\inverttriangle(\pi_1\otimes'\pi_1)\triangle\Big(\mathbb{I}|^2\s_1\F +\m_2|_0\F\Xi|^2\s_1+\F\Xi|^2\s_1\m_2|^0\Big)\nonumber\\
\lineup = -\pi_1\m_2|^2\Big(\mathbb{I}|^2\s_1\F +\m_2|_0\F\Xi|^2\s_1+\F\Xi|^2\s_1\m_2|^0\Big)\nonumber\\
\lineup = \pi_1\Big(-\m_2|^2\mathbb{I}|^2\s_1\F -\m_2|^2\m_2|_0\F\Xi|^2\s_1-{\bf d}\Xi|^2\s_1\m_2|^0\Big).
\end{eqnarray}
Recalling the argument of \eq{app8} and \eq{app9} we have the relation
\begin{equation}\pi_1\m_2|^2\m_2|_0\F =-\pi_1\m_2|^0{\bf d}.\end{equation}
So we get
\begin{equation}
\pi_1[\n,{\bf d}\Xi|^2\s_1]=\pi_1\Big(-\m_2|^2\mathbb{I}|^2\s_1\F +\m_2|^0{\bf d}\Xi|^2\s_1-{\bf d}\Xi|^2\s_1\m_2|^0\Big).
\end{equation}
Returning to \eq{appsusy4}, we can go back and finally derive $\pi_1[\n,\Ssigma]$. From \eq{appsusy325} we get
\begin{eqnarray}
\pi_1[\n,\Ssigma] \lineup = \pi_1\Big(\s_1\mathbb{I}|^2{\bf d}-\s_1\Xi|^2{\bf d}\m_2|^0-\m_2|^2\mathbb{I}|^2\s_1\F +\m_2|^0{\bf d}\Xi|^2\s_1-{\bf d}\Xi|^2\s_1\m_2|^0\Big)\nonumber\\
\lineup = \pi_1\Big(-\big(\s_1 +{\bf d}\Xi|^2\s_1 + \s_1\Xi|^2{\bf d}\big)\m_2|^0+\m_2|^0\big(\s_1\F +{\bf d}\Xi|^2\s_1)\nonumber\\
\lineup\ \ \ \ \ \ \ \ \ +\s_1\m_2|^0-\m_2|^0\s_1 \F +\s_1\mathbb{I}|^2{\bf d}-\m_2|^2\mathbb{I}|^2\s_1\F\Big)\nonumber\\
\lineup = \pi_1\Big(-\Ssigma \m_2|^0+\m_2|^0\Ssigma+\s_1\m_2|^0\F-\m_2|^0\s_1 \F +\s_1\mathbb{I}|^2\m_2|^2\F-\m_2|^2\mathbb{I}|^2\s_1\F\Big)\nonumber\\
\lineup = \pi_1\Big([\m_2|^0,\Ssigma] +\big([\s_1,\m_2|^0]+\s_1\mathbb{I}|^2\m_2|^2-\m_2|^2\mathbb{I}|^2\s_1\big)\F\Big).
\end{eqnarray}
In the second step we added and subtracted terms; in the third step we substituted the definition of $\Ssigma$, noted that $\F$ acts as the identity on the input of $m_2|^0$, and substituted \eq{appd} for ${\bf d}$; in the last step we collected terms. This result is almost what we want, but we must show that
\begin{equation}\pi_1\Big([\s_1,\m_2|^0]+\s_1\mathbb{I}|^2\m_2|^2-\m_2|^2\mathbb{I}|^2\s_1\Big)=0.\label{eq:appsusy5}\end{equation}
To show this, we write
\begin{eqnarray}
\pi_1\Big(\s_1\mathbb{I}|^2\m_2|^2-\m_2|^2\mathbb{I}|^2\s_1\Big)\lineup = \pi_1\Big(\mathbb{I}|^0 \s_1\m_2-(\m_2|_2+\mathbb{I}|^2\m_2|_0)\s_1|_{-1}\Big).
\end{eqnarray}
Here we noted that $m_2|^2$ is equivalent to $m_2$ when it produces a Ramond output, we noted $s_1\mathbb{I}|^2 = \mathbb{I}|^0s_1$, we substituted
\begin{equation}m_2|^2 = m_2|_2 +\mathbb{I}|^2m_2|_0\end{equation}
and noted that $\mathbb{I}|^2s_1$ is equivalent to the Ramond number $-1$ component of $s_1$. Taking the Ramond number $-1$ component of $[\s_1,\m_2]=0$ implies
\begin{equation}[\s_1|_{-1},\m_2|_0]=0.\end{equation}
Moreover, $m_2|_2$ can be replaced with $\mathbb{I}|^0m_2$ since both receive at least one Ramond input. Then
\begin{eqnarray}
\pi_1\Big(\s_1\mathbb{I}|^2\m_2|^2-\m_2|^2\mathbb{I}|^2\s_1\Big) \lineup = 
\pi_1\Big(\mathbb{I}|^0 \s_1\m_2-\mathbb{I}|^0\m_2\s_1|_{-1}-\mathbb{I}|^2\s_1|_{-1}\m_2|_0\Big)\nonumber\\
\lineup = 
\pi_1\Big(\mathbb{I}|^0 \s_1\m_2-\mathbb{I}|^0\m_2(\s_1-\mathbb{I}|^0\s_1)-\s_1\mathbb{I}|^0\m_2|_0\Big)\nonumber\\
\lineup = \pi_1\Big(\mathbb{I}|^0\m_2\mathbb{I}|^0\s_1-\s_1\mathbb{I}|^0\m_2|_0\Big)\nonumber\\
\lineup = \pi_1\Big(\m_2|^0\s_1-\s_1\m_2|^0\Big).
\end{eqnarray}
In the second step we replaced $s_1|_{-1}$ with $s_1-\mathbb{I}|^0s_1$ and replaced $\mathbb{I}|^2s_1|_{-1}$ with $s_1\mathbb{I}|^0$; in the third step we used $[\s_1,\m_2]=0$; in the final step we noted that in both terms the star product will only multiply NS states. This establishes \eq{appsusy5} and supersymmetry of the constraint $A_\infty$ structure.

\end{appendix}

\end{document}